\documentclass[11pt]{article}
\pdfoutput=1
\usepackage{jheppub}
\usepackage{empheq}
\usepackage{physics}
\usepackage{amsmath}
\usepackage{footmisc}
\usepackage{mathtools}
\usepackage{amssymb}

\newcommand\fft[2]{{\frac{#1}{#2}}}

\newcommand\nn{\nonumber}

\makeatletter
\newcommand*{\rom}[1]{\expandafter\@slowromancap\romannumeral #1@}
\makeatother

\renewcommand{\Re}{\operatorname{Re}}
\renewcommand{\Im}{\operatorname{Im}}

\begin{document}

\preprint{LCTP-21-20}

\title{\boldmath The topologically twisted index of $\mathcal N=4$ SU($N$) Super-Yang-Mills theory and a black hole Farey tail}

\author{Junho Hong}
\affiliation{Leinweber Center for Theoretical Physics, Randall Laboratory of Physics,\\
The University of Michigan, Ann Arbor, MI 48109-1040, USA}
\emailAdd{junhoh@umich.edu}

\abstract{We investigate the large-$N$ asymptotics of the topologically twisted index of $\mathcal N=4$ SU($N$) Super-Yang-Mills (SYM) theory on $T^2\times S^2$ and provide its holographic interpretation based on the black hole Farey tail \cite{Dijkgraaf:2000fq}. In the field theory side, we use the Bethe-Ansatz (BA) formula, which gives the twisted index of $\mathcal N=4$ SYM theory as a discrete sum over Bethe vacua, to compute the large-$N$ asymptotics of the twisted index. In a dual $\mathcal N=2$ gauged STU model, we construct a family of 5d extremal solutions uplifted from the 3d black hole Farey tail, and compute the regularized on-shell actions. The gravitational partition function given in terms of these regularized on-shell actions is then compared with a canonical partition function derived from the twisted index by the inverse Laplace transform, in the large-$N$ limit. This extends the previous microstate counting of an AdS$_5$ black string by the twisted index and thereby improves holographic understanding of the twisted index.}

\maketitle \flushbottom

\section{Introduction}
Microscopic origin of black hole entropy has been one of the most important questions in quantum gravity since Bekenstein and Hawking found that a black hole has macroscopic entropy proportional to the area of its event horizon \cite{Bekenstein:1973ur,Hawking:1974sw}. In this context, microstates associated with the entropy of an asymptotically flat $\fft14$-BPS black hole in 5d $\mathcal N=4$ supergravity have been successfully addressed by counting supersymmetric ground states of the corresponding $D$-brane world volume theory in the pioneering work \cite{Strominger:1996sh}. For black holes in asymptotically Anti-de Sitter (AdS) backgrounds, the first successful microstate counting has been done by investigating asymptotic symmetries of the near-horizon limit of a 3d Ba\~nados-Teitelboim-Zanelli (BTZ) black hole \cite{Banados:1992wn,Strominger:1997eq}.

Recently, more systematic approach towards a microstate counting of black holes has been introduced for higher dimensional supersymmetric AdS black holes based on the AdS/CFT correspondence. In this approach initiated in \cite{Benini:2015eyy}, the Bekenstein-Hawking entropy of a supersymmetric AdS black hole is matched with the logarithm of the micro-canonical partition function\,(=the number of microstates) of a dual superconformal field theory (SCFT) on the conformal boundary of the bulk, in the large-$N$ limit where $N$ denotes the rank of the gauge group. Precise matching between the two quantities, namely the black hole entropy and the logarithm of the micro-canonical partition function of a dual SCFT, implies that microstates associated with the entropy of a supersymmetric AdS black hole correspond to the ensemble of BPS states in a dual SCFT counted by a micro-canonical partition function.

In the above mentioned black hole microstate counting, one of the key techniques is supersymmetric localization \cite{Pestun:2007rz}. Thanks to this powerful technique, grand-canonical partition functions of various SCFTs could be computed \cite{Benini:2015noa} and then used to derive micro-canonical partition functions that count microstates associated with dual AdS black holes as explained above: to be specific, the logarithm of a grand-canonical partition function is Legendre transformed with respect to chemical potentials to yield the logarithm of a micro-canonical partition function that matches dual AdS black hole entropy (this process is called $I$-extremization \cite{Benini:2015eyy,Hosseini:2017mds}). In particular, grand-canonical partition functions of various SCFTs defined on curved manifolds with a topological twist have been studied extensively in this context of black hole microstate counting, since conformal boundaries of numerous AdS black holes, where dual SCFTs live, are given as curved manifolds with a topological twist. These grand-canonical partition functions of topologically twisted SCFTs are dubbed as a topologically twisted index \cite{Benini:2015noa}.

As a first concrete example, the topologically twisted index of 3d ABJM theory on $S^2\times S^1$ has been computed in the large-$N$ limit. Then the micro-canonical partition function, which was derived from the twisted index by the $I$-extremization described above, successfully counted microstates associated with the entropy of a dual $\fft14$-BPS magnetic AdS$_4$ black hole in 4d $\mathcal N=2$ gauged STU model \cite{Benini:2015eyy}. See \cite{Benini:2016rke,Benini:2016hjo,Cabo-Bizet:2017jsl,Hosseini:2017fjo,Benini:2017oxt,Hosseini:2016ume,Hosseini:2016tor} and \cite{Hosseini:2018uzp,Fluder:2019szh,Suh:2018tul,Suh:2018szn} for other examples in AdS$_4$/CFT$_3$ and AdS$_6$/CFT$_5$ respectively.

Of particular interest is the topologically twisted index of $\mathcal N=4$ SU($N$) Super-Yang-Mills (SYM) theory on $T^2\times S^2$ \cite{Benini:2015noa,Honda:2015yha}. The twisted index of $\mathcal N=4$ SYM theory was first computed in the Cardy-like limit where the modular parameter of the torus $T^2$, namely $\tau$, is sent to
\begin{equation}
	|\tau|\to0^+\quad\text{with fixed}\quad\arg\tau\in(0,\pi),\label{Cardy}
\end{equation}
by the Bethe-Ansatz (BA) formula that gives the index as a discrete sum over Bethe vacua \cite{Hosseini:2016cyf}. The resulting Cardy-like asymptotics of the twisted index was then matched in the large-$N$ limit with the central charge associated with the AdS$_3\times S^2$ near-horizon geometry of a dual magnetic AdS$_5$ black string in 5d $\mathcal N=2$ gauged STU model. In this first investigation, however, the twisted index was not used to count microstates associated with the entropy of a dual AdS$_5$ black string. It is later in \cite{Hosseini:2019lkt,Hosseini:2018tha,Zaffaroni:2019dhb,Hosseini:2020vgl} that the Legendre transform of the logarithm of the twisted index ($I$-extremization) was matched in the large-$N$ \underline{after} the Cardy-like limit with the entropy of a 4d black hole obtained by the Kaluza-Klein (KK) compactification of a dual AdS$_5$ black string along the string direction \cite{Hristov:2014eza}. Since the entropy of a 4d black hole from the KK compactification is identical to that of an original AdS$_5$ black string \cite{Hong:2021dja}, this shows that the twisted index of $\mathcal N=4$ SYM theory can be used to count microstates associated with the entropy of a dual AdS$_5$ black string through $I$-extremization. Refer to \cite{David:2021qaa} for a recent analysis of $\log N$ corrections, which are sub-leading to $N^2$-leading order contributions discussed in the literature \cite{Hosseini:2016cyf,Hosseini:2019lkt,Hosseini:2018tha,Zaffaroni:2019dhb,Hosseini:2020vgl}. 

In this paper, we explore the twisted index of $\mathcal N=4$ SU($N$) SYM theory and its holography in the large-$N$ limit with a finite modular parameter $\tau$ instead, which is a more appropriate parameter regime in the context of AdS/CFT correspondence. In due process, we will address the following two issues overlooked in the Cardy-like limit.

First, the BA formula for the twisted index of $\mathcal N=4$ SYM theory on $T^2\times S^2$ contains multiple contributions to the twisted index with the SL(2,$\mathbb Z$) modular structure \cite{Hong:2018viz}, whose holographic duals have not yet been fully understood. The black hole microstate counting in \cite{Hosseini:2019lkt,Hosseini:2018tha,Zaffaroni:2019dhb,Hosseini:2020vgl} has been done by focusing on a particular contribution among them, which is expected to be dominant in the Cardy-like limit. Complete holographic understanding of the twisted index in the pure large-$N$ limit with a finite modular parameter $\tau$ is therefore still an open problem. Note that a similar problem for the superconformal index of $\mathcal N=4$ SYM theory has recently been addressed by considering orbifold solutions in the gravity side \cite{Aharony:2021zkr}, whose field theory duals have been investigated under the generalized Cardy-like limit where a chemical potential approaches a rational number \cite{ArabiArdehali:2021nsx}.

We resolve this first issue by providing a family of gravity duals for multiple contributions to the BA formula of the twisted index. This family of gravity duals are obtained by imposing the extremal limit on the SL(2,$\mathbb Z$) family of black holes \cite{Maldacena:1998bw} and then uplifting it to extremal solutions of 5d $\mathcal N=2$ gauged STU model. The SL(2,$\mathbb Z$) family of black holes was dubbed as a black hole Farey tail in \cite{Dijkgraaf:2000fq}. 

Another important remark is that the holographic relation between the twisted index of $\mathcal N=4$ SYM theory on $T^2\times S^2$ and a dual AdS$_5$ black string entropy has been investigated by treating a modular parameter of the torus $T^2$, $\tau$, as a chemical potential associated with the momentum along the string direction \cite{Hosseini:2019lkt,Hosseini:2018tha,Zaffaroni:2019dhb,Hosseini:2020vgl}. This means, in the step of $I$-extremization where the logarithm of a twisted index is Legendre transformed with respect to chemical potentials, a modular parameter $\tau$ is fixed in terms of a string momentum. Consequently, the logarithm of a micro-canonical partition function is matched with a dual black string entropy in micro-canonical ensemble with a fixed $\tau$. Therefore one cannot explore holographic duals of interesting modular properties of the twisted index involving $T$ and $S$ transformations of a modular parameter $\tau$ \cite{Hong:2018viz} in this approach. It is then natural to ask if the twisted index and its holographic dual can be compared with each other in canonical ensemble where $\tau$ is treated as a free modular parameter.

We tackle this second issue by evaluating the regularized on-shell action $S^E_\text{reg}$ of the aforementioned family of gravity duals as a function of a modular parameter $\tau$, rather than computing the entropy in terms of a string momentum following \cite{Hosseini:2019lkt,Hosseini:2018tha,Zaffaroni:2019dhb,Hosseini:2020vgl}. The gravitational partition function is then approximated as a sum of $\exp[-S^E_\text{reg}]$ over these gravity duals. The resulting gravitational partition function matches the canonical partition function derived from the twisted index by the inverse Laplace transform treating $\tau$ as a free modular parameter. 

This paper is organized as follows. In the remaining part of introduction, we summarize the main results. In section \ref{sec:TTI}, we review the BA formula of the topologically twisted index of $\mathcal N=4$ SU($N$) SYM theory on $T^2\times S^2$, and compute the large-$N$ asymptotics of the twisted index based on it. Then we derive a canonical partition function from the twisted index by the inverse Laplace transform treating $\tau$ as a free modular parameter of the torus $T^2$. In section \ref{sec:BS}, we construct a family of 5d extremal solutions in a dual $\mathcal N=2$ gauged STU model and compute the corresponding regularized on-shell actions. A gravitational partition function is then easily approximated in the large-$N$ limit. In section \ref{sec:compare}, we compare the canonical partition function from the field theory side with the gravitational partition function in the large-$N$ limit, and thereby improve holographic understanding of the twisted index of $\mathcal N=4$ SU($N$) SYM theory.

\subsection*{Summary of the main results}
The BA formula gives the topologically twisted index of $\mathcal N=4$ SU($N$) SYM theory on $T^2\times S^2$ as a sum over Bethe vacua, or solutions to the Bethe Ansatz Equations (BAE). The result reads
\begin{equation}
	\mathcal Z(\tau,\Delta_a,\mathfrak n_a)=\sum_{n|N}\sum_{r=0}^{n-1}Z_{\{m,n,r\}}(\tau,\Delta_a,\mathfrak n_a)+\mathcal Z_\text{non-st}(\tau,\Delta_a,\mathfrak n_a),\label{index:BA}
\end{equation}
where $\tau$ is a modular parameter of $T^2$ and $\Delta_a,\mathfrak n_a$ are chemical potentials and magnetic charges associated with flavor symmetries respectively. $Z_{\{m,n,r\}}$ denotes the contribution from a `standard' BAE solution labelled by three integers $\{m,n,r\}$ with $N=mn$ and $r\in\{0,1,\cdots,n-1\}$ and $\mathcal Z_\text{non-st}$ denotes the contribution from all the other `non-standard' BAE solutions. Refer to subsection \ref{sec:TTI:BA} for details. In the following subsections, we focus on a standard contribution and leave a conjecture for a non-standard one, which has not yet been investigated thoroughly. 

In the large-$N$ limit, we found the large-$N$ asymptotics of a standard contribution $Z_{\{m,n,r\}}$ for generic complex chemical potentials $\Delta_a\in\mathbb C$, which is computed for the first time to the best of our knowledge. For real chemical potentials $\Delta_a\in\mathbb R$, the results are simplified as
\begin{equation}
\begin{split}
	&\log Z_{\{m,n,r\}}(\tau,\Delta_a,\mathfrak n_a)\\
	&\simeq\begin{cases}
		\fft{N^2\pi i}{n}\sum_{a=1}^3(1-\mathfrak n_a)\Delta_a& (m\neq\mathcal O(N^0))\\
		-\fft{N^2\pi i}{(mp)^2(\tau+s)}\sum_{a=1}^3(1-\mathfrak n_a)\{mp\Delta_a\}(1-\{mp\Delta_a\}) & (m=\mathcal O(N^0),~p=o(N))
	\end{cases}
\end{split}
\end{equation}
upto sub-leading orders of the form $o(N^2)$, ignoring pure imaginary terms defined modulo $2\pi i\mathbb Z$. Note that the large-$N$ asymptotics of $\log Z_{\{m,n,r\}}$ is given for an infinite $m\neq\mathcal O(N^0)$ and for a finite $m=\mathcal O(N^0)$ separately. For the case with a finite $m$, a finite fraction $s$ is chosen to minimize an integer $p$ determined by the relation $\fft{r-ms}{n}=\fft{q}{p}$ with $\gcd(p,q)=1$. Refer to subsection \ref{sec:TTI:large-N} for details, in particular for the large-$N$ asymptotics with generic complex chemical potentials $\Delta_a\in\mathbb C$.

The canonical partition function is then given from the twisted index by the inverse Laplace transform in the large-$N$ limit as
\begin{equation}
\begin{split}
	\Omega(\tau,\mathfrak n_a)&=\int_0^1(\prod_{a=1}^3d\Re\Delta_a)\delta(\sum_{a=1}^3\Delta_a)\mathcal Z(\tau,\Delta_a,\mathfrak n_a)\\
	&=\sum_{c\in\mathbb N}\sum_{d\in\mathbb Z}\exp[\fft{N^2\pi i}{4c^2(\tau+d/c)}\fft{\mathfrak n_1\mathfrak n_2\mathfrak n_3}{1-\mathfrak n_1\mathfrak n_2-\mathfrak n_2\mathfrak n_3-\mathfrak n_3\mathfrak n_1}+o(N^2)]+\Omega_\text{non-st}'(\tau,\mathfrak n_a),
\end{split}\label{Omega:field}%
\end{equation}
where we have treated $\tau$ as a free modular parameter. Refer to subsection \ref{sec:TTI:micro} for details, in particular we explain there how the large-$N$ asymptotics of various contributions to the twisted index through the BA formula (\ref{index:BA}) can be labelled by a set of integers $(c,d)\in\mathbb N\times\mathbb Z$ rather than by three integers $\{m,n,r\}$.

The family of extremal solutions to the BPS equations of 5d $\mathcal N=2$ gauged STU model (\ref{N=2:sugra:BPS}), whose element is labelled by $(c,d)\in\mathbb N\times\mathbb Z$ with $\gcd(c,d)=1$, is given as
\begin{equation}
\begin{split}
	ds^2&=\left(\fft{8p^1p^2p^3\Pi}{\Theta^3}\right)^\fft23\left(-\fft14(r^2-1)dt'{}^2+\rho_+^2\left(d\phi'-\fft{r-1}{2\rho_+}dt'\right)^2+\fft{dr^2}{4(r^2-1)}\right)\\
	&\quad+\left(\fft{(p^1p^2p^3)^2}{\Pi}\right)^\fft13e^{2h_{\mathfrak g}(x,y)}(dx^2+dy^2),\\
	A^I&=-p^I\omega_{\mathfrak g}\quad\to\quad F^I=-p^Ie^{2h_{\mathfrak g}(x,y)}dx\wedge dy,\\
	X^I&=\fft{p^I(p^1+p^2+p^3-2p^I)}{(p^1p^2p^3\Pi)^\fft13},\\
	-\kappa&=p^1+p^2+p^3,
\end{split}\label{family:sol:repeat}
\end{equation}
where $\rho_+=\fft{i}{2(c\tau+d)}$ and $(t',\phi')$ coordinates are periodic as
\begin{equation}
	(t',\phi')\sim(t',\phi')+2\pi(0,1)\sim(t',\phi')-2\pi(i/c,d/c).
\end{equation}
This family of extremal solutions is constructed based on the black hole Farey tail \cite{Maldacena:1998bw,Dijkgraaf:2000fq,Murthy:2009dq}, and expected to arise in the near-horizon limit of a supersymmetric magnetically charged AdS$_5$ black string. Refer to subsections \ref{sec:BS:STU} and \ref{sec:BS:sl2z} for details.

The regularized Euclidean on-shell actions of the above extremal solutions (\ref{family:sol:repeat}) read
\begin{equation}
	S^E_\text{reg}=-\fft{\pi^2 i}{c^2(\tau+d/c)G^{(5)}_N}\fft{p^1p^2p^3}{\Theta}
\end{equation}
for $\Sigma_{\mathfrak g}=S^2~(\mathfrak g=0)$. The gravitational partition function is then approximated in the large-$N$ limit (or in the small Newton's constant limit) as
\begin{equation}
	\mathcal I(\tau,p^a)=\sum_{c\in\mathbb N}\sum_{d\in\mathbb Z}^{\gcd(c,d)=1}\exp[\fft{\pi^2 i}{c^2(\tau+d/c)G^{(5)}_N}\fft{p^1p^2p^3}{\Theta}+\mathcal O((G^{(5)}_N)^0)].\label{I:gr}
\end{equation}
Refer to subsection \ref{sec:BS:on-shell} for details. 

Finally, the gravitational partition function (\ref{I:gr}) is compared with the canonical partition function in the field theory side (\ref{Omega:field}) under the AdS/CFT dictionary provided in section \ref{sec:compare}.

\section{Topologically twisted index on $T^2\times S^2$}\label{sec:TTI}
In \cite{Benini:2015noa,Honda:2015yha}, the path integral for the grand-canonical partition function of $\mathcal N=4$ SU($N$) SYM theory on $T^2\times S^2$ with a topological twist on $S^2$, dubbed as the topologically twisted index, was reduced to a matrix integral via supersymmetric localization \cite{Pestun:2007rz}. The resulting matrix integral was then rewritten by the BA formula, which simplifies the matrix integral for the twisted index into a discrete sum over Bethe vacua. Based on the BA formula, the twisted index of $\mathcal N=4$ SYM theory has been studied mainly in the Cardy-like limit (\ref{Cardy}), where the relation with a dual central charge \cite{Hosseini:2016cyf} and a dual black hole entropy \cite{Hosseini:2019lkt,Hosseini:2020vgl} have been addressed.

In this section, we investigate the large-$N$ asymptotics of the twisted index of $\mathcal N=4$ SU($N$) SYM theory on $T^2\times S^2$ instead, which is more appropriate to be compared with holographic duals in the context of AdS/CFT correspondence. To begin with, in subsection \ref{sec:TTI:BA}, we briefly review the BA formula of the twisted index. In subsection \ref{sec:TTI:large-N}, we explore the large-$N$ asymptotics of the twisted index based on the BA formula. In subsection \ref{sec:TTI:micro}, we derive a canonical partition function from the twisted index in the large-$N$ limit by the inverse Laplace transform treating $\tau$ as a free modular parameter, which will then be identified with a dual gravitational partition function in section \ref{sec:compare}.

\subsection{The Bethe Ansatz formula}\label{sec:TTI:BA}
The matrix integral for the twisted index of $\mathcal N=4$ SU($N$) SYM theory on $T^2\times S^2$ is given as \cite{Benini:2015noa,Honda:2015yha,Hosseini:2016cyf}
\begin{equation}
\begin{split}
	\mathcal Z(\tau,\Delta_a,\mathfrak n_a)=\fft{\mathcal A}{N!}\sum_{\{\mathfrak m_i\}\in\mathbb Z^N}^{\sum_i\mathfrak m_i=0}\int_{\mathcal C}(\prod_{i=1}^{N-1}\fft{dz_i}{2\pi iz_i})\prod_{i,j=1\,(i\neq j)}^N\left[\fft{\theta_1(u_{ij};\tau)}{i\eta(\tau)}\prod_{a=1}^3\left(\fft{i\eta(\tau)}{\theta_1(u_{ij}+\Delta_a;\tau)}\right)^{\mathfrak m_{ij}-\mathfrak n_a+1}\right],\label{TTI:matrix}
\end{split}
\end{equation}
where $z_i=e^{2\pi i u_i}$, $u_{ij}\equiv u_i-u_j$, $\mathfrak m_{ij}=\mathfrak m_i-\mathfrak m_j$, and we have defined the prefactor $\mathcal A$ as
\begin{equation}
	\mathcal A=\eta(\tau)^{2(N-1)}\prod_{a=1}^3\left(\fft{i\eta(\tau)}{\theta_1(\Delta_a;\tau)}\right)^{(N-1)(1-\mathfrak n_a)}.\label{prefactor}
\end{equation}
Refer to Appendix \ref{App:elliptic} for definitions of elliptic functions and their properties. In (\ref{TTI:matrix}), $\tau$ is the modular parameter of a torus $T^2$. Chemical potentials $\Delta_a$ and magnetic fluxes $\mathfrak n_a$ are associated with flavor symmetries of the theory and satisfy the constraints
\begin{equation}
	\sum_{a=1}^3\Delta_a\in\mathbb Z,\qquad \sum_{a=1}^3\mathfrak n_a=2,\label{constraint:chemical}
\end{equation}
where the former is from the invariance of a superpotential under flavor symmetries and the latter is from supersymmetry. The holonomies of a gauge field in the vector multiplet along both temporal and spatial cycles of $T^2$, namely
\begin{equation}
	u_i=2\pi\oint_{A\text{-cycle}}A_i-2\pi\tau\oint_{B\text{-cycle}}A_i,
\end{equation}
satisfy the SU($N$) constraint
\begin{equation}
	\sum_{i=1}^Nu_i\in\mathbb Z.\label{constraint:SU(N)}
\end{equation}
They parametrize the localization locus together with gauge magnetic fluxes $\mathfrak m_i$, which explains why the formula (\ref{TTI:matrix}) contains both an integral over holonomies and a sum over gauge magnetic fluxes. The integration contour $\mathcal C$ for holonomies is chosen to yield Jefferey-Kirwan (JK) residues at a certain subset of poles of 1-loop contribution from chiral multiplets in the SYM theory. See \cite{Honda:2015yha} for more details and \cite{Benini:2013nda,Benini:2013xpa} for previous application of JK residues in the matrix model of 2d elliptic genera.

The matrix integral form of the twisted index (\ref{TTI:matrix}) can be rewritten by introducing a contour integral for $w$ as \cite{Hosseini:2016cyf}
\begin{equation}
	\begin{split}
		\mathcal Z(\tau,\Delta_a,\mathfrak n_a)&=\fft{\mathcal A}{N!}\int_{\mathcal B}\fft{dw}{2\pi iw}\int_{\mathcal C}(\prod_{i=1}^{N-1}\fft{dz_i}{2\pi iz_i})(\prod_{i=1}^N\fft{Q_i^M}{Q_i-1})\\
		&\kern8em\times\prod_{i,j=1\,(i\neq j)}^N\left[\fft{\theta_1(u_{ij};\tau)}{i\eta(\tau)}\prod_{a=1}^3\left(\fft{i\eta(\tau)}{\theta_1(u_{ij}+\Delta_a;\tau)}\right)^{1-\mathfrak n_a}\right]\label{TTI:sum}
	\end{split}
\end{equation}
with a large positive integer $M$ and the integration contour $\mathcal B$ encircling the origin $w=0$. Here the BA operator $Q_i$ is defined as
\begin{equation}
	Q_i(\{u_j\};\tau,\Delta_a)\equiv e^{2\pi i\lambda}\prod_{j=1}^N\prod_{a=1}^3\fft{\theta_1(u_{ji}+\Delta_a;\tau)}{\theta_1(u_{ij}+\Delta_a;\tau)},\label{twisted:Q}
\end{equation}
where we have introduced a parameter $\lambda$ as $w=e^{2\pi i\lambda}$. A set of holonomies $\{u_j\}$ is a shorthand notation of $\{u_j\,|\,j=1,\cdots N\}$. The integral in (\ref{TTI:sum}) can be evaluated by picking residues of the integrand determined by the Bethe-Ansatz Equations (BAE), namely
\begin{equation}
	Q_i(\{u_j\};\tau,\Delta_a)=1.\label{BAE}
\end{equation}
The twisted index (\ref{TTI:sum}) then reduces to a sum over Bethe vacua, or BAE solutions, as
\begin{equation}
	\mathcal Z(\tau,\Delta_a,\mathfrak n_a)=\sum_{\{u_i\}\in\text{BAE}}\fft{\mathcal A}{\mathcal H(\{u_j\};\tau,\Delta_a)}\prod_{i,j=1\,(i\neq j)}^N\prod_{a=1}^3\left(\fft{\theta_1(u_{ij};\tau)}{\theta_1(u_{ij}+\Delta_a;\tau)}\right)^{1-\mathfrak n_a}\label{TTI:BA}
\end{equation}
in terms of the prefactor (\ref{prefactor}) and the Jacobian
\begin{equation}
\begin{split}
	H(\{u_i\};\tau,\Delta_a)&\equiv\fft{1}{2\pi i}\left(\fft{\partial(Q_1,\cdots Q_N)}{\partial(u_1,\cdots,u_{N-1},\lambda)}\right),\\
	\mathcal H(\{u_j\};\tau,\Delta_a)&\equiv\det[H(\{u_j\};\tau,\Delta_a)].
\end{split}\label{eq:H}
\end{equation}
We call (\ref{TTI:BA}) the BA formula of the twisted index of $\mathcal N=4$ SU($N$) SYM theory on $T^2\times S^2$. Note that the factor of $N!$ in the denominator of (\ref{TTI:sum}) is canceled by identifying BAE solutions equivalent under the permutation of holonomies $\{u_i\}$.

The BA formula (\ref{TTI:BA}) is incomplete, however, for the following two reasons:
\begin{itemize}
	\item The BA formula (\ref{TTI:BA}) does not take continuous family of BAE solutions into account. Note that the BA formula (\ref{TTI:BA}) is derived by applying the Cauchy's integral formula to (\ref{TTI:sum}) under the assumption that all the solutions to the BAE (\ref{BAE}) are isolated. As it has been observed in the literature \cite{ArabiArdehali:2019orz,Lezcano:2021qbj,Benini:2021ano}, however, the BAE (\ref{BAE}) have continuous family of solutions and therefore the BA formula (\ref{TTI:BA}) has to be refined to incorporate their contributions. 
	\item The BA formula (\ref{TTI:BA}) does not clarify `relevant' BAE solutions that contribute to the twisted index. Note that the discrete sum in (\ref{TTI:BA}) should be taken for solutions to the BAE (\ref{BAE}) enclosed by the contour $\mathcal C$ only. The contour $\mathcal C$ is introduced rather implicitly in (\ref{TTI:matrix}) to capture the JK residues, however, and therefore it is not trivial to decide if a given BAE solution truly contributes to the twisted index through the BA formula (\ref{TTI:BA}). 
\end{itemize}
To circumvent the first issue about continuous family of BAE solutions, in this paper we focus on the contribution from isolated BAE solutions to the twisted index so that the BA formula (\ref{TTI:BA}) can be used without modifications. In particular, we focus on \underline{standard} BAE solutions \cite{Hong:2018viz,ArabiArdehali:2019orz,Lezcano:2021qbj}, which we will review in the remaining part of this subsection. Then, to take care of the second issue about `relevant' BAE solutions that contribute to the twisted index, we assume that any standard BAE solution is enclosed by the contour $\mathcal C$ in (\ref{TTI:matrix}) and therefore contributes to the twisted index. We leave more complete analysis of the BA formula that resolves the above mentioned issues for future research.

\subsubsection*{Standard BAE solutions}
The first step to compute the twisted index using the BA formula (\ref{TTI:BA}) is to find the most general solutions to the BAE (\ref{BAE}). Since the BAE is a highly involved system of transcendental equations, however, it is difficult to derive its general solutions. Hence we focus on a particular set of BAE solutions found in \cite{Hong:2018viz} and dubbed as \underline{standard} solutions in \cite{ArabiArdehali:2019orz,Lezcano:2021qbj}, namely
\begin{equation}
	\{u_i\,|\,i=1,\cdots,N\}_{\{m,n,r\}}=\left\{u_{\hat j\hat k}=\bar u+\fft{\hat j}{m}+\fft{\hat k}{n}(\tau+\fft{r}{m})\,\bigg|\,(\hat j,\hat k)\in\mathbb Z_m\times\mathbb Z_n\right\}\label{BAE:sol:st}
\end{equation}
labelled by three integers $\{m,n,r\}$ where $m$ and $n$ are positive divisor of $N$ such that $N=mn$ and $r\in\{0,\cdots,n-1\}$. In (\ref{BAE:sol:st}), $\bar u$ is introduced to satisfy the SU($N$) constraint (\ref{constraint:SU(N)}) and we have also introduced a double-index notation
\begin{equation}
	u_{n\hat j+\hat k}\quad\to\quad u_{\hat j\hat k}.\label{double:index}
\end{equation}
with the identification $u_N=u_0$. Each standard BAE solution $\{u_i\}_{\{m,n,r\}}$ distributes holonomies $u_i$'s evenly over the torus $T^2$ and thereby define a freely acting orbifold $T^2/\mathbb Z_m\times\mathbb Z_n$. All the other solutions to the BAE (\ref{BAE}) are called \underline{non-standard} \cite{ArabiArdehali:2019orz,Lezcano:2021qbj}.

The contribution from various standard BAE solutions to the twisted index through the BA formula (\ref{TTI:BA}) has been written explicitly in \cite{Hong:2018viz}. The result can be summarized as
\begin{subequations}
\begin{align}
	\mathcal Z(\tau,\Delta_a,\mathfrak n_a)&=\mathcal Z_\text{st}(\tau,\Delta_a,\mathfrak n_a)+\mathcal Z_\text{non-st}(\tau,\Delta_a,\mathfrak n_a),\label{TTI:BA:total}\\
	\mathcal Z_\text{st}(\tau,\Delta_a,\mathfrak n_a)&=\sum_{n|N}\sum_{r=0}^{n-1}Z_{\{m,n,r\}}(\tau,\Delta_a,\mathfrak n_a),\label{TTI:BA:st}
\end{align}\label{TTI:BA:split}%
\end{subequations}
where $\mathcal Z_{\{m,n,r\}}(\tau,\Delta_a,\mathfrak n_a)$ stands for the contribution from a particular standard BAE solution labelled by $\{m,n,r\}$ (\ref{BAE:sol:st}) in the BA formula (\ref{TTI:BA}), namely \cite{Hong:2018viz}
\begin{equation}
\begin{split}
	Z_{\{m,n,r\}}(\tau,\Delta_a,\mathfrak n_a)&=\fft{\mathcal A}{\mathcal H(\{u_j\};\tau,\Delta_a)}\prod_{i,j=1\,(i\neq j)}^N\prod_{a=1}^3\left(\fft{\theta_1(u_{ij};\tau)}{\theta_1(u_{ij}+\Delta_a;\tau)}\right)^{1-\mathfrak n_a}\bigg|_{\{u_i\}=\{u_i\}_{\{m,n,r\}}}\\
	&=\fft{i^{N-1}}{\mathcal H_{\{m,n,r\}}(\tau,\Delta_a)}\prod_{a=1}^3\left[\left(\fft{\theta_1(\Delta_a;\tau)}{\eta(\tau)^3}\right)\left(\fft{m\eta(\tilde\tau)^3}{\theta_1(m\Delta_a;\tilde\tau)}\right)^N\right]^{1-\mathfrak n_a}.
\end{split}\label{TTI:BA:mnr}
\end{equation}
Here we have introduced the modular parameter of a freely acting orbifold $T^2/\mathbb Z_m\times\mathbb Z_n$ as
\begin{equation}
	\tilde\tau=\fft{m\tau+r}{n}.\label{lattice:tau}
\end{equation}
We have also introduced a shorthand notation for the Jacobian determinant of a standard BAE solution as $\mathcal H_{\{m,n,r\}}(\tau,\Delta_a)\equiv\mathcal H(\{u_i\}_{\{m,n,r\}};\tau,\Delta_a)$.

Following \cite{Hong:2018viz}, one can rewrite the Jacobian determinant $\mathcal H_{\{m,n,r\}}$ in (\ref{TTI:BA:mnr}) by factoring out its universal diagonal element. We start from the matrix elements of the Jacobian (\ref{eq:H}) of a standard BAE solution (\ref{BAE:sol:st}), namely ($\mu,\nu\in\{1,\cdots,N-1\}$)
\begin{subequations}
\begin{align}
	[H_{\{m,n,r\}}]_{\mu,\nu}=\fft{1}{2\pi i}\fft{\partial Q_\mu}{\partial u_\nu}&=\delta_{\mu\nu}\sum_{\hat j=0}^{m-1}\sum_{\hat k=0}^{n-1}\mathcal G(\hat j_\mu-\hat j,\hat k_\mu-\hat k;\tau,\Delta_a)+\mathcal G(\hat j_\mu,\hat k_\mu;\tau,\Delta_a)\nn\\
	&\quad-\mathcal G(\hat j_\mu-\hat j_\nu,\hat k_\mu-\hat k_\nu;\tau,\Delta_a),\label{eq:H:element:N-1}\\
	[H_{\{m,n,r\}}]_{N,\nu}=\fft{1}{2\pi i}\fft{\partial Q_N}{\partial u_\nu}&=-\sum_{\hat j=0}^{m-1}\sum_{\hat k=0}^{n-1}\mathcal G(\hat j,\hat k;\tau,\Delta_a)+\mathcal G(0,0;\tau,\Delta_a)\nn\\
	&\quad-\mathcal G(\hat j_\nu,\hat k_\nu;\tau,\Delta_a),\\
	[H_{\{m,n,r\}}]_{\mu,N}=\fft{1}{2\pi i}\fft{\partial Q_\mu}{\partial\lambda}&=1,\\
	[H_{\{m,n,r\}}]_{N,N}=\fft{1}{2\pi i}\fft{\partial\lambda}{\partial u_\nu}&=1,
\end{align}\label{eq:H:element}%
\end{subequations}
where hatted indices are introduced following the double-index notation (\ref{double:index}) as $(\mu,\nu)=(n\hat j_\mu+\hat k_\mu,n\hat j_\nu+\hat k_\nu)$ and we have defined the $\mathcal G$-function as
\begin{equation}
	\mathcal G_{\{m,n,r\}}(\hat j,\hat k;\tau,\Delta_a)\equiv\fft{i}{2\pi}\sum_{a=1}^3\fft{\partial}{\partial\Delta_a}\log\bigg[\theta_1(\Delta_a+\fft{\hat j+\hat k\tilde\tau}{m};\tau)\theta_1(\Delta_a-\fft{\hat j+\hat k\tilde\tau}{m};\tau)\bigg].\label{eq:G}
\end{equation}
The matrix elements (\ref{eq:H:element}) imply that the Jacobian determinant $\mathcal H_{\{m,n,r\}}$ is related to the determinant of an $(N-1)\times(N-1)$ submatrix as
\begin{equation}
	\mathcal H_{\{m,n,r\}}(\tau,\Delta_a)=N\det[\fft{1}{2\pi i}\fft{\partial(Q_1,\cdots,Q_{N-1})}{\partial(u_1,\cdots,u_{N-1})}].\label{eq:H:mnr:1}
\end{equation}
Using the identity \cite{Hong:2018viz}
\begin{equation}
	\sum_{\hat j=0}^{m-1}\sum_{\hat k=0}^{n-1}\mathcal G(\hat j_\mu-\hat j,\hat k_\mu-\hat k;\tau,\Delta_a)=\fft{i}{\pi}\sum_{a=1}^3\fft{\partial}{\partial\Delta_a}\log\theta_1(m\Delta_a;\tilde\tau),
\end{equation}
one can rewrite the Jacobian determinant (\ref{eq:H:mnr:1}) by factoring out a universal diagonal element as
\begin{equation}
	\mathcal H_{\{m,n,r\}}(\tau,\Delta_a)=N\left(\fft{i}{\pi}\sum_{a=1}^3\fft{\partial}{\partial\Delta_a}\log\theta_1(m\Delta_a;\tilde\tau)\right)^{N-1}\det[I_{N-1}+\widetilde H_{\{m,n,r\}}(\tau,\Delta_a)],\label{eq:H:mnr:2}
\end{equation}
where we have defined an $(N-1)\times(N-1)$ square matrix $\widetilde H_{\{m,n,r\}}$ as
\begin{equation}
	[\widetilde H_{\{m,n,r\}}(\tau,\Delta_a)]_{\mu,\nu}\equiv\fft{\mathcal G_{\{m,n,r\}}(\hat j_\mu,\hat k_\mu;\tau,\Delta_a)-\mathcal G_{\{m,n,r\}}(\hat j_\mu-\hat j_\nu,\hat k_\mu-\hat k_\nu;\tau,\Delta_a)}{\fft{i}{\pi}\sum_{a=1}^3\fft{\partial}{\partial\Delta_a}\log\theta_1(m\Delta_a;\tilde\tau)}.
\end{equation}
Substituting the expression (\ref{eq:H:mnr:2}) into (\ref{TTI:BA:mnr}) then gives
\begin{equation}
	Z_{\{m,n,r\}}(\tau,\Delta_a,\mathfrak n_a)=\frac{\prod_{a=1}^3\left[\left(\fft{\theta_1(\Delta_a;\tau)}{\eta(\tau)^3}\right)\left(\fft{m\eta(\tilde\tau)^3}{\theta_1\left(m\Delta_a;\tilde\tau\right)}\right)^N\right]^{1-\mathfrak n_a}}{N\det(I_{N-1}+\widetilde H_{\{m,n,r\}})\left[\fft1\pi\sum_{a=1}^3\fft{\partial}{\partial\Delta_a}\log\theta_1\left(m\Delta_a;\tilde\tau\right)\right]^{N-1}}.\label{TTI:mnr}
\end{equation}
The expressions (\ref{TTI:BA:mnr}) and (\ref{TTI:mnr}) are equivalent to each other and give the contribution to the twisted index from a standard BAE solution -- labelled by $\{m,n,r\}$ as (\ref{BAE:sol:st}) -- through the BA formula (\ref{TTI:BA}) in terms of elliptic functions.

The standard contribution $\mathcal Z_\text{st}$ in (\ref{TTI:BA:total}) has received more attention compared to the non-standard contribution $\mathcal Z_\text{non-st}$ since it can be written explicitly as (\ref{TTI:BA:st}) equipped with (\ref{TTI:BA:mnr}) or equivalently with (\ref{TTI:mnr}) so one can explore its property more concretely. In particular, $\mathcal Z_\text{st}$ has been investigated mainly in the Cardy-like limit (\ref{Cardy}), where the asymptotic expansion of $Z_{\{m,n,r\}}$ (\ref{TTI:BA:mnr}) was computed and a dominant sector $Z_{\{1,N,0\}}$ was compared with an appropriate holographic dual \cite{Hosseini:2016cyf,Hosseini:2019lkt,Hosseini:2020vgl}.

In this paper, we also focus on the standard contribution $\mathcal Z_\text{st}$ (\ref{TTI:BA:st}) but in the large-$N$ limit with a finite $\tau$, since it is more appropriate to be compared with holographic duals in the context of AdS/CFT correspondence as pointed out in the introduction. In the following subsection, we therefore investigate the large-$N$ asymptotics of a particular standard contribution $\mathcal Z_{\{m,n,r\}}$ (\ref{TTI:BA:mnr}), or (\ref{TTI:mnr}), that composes the total standard contribution $\mathcal Z_\text{st}$ through (\ref{TTI:BA:st}).

\subsection{Large-$N$ asymptotics}\label{sec:TTI:large-N}
Here we expand a particular standard contribution to the twisted index, namely $Z_{\{m,n,r\}}$ given in terms of elliptic functions as (\ref{TTI:BA:mnr}) or equivalently as (\ref{TTI:mnr}) in the large-$N$ limit. We will investigate the `basic' sector $Z_{\{1,N,0\}}$ first, and then move on to general sectors $Z_{\{m,n,r\}}$.

\subsubsection{Basic contribution}\label{sec:TTI:large-N:basic}
For the basic sector with $\{m,n,r\}=\{1,N,0\}$, we use the logarithm of (\ref{TTI:mnr})
\begin{equation}
\begin{split}
	\log Z_{\{1,N,0\}}(\tau,\Delta_a,\mathfrak n_a)&=N\sum_{a=1}^3(1-\mathfrak n_a)\log\fft{\eta(\fft{\tau}{N})^3}{\theta_1(\Delta_a;\fft{\tau}{N})}-\log\det[I_{N-1}+\widetilde H_{\{1,N,0\}}(\tau,\Delta_a)]\\
	&\quad-\log N-(N-1)\log(\fft1\pi\sum_{a=1}^3\fft{\partial}{\partial\Delta_a}\log\theta_1(\Delta_a;\fft{\tau}{N}))+\mathcal O(N^0).\label{TTI:1N0:large-N}
\end{split}
\end{equation}
To evaluate the large-$N$ asymptotics of (\ref{TTI:1N0:large-N}), first we derive 
\begin{subequations}
\begin{align}
	\log\fft{\eta(\fft{\tau}{N})^3}{\theta_1(\Delta_a;\fft{\tau}{N})}&=-\fft{N\pi i}{\tau}\{\Delta_a\}_\tau(1-\{\Delta_a\}_\tau)-\pi i\lfloor\Re\Delta_a-\cot(\arg\tau)\Im\Delta_a\rfloor\nn\\
	&\quad-\log\fft{\tau}{N}+\fft{\pi i}{2}+\mathcal O(e^{-\fft{2\pi N\sin(\arg\tau)}{|\tau|}\min[\{\tilde\Delta_a\},1-\{\tilde\Delta_a\}]}),\\
	\fft1\pi\sum_{a=1}^3\fft{\partial}{\partial\Delta_a}\log\theta_1(\Delta_a;\fft{\tau}{N})&=-\fft{iN\eta_{1,0}}{\tau}+\mathcal O(e^{-\fft{2\pi N\sin(\arg\tau)}{|\tau|}\min[\{\tilde\Delta_a\},1-\{\tilde\Delta_a\}]}),\label{1N0:large-N:1:det}
\end{align}\label{1N0:large-N:1}%
\end{subequations}
using the asymptotic expansions of elliptic functions (\ref{eta:asymp}) and (\ref{theta1:asymp}). See (\ref{u:component}) for the definition of the `tilde' component of chemical potentials, namely $\tilde\Delta_a$. For $\partial_{\Delta_a}$ in (\ref{1N0:large-N:1:det}) to be well-defined, we assume generic chemical potentials where $\Re\Delta_a-\cot(\arg\tau)\Im\Delta_a\notin\mathbb Z$. We have also introduced $\eta_{c,s}$ for generic chemical potentials as\footnote{Note $\eta_{c,s}=3$ if $\tilde\Delta_a-\check\Delta_as\in\mathbb Z$ for all $a\in\{1,2,3\}$, but this is not the case for generic chemical potentials.}
\begin{equation}
	\eta_{c,s}\equiv2\sum_{a=1}^3\{c\Delta_a\}_{\tau+s}-3\in\{\pm1\}\qquad(c,s\in\mathbb R)\label{eq:eta}
\end{equation}
from the constraint on chemical potentials $\sum_{a=1}^3\Delta_a\in\mathbb Z$ in (\ref{constraint:chemical}). 

Next we consider the Jacobian determinant in (\ref{TTI:1N0:large-N}). In \cite{GonzalezLezcano:2020yeb}, it has been shown analytically using the Gershgorin Circle Theorem that the Jacobian determinant scales like $\mathcal O(N^0)$ for a finite but small enough $|\tau|$. For a general finite $\tau$, the authors used numerical analysis to support the same scaling
\begin{equation}
	\log\det[I_{N-1}+\widetilde H_{\{1,N,0\}}(\tau,\Delta_a)]=\mathcal O(N^0).\label{1N0:large-N:2}
\end{equation}
We use this result to provide an upper bound to the large-$N$ asymptotics of the Jacobian determinant in (\ref{TTI:1N0:large-N}).

Substituting (\ref{1N0:large-N:1}) and (\ref{1N0:large-N:2}) into (\ref{TTI:1N0:large-N}) then gives the large-$N$ asymptotics of the basic sector $Z_{\{1,N,0\}}$ as
\begin{equation}
\begin{split}
	\log Z_{\{1,N,0\}}(\tau,\Delta_a,\mathfrak n_a)&=-\fft{N^2\pi i}{\tau}\sum_{a=1}^3(1-\mathfrak n_a)\{\Delta_a\}_\tau(1-\{\Delta_a\}_\tau)+\fft{N(1+\eta_{1,0})\pi i}{2}\\
	&\quad-N\pi i\sum_{a=1}^3(1-\mathfrak n_a)\lfloor\Re\Delta_a-\cot(\arg\tau)\Im\Delta_a\rfloor+\mathcal O(N^0).\label{TTI:1N0:large-N:simple}
\end{split}
\end{equation}
Note that linear $N$ sub-leading order terms are pure imaginary in (\ref{TTI:1N0:large-N:simple}) and therefore they affect the phase of the large-$N$ asymptotics of $Z_{\{1,N,0\}}$ at most. 

It is remarkable that the $N^2$-leading order in (\ref{TTI:1N0:large-N:simple}) does not have any sub-leading corrections of the form $\sim N^2\tau^\#$ with a non-negative integer $\#$ in the Cardy-like limit (\ref{Cardy}). This is not obvious from the Cardy-like asymptotics computed in \cite{Hosseini:2016cyf,Hong:2018viz}, which is the same as the first term of (\ref{TTI:1N0:large-N:simple}) with a simple replacement $N^2\to N^2-1$, since they did not keep track of sub-leading orders in the Cardy-like limit. Thanks to this result, the microstate counting of a dual AdS$_4$ black hole from the KK compactification by the twisted index in the Cardy-like limit \cite{Hosseini:2018tha,Zaffaroni:2019dhb,Hosseini:2019lkt,Hosseini:2020vgl} can be reproduced in the large-$N$ limit with a finite $\tau$.

\subsubsection{Other contributions}\label{sec:TTI:large-N:other}
Next we investigate the large-$N$ asymptotics of a general sector $Z_{\{m,n,r\}}$. In this case, we use the logarithm of (\ref{TTI:BA:mnr})
\begin{equation}
	\log Z_{\{m,n,r\}}(\tau,\Delta_a,\mathfrak n_a)=N\sum_{a=1}^3(1-\mathfrak n_a)\log\fft{m\eta(\tilde\tau)^3}{\theta_1(m\Delta_a;\tilde\tau)}-\log\mathcal H_{\{m,n,r\}}(\tau,\Delta_a)+\mathcal O(N).\label{TTI:mnr:large-N:H}
\end{equation}
The Jacobian determinant in (\ref{TTI:mnr:large-N:H}) is bounded from above in the large-$N$ limit as\footnote{One subtle issue is that (\ref{eq:H:mnr}) does not exclude the possibility for the Jacobian determinant to vanish asymptotically in the large-$N$ limit. In that case, particularly when it approaches zero exponentially fast as $\sim e^{-N^2}$, the logarithm of the Jacobian determinant may contribute to the $N^2$-leading order. We assume this is not the case for generic chemical potentials $\Delta_a$.}
\begin{equation}
\begin{split}
	\log\mathcal H_{\{m,n,r\}}&=\log(N\det[\fft{1}{2\pi i}\fft{\partial(Q_1,\cdots,Q_{N-1})}{\partial(u_1,\cdots,u_{N-1})}])\\
	&=\log\relax\bigg(N(2\pi i)^{-(N-1)}\sum_{\sigma\in S_{N-1}}\text{sign}[\sigma](\partial_{u_1}Q_{\sigma(1)})\cdots(\partial_{u_{N-1}}Q_{\sigma(N-1)})\bigg)\\
	&=\mathcal O(\log(N!\times N^{N-1}))=\mathcal O(N\log N),
\end{split}\label{eq:H:mnr}
\end{equation}
where we have used the expression (\ref{eq:H:mnr:1}) in the 1st line and that the matrix element (\ref{eq:H:element:N-1}) scales like $\mathcal O(N)$ at most in the 3rd line. To be specific, since the $\mathcal G$-function (\ref{eq:G}) is of order $\mathcal O(N^0)$, diagonal elements scale like $\mathcal O(N)$ at most and off-diagonal elements are finite in (\ref{eq:H:element:N-1}). This is exactly the same argument used in \cite{Benini:2018ywd} to estimate the upper bound for the Jacobian determinant of the basic $\{1,N,0\}$ standard solution. We have just applied the same logic to the Jacobian determinant of a general $\{m,n,r\}$ standard solution. Substituting (\ref{eq:H:mnr}) into (\ref{TTI:mnr:large-N:H}) then gives
\begin{equation}
	\log Z_{\{m,n,r\}}(\tau,\Delta_a,\mathfrak n_a)=N\sum_{a=1}^3(1-\mathfrak n_a)\log\fft{\eta(\tilde\tau)^3}{\theta_1(m\Delta_a;\tilde\tau)}+\mathcal O(N\log N).\label{TTI:mnr:large-N}
\end{equation}
To extract the $N^2$-leading order from the expression (\ref{TTI:mnr:large-N}), we split various configurations of three integers $\{m,n,r\}$ into two cases based on the finiteness of $m$ in the large-$N$ limit. 

\subsubsection*{Infinite $m$}
First we consider the case where $m$ is infinite in the large-$N$ limit as
\begin{equation}
	\lim_{N\to\infty}\fft{1}{m}=0\quad\Leftrightarrow\quad m\neq\mathcal O(N^0).\label{m:infinite}
\end{equation}
The goal is to figure out the $N^2$-leading order of $\log Z_{\{m,n,r\}}$ (\ref{TTI:mnr:large-N}) in the large-$N$ limit under the condition (\ref{m:infinite}). To begin with, we derive
\begin{equation}
	\fft{\eta(\tilde\tau)^3}{\theta_1(m\Delta_a;\tilde\tau)}=e^{\fft{N\pi i\Delta_a^2}{\tau}}e^{-\fft{m\pi i\tau}{n}\{\fft{n\Delta_a}{\tau}\}_{-\fft1\tau}^2}e^{-\pi ij_a-\pi ij_a^2\fft{r}{n}}\fft{\eta(\tilde\tau)^3}{\theta_1(\fft{m\tau}{n}\{\fft{n\Delta_a}{\tau}\}_{-\fft1\tau}-\fft{r}{n}j_a;\tilde\tau)}\label{m:not-finite:1:shift}
\end{equation}
using the quasi-periodicity (\ref{theta1:periodic}), where the integer $j_a$ is defined as
\begin{equation}
	j_a\equiv\left\lfloor\Re\fft{n\Delta_a}{\tau}-\cot\relax(\arg(-\fft1\tau))\Im\fft{n\Delta_a}{\tau}\right\rfloor=\left\lfloor\fft{\Im n\Delta_a}{\Im\tau}\right\rfloor.\label{ja}
\end{equation}
Substituting the product forms (\ref{def:eta}) and (\ref{def:theta1}) into (\ref{m:not-finite:1:shift}) then gives
\begin{equation}
\begin{split}
	\fft{\eta(\tilde\tau)^3}{\theta_1(m\Delta_a;\tilde\tau)}&=-ie^{\fft{N\pi i\Delta_a^2}{\tau}}e^{\fft{m\pi i\tau}{n}\{\fft{n\Delta_a}{\tau}\}_{-\fft1\tau}(1-\{\fft{n\Delta_a}{\tau}\}_{-\fft1\tau})}e^{-\pi ij_a(1+(j_a+1)\fft{r}{n})}\\
	&\quad\times\prod_{l=1}^\infty\fft{(1-e^{2\pi il\fft{m\tau+r}{n}})^2}{(1-e^{2\pi i((l-1)\fft{m\tau+r}{n}+\fft{m\tau}{n}\{\fft{n\Delta_a}{\tau}\}_{-\fft1\tau}-\fft{r}{n}j_a)})(1-e^{2\pi i(l\fft{m\tau+r}{n}-\fft{m\tau}{n}\{\fft{n\Delta_a}{\tau}\}_{-\fft1\tau}+\fft{r}{n}j_a)})}.\label{m:not-finite:1}
\end{split}
\end{equation}
Using the upper bound for the infinite product term given in (\ref{m:not-finite:1:bound}), one can estimate the logarithm of (\ref{m:not-finite:1}) as
\begin{equation}
\begin{split}
	\log\fft{\eta(\tilde\tau)^3}{\theta_1(m\Delta_a;\tilde\tau)}&=\fft{N\pi i\Delta_a^2}{\tau}+\fft{m\pi i\tau}{n}\{\fft{n\Delta_a}{\tau}\}_{-\fft1\tau}(1-\{\fft{n\Delta_a}{\tau}\}_{-\fft1\tau})-\pi ij_a(1+(j_a+1)\fft{r}{n})\\
	&\quad+\log(-i)+\mathcal O(\fft{n}{m}).\label{m:not-finite:1:large-N}
\end{split}
\end{equation}
Finally, substituting (\ref{m:not-finite:1:large-N}) into (\ref{TTI:mnr:large-N}) gives the large-$N$ asymptotics of a general sector $Z_{\{m,n,r\}}$ under the condition (\ref{m:infinite}) as
\begin{equation}
\begin{split}
	\log Z_{\{m,n,r\}}(\tau,\Delta_a,\mathfrak n_a)&=\fft{N^2\pi i}{\tau}\sum_{a=1}^3(1-\mathfrak n_a)\Delta_a^2+m^2\pi i\tau\sum_{a=1}^3(1-\mathfrak n_a)\{\fft{n\Delta_a}{\tau}\}_{-\fft1\tau}(1-\{\fft{n\Delta_a}{\tau}\}_{-\fft1\tau})\\
	&\quad-N\pi i\sum_{a=1}^3(1-\mathfrak n_a)j_a(1+(j_a+1)\fft{r}{n})+\mathcal O(N\log N,n^2).
\end{split}\label{TTI:mnr:large-N:m:not-finite}
\end{equation}
We can simplify (\ref{TTI:mnr:large-N:m:not-finite}) further as
\begin{empheq}[box=\fbox]{equation}
\begin{split}
	\log Z_{\{m,n,r\}}(\tau,\Delta_a,\mathfrak n_a)&=N\pi i\sum_{a=1}^3(1-\mathfrak n_a)\left(m\Delta_a(1+2j_a)-j_a(1+(j_a+1)\tilde\tau)\right)\\
	&\quad+\mathcal O(N\log N,n^2).
\end{split}\label{TTI:mnr:large-N:m:not-finite:alter}
\end{empheq}
Note that one can determine the $N^2$-leading order from the expression (\ref{TTI:mnr:large-N:m:not-finite}) or (\ref{TTI:mnr:large-N:m:not-finite:alter}) since the last term of order $\mathcal O(N\log N,n^2)$ does not contribute to the $N^2$-leading order for an infinite $m$ satisfying the condition (\ref{m:infinite}). 

\subsubsection*{Finite $m$}
The case with a finite $m$ is more subtle to explore in general. To begin with, we introduce a finite fraction $s$ in terms of which the modular parameter of an orbifold (\ref{lattice:tau}) is written as
\begin{equation}
	\tilde\tau=\fft{m\tau+r}{n}=\fft{m(\tau+s)}{n}+\fft{r-ms}{n}=\fft{m(\tau+s)}{n}+\fft{q}{p},\label{tau:e}
\end{equation}
where we have also introduced relatively prime integers $p$ and $q$ such that $0\leq q<p$. Note that $(s,p)\in\fft{1}{mp}\mathbb Z\times\mathbb N$ is not uniquely determined for a given configuration $\{m,n,r\}$ with a finite $m$ though. 

To fix $(s,p)\in\fft{1}{mp}\mathbb Z\times\mathbb N$ uniquely for a given configuration $\{m,n,r\}$ with a finite $m$, we \emph{choose a finite fraction $s$ that \underline{minimizes} a natural number $p$ determined by the relation (\ref{tau:e})}. In this way, a natural number $p$ is uniquely determined. Then we further assume
\begin{equation}
	\lim_{N\to\infty}\fft{p}{N}=0\quad\Leftrightarrow\quad p=o(N),\label{m:finite}
\end{equation}
which is an extra condition we need to explore the large-$N$ asymptotics of a general sector $Z_{\{m,n,r\}}$ with a finite $m$. Under this assumption (\ref{m:finite}), one can prove that there is a unique finite fraction $s$ that minimizes a natural number $p$ determined by the relation (\ref{tau:e}) for a given configuration $\{m,n,r\}$ with a finite $m$:
\begin{equation}
	\fft{r-ms}{n}=\fft{q}{p}~\&~\fft{r-ms'}{n}=\fft{q'}{p}~~\Rightarrow~~|s-s'|=\fft{n|q-q'|}{mp}\begin{cases}
		=0 & (s=s')\\
		\geq\fft{N}{m^2p}\to\infty & (s\neq s')
	\end{cases},
\end{equation}
where the second case contradicts that $s$ and $s'$ are finite. As a result, $(s,p)\in\fft{1}{mp}\mathbb Z\times\mathbb N$ is uniquely determined for a given configuration $\{m,n,r\}$ with a finite $m$. Several examples are as follows:
\begin{equation}
\begin{split}
	\{m,n,r\}&=\{2,N/2,1\}\\
	\{m,n,r\}&=\{3,N/3,(N-5)/7\}\\
	\{m,n,r\}&=\{1,N,\sqrt N\}
\end{split}
\qquad
\begin{split}
	\to\\\to\\\to
\end{split}
\qquad
\begin{split}
	(s,p,q)&=(1/2,1,0)\\
	(s,p,q)&=(-5/21,7,3)\\
	(s,p,q)&=(0,\sqrt{N},1)
\end{split}~.
\end{equation}

It is subtle to test if, for any given configuration of $\{m,n,r\}$ with a finite $m$, one can always find a finite fraction $s$ such that a natural number $p$ determined by the relation (\ref{tau:e}) satisfies the condition (\ref{m:finite}) in the large-$N$ limit. For simplicity, we focus on sectors $Z_{\{m,n,r\}}$ that allow for such $s$ and leave a complete analysis for future research.

As we have done for the case with an infinite $m$ (\ref{m:infinite}), here we investigate the $N^2$-leading order of $\log Z_{\{m,n,r\}}$ (\ref{TTI:mnr:large-N}) in the large-$N$ limit under the condition (\ref{m:finite}) with a finite $m$. To begin with, using the SL(2,$\mathbb Z$) transformation of elliptic functions (\ref{eq:sl2z}) with
\begin{equation}
	\begin{pmatrix}
		a & b\\
		c & d
	\end{pmatrix}\in\text{SL}(2,\mathbb Z),\qquad (c,d)=(p,-q),
\end{equation}
we derive
\begin{equation}
	\fft{\eta(\tilde\tau)^3}{\theta_1(m\Delta_a;\tilde\tau)}=\fft{n}{mp(\tau+s)}e^\fft{N\pi i\Delta_a^2}{\tau+s}\fft{\eta(-\fft{N}{(mp)^2(\tau+s)}+\fft{a}{p})^3}{\theta_1(\fft{N\Delta_a}{mp(\tau+s)};-\fft{N}{(mp)^2(\tau+s)}+\fft{a}{p})}.\label{m:finite:1:sl2z}
\end{equation}
Using the quasi-periodicity (\ref{theta1:periodic}), (\ref{m:finite:1:sl2z}) can be rewritten as
\begin{equation}
\begin{split}
	\fft{\eta(\tilde\tau)^3}{\theta_1(m\Delta_a;\tilde\tau)}&=(-1)^{k_a}\fft{n}{mp(\tau+s)}e^{\fft{N\pi i}{(mp)^2(\tau+s)}\{mp\Delta_a\}_{\tau+s}^2-\pi ik_a^2\fft{a}{p}}\\
	&\quad\times\fft{\eta(-\fft{N}{(mp)^2(\tau+s)}+\fft{a}{p})^3}{\theta_1(\fft{N}{(mp)^2(\tau+s)}\{mp\Delta_a\}_{\tau+s}+\fft{a}{p}k_a;-\fft{N}{(mp)^2(\tau+s)}+\fft{a}{p})},\label{m:finite:1:shift}
\end{split}
\end{equation}
where the integer $k_a$ is defined as
\begin{equation}
	k_a\equiv\left\lfloor\Re mp\Delta_a-\cot(\arg(\tau+s))\Im mp\Delta_a\right\rfloor=\lfloor mp(\tilde\Delta_a-\check\Delta_as)\rfloor.\label{ka}
\end{equation}
Substituting the product forms (\ref{def:eta}) and (\ref{def:theta1}) into (\ref{m:finite:1:shift}) then gives
\begin{equation}
\begin{split}
	\fft{\eta(\tilde\tau)^3}{\theta_1(m\Delta_a;\tilde\tau)}
	&=i\fft{n}{mp(\tau+s)}e^{\fft{N\pi i}{(mp)^2(\tau+s)}\{mp\Delta_a\}_{\tau+s}(\{mp\Delta_a\}_{\tau+s}-1)-\pi ik_a(1+(k_a+1)\fft{a}{p})}\\
	&\quad\times\prod_{l=1}^\infty\left(1-e^{2\pi il(-\fft{N}{(mp)^2(\tau+s)}+\fft{a}{p})}\right)^2\\
	&\kern3em~~\times\left(1-e^{2\pi i(l(-\fft{N}{(mp)^2(\tau+s)}+\fft{a}{p})+\fft{N}{(mp)^2(\tau+s)}\{mp\Delta_a\}_{\tau+s}+\fft{a}{p}k_a)}\right)^{-1}\\
	&\kern3em~~\times\left(1-e^{2\pi i((l-1)(-\fft{N}{(mp)^2(\tau+s)}+\fft{a}{p})-\fft{N}{(mp)^2(\tau+s)}\{mp\Delta_a\}_{\tau+s}-\fft{a}{p}k_a)}\right)^{-1}.
\end{split}\label{m:finite:1}
\end{equation}
Using the upper bound for the infinite product term given in (\ref{m:finite:1:bound}), one can rewrite the logarithm of (\ref{m:finite:1}) as
\begin{equation}
\begin{split}
	\log\fft{\eta(\tilde\tau)^3}{\theta_1(m\Delta_a;\tilde\tau)}&=\fft{N\pi i}{(mp)^2(\tau+s)}\{mp\Delta_a\}_{\tau+s}(\{mp\Delta_a\}_{\tau+s}-1)-\pi ik_a(1+(k_a+1)\fft{a}{p})\\
	&\quad+\log(i\fft{n}{mp(\tau+s)})+\mathcal O(\fft{p^2}{N}).\label{m:finite:1:large-N}
\end{split}
\end{equation}
Finally, substituting (\ref{m:finite:1:large-N}) into (\ref{TTI:mnr:large-N}) gives
\begin{empheq}[box=\fbox]{equation}
\begin{split}
	\log Z_{\{m,n,r\}}(\tau,\Delta_a,\mathfrak n_a)&=-\fft{N^2\pi i}{(mp)^2(\tau+s)}\sum_{a=1}^3(1-\mathfrak n_a)\{mp\Delta_a\}_{\tau+s}(1-\{mp\Delta_a\}_{\tau+s})\\
	&\quad-N\pi i\sum_{a=1}^3(1-\mathfrak n_a)k_a(1+(k_a+1)\fft{a}{p})+\mathcal O(N\log N,p^2).
\end{split}\label{TTI:mnr:large-N:m:finite}
\end{empheq}
Note that we can determine the $N^2$-leading order from the expression (\ref{TTI:mnr:large-N:m:finite}), since the last term of order $\mathcal O(N\log N,p^2)$ does not contribute to the $N^2$-leading order under the condition (\ref{m:finite}). The large-$N$ asymptotics (\ref{TTI:mnr:large-N:m:finite}) is consistent with the result for the basic sector $Z_{\{1,N,0\}}$ (\ref{TTI:1N0:large-N:simple}), which can be confirmed by choosing $(s,p,q)=(0,1,0)$ and $(a,b,c,d)=(0,-1,1,0)$.

\subsubsection*{Consistency under modular transformations}
In \cite{Hong:2018viz}, the standard contribution $\mathcal Z_\text{st}$ to the twisted index (\ref{TTI:BA:st}) has been investigated in detail and, in particular, its modular property as a weak Jacobi form was found. To be specific, it has been shown that generators of modular transformations, namely
\begin{equation}
	T:~\tau\to\tau+1,\qquad S:~\tau\to-\fft1\tau,
\end{equation}
permute standard contributions $Z_{\{m,n,r\}}$ (\ref{TTI:BA:mnr}) as
\begin{subequations}
\begin{align}
	T:&~~Z_{\{m,n,r\}}(\tau+1;\Delta_a,\mathfrak n_a)=Z_{\{m,n,r+m\}}(\tau,\Delta_a,\mathfrak n_a),\label{T:mnr}\\
	S:&~~Z_{\{m,n,r\}}(-1/\tau,\Delta_a/\tau,\mathfrak n_a)=e^{-\fft{(N^2-1)\pi i}{\tau}\sum_{a=1}^3(1-\mathfrak n_a)\Delta_a^2}Z_{\{g,N/g,-dm\}}(\tau,\Delta_a,\mathfrak n_a),\label{S:mnr}
\end{align}\label{modular:mnr}%
\end{subequations}
where $g=\gcd(n,r)$ and an integer $d$ is determined by a relation $(r/g)d-(n/g)b=1~(d,b\in\mathbb Z)$. In (\ref{modular:mnr}), the third entry in the three-integer notation $\{\cdot,\cdot,\cdot\}$ is defined modulo the second entry. Due to the permutation (\ref{modular:mnr}), the total standard contribution $\mathcal Z_\text{st}$ (\ref{TTI:BA:st}) satisfies
\begin{subequations}
\begin{align}
	T:&~~\mathcal Z_\text{st}(\tau+1;\Delta_a,\mathfrak n_a)=\mathcal Z_\text{st}(\tau,\Delta_a,\mathfrak n_a),\\
	S:&~~\mathcal Z_\text{st}(-1/\tau,\Delta_a/\tau,\mathfrak n_a)=e^{-\fft{(N^2-1)\pi i}{\tau}\sum_{a=1}^3(1-\mathfrak n_a)\Delta_a^2}\mathcal Z_\text{st}(\tau,\Delta_a,\mathfrak n_a).
\end{align}
\end{subequations}
Here we will check if the large-$N$ asymptotics (\ref{TTI:mnr:large-N:m:not-finite}) and (\ref{TTI:mnr:large-N:m:finite}) are consistent with modular properties among various sectors given in (\ref{modular:mnr}). 

First we consider the $T$-transformation (\ref{T:mnr}). Note that the orbifold modular parameter $\tilde\tau$ (\ref{lattice:tau}) transforms in the same way under the $T$-transformation and the shift of $r$ by $m$ as
\begin{equation}
	\tilde\tau|_{\tau\to\tau+1}=\fft{m(\tau+1)+r}{n}=\tilde\tau|_{r\to r+m}
\end{equation}
Using this property, one can easily check that the large-$N$ asymptotics of $\log Z_{\{m,n,r\}}$, both (\ref{TTI:mnr:large-N:m:not-finite:alter}) for an infinite $m$ and (\ref{TTI:mnr:large-N:m:finite}) for a finite $m$, satisfy the $T$-transformation (\ref{T:mnr}). 

Next we consider the $S$-transformation (\ref{S:mnr}). For simplicity, let us focus on the case with $r=0$ that gives $g=n$, $d=0$, and $b=-1$ in (\ref{S:mnr}). The $S$-transformation (\ref{S:mnr}) then reduces to
\begin{equation}
	\log Z_{\{m,n,0\}}(-1/\tau,\Delta_a/\tau,\mathfrak n_a)=-\fft{(N^2-1)\pi i}{\tau}\sum_{a=1}^3(1-\mathfrak n_a)\Delta_a^2+\log Z_{\{n,m,0\}}(\tau,\Delta_a,\mathfrak n_a).\label{S:mnr:large-N}
\end{equation}
in the large-$N$ limit. For a[an] finite[infinite] $m$, one can confirm the identity (\ref{S:mnr:large-N}) by substituting (\ref{TTI:mnr:large-N:m:not-finite}) and (\ref{TTI:mnr:large-N:m:finite}) into the RHS[LHS] and the LHS[RHS] of (\ref{S:mnr:large-N}) respectively. Hence the large-$N$ asymptotics (\ref{TTI:mnr:large-N:m:not-finite}) and (\ref{TTI:mnr:large-N:m:finite}) are consistent with the $S$-transformation (\ref{S:mnr}).

\subsection{Canonical partition function}\label{sec:TTI:micro}
Combining (\ref{TTI:BA:split}) with (\ref{TTI:mnr:large-N:m:not-finite:alter}) and (\ref{TTI:mnr:large-N:m:finite}), we obtain the twisted index in the large-$N$ limit (leaving non-standard contributions implicit). We will not compare the resulting twisted index directly with a gravitational partition function studied in section \ref{sec:BS}, however, because the former is a \underline{grand-canonical} partition function and the latter corresponds to a \underline{canonical} one. There are two different ways for a correct comparison:
\begin{enumerate}
	\item In the field theory side, derive a canonical partition function from the twisted index. Then one can compare the resulting canonical partition function with a dual gravitational partition function given in section \ref{sec:BS}. This process is similar to the $I$-extremization \cite{Hosseini:2017mds}, which was interpreted as relating micro-canonical and grand-canonical partition functions in \cite{Cabo-Bizet:2018ehj}. One difference is that here we do not fix a modular parameter $\tau$ in terms of the associated string momentum, and therefore the resulting partition function is in canonical ensemble, not in micro-canonical ensemble. 
	\item In the gravity side, find a larger set of Euclidean BPS solutions dubbed as Euclidean black saddles \cite{Bobev:2020pjk}, which includes a dual supersymmetric magnetically charged AdS$_5$ black string as a special case. One can use the regularized on-shell actions of these black saddles to derive a gravitational partition function in grand-canonical emsemble, which can be compared directly with the dual twisted index as a function of chemical potentials $\Delta_a$. This has been done for a supersymmetric AdS$_4$ black hole and its dual topologically twisted ABJM theory in \cite{Bobev:2020pjk}.
\end{enumerate}
In this subsection we take the 1st approach and leave the 2nd approach for future research.

To begin with, we explain how the inverse Laplace transform gives a canonical partition function from the grand-canonical one in a simple toy model. For a given theory, one can write the grand-canonical partition function $\mathcal Z(\tau,\mu)$ schematically as
\begin{equation}
	\mathcal Z(\tau,\mu)=\sum_Q \Omega(\tau,Q)e^{2\pi i\mu Q}.\label{grand:ptn:fct}
\end{equation}
Here $Q$ represents a set of charges that labels quantum states and $\mu$ stands for associated chemical potentials. $\tau$ is also a chemical potential but we will not fix this parameter through the inverse Laplace transform. $\Omega(\tau,Q)$ is a canonical partition function labelled by a chemical potential $\tau$ and charges $Q$. Now, applying the Cauchy's integral formula to (\ref{grand:ptn:fct}), one can derive $\Omega(\tau,Q)$ from $\mathcal Z(\tau,\mu)$ as
\begin{equation}
	\Omega(\tau,Q)=\int_0^1 d\Re\mu\,\mathcal Z(\tau,\mu)e^{-2\pi i\mu Q}.\label{micro:ptn:fct}
\end{equation}
This is the inverse Laplace transform we will use to derive the canonical partition function from the grand-canonical one.

\subsubsection{Canonical partition function from the twisted index}\label{sec:TTI:micro:setup}
Now we come back to the original problem. The twisted index of $\mathcal N=4$ SU($N$) SYM theory on $T^2\times S^2$, which is a grand-canonical partition function of the theory with a topological twist on $S^2$, can be written as
\begin{equation}
	\mathcal Z(\tau,\Delta_a,\mathfrak n_a)=\sum_{Q_a}\Omega(\tau,Q_a,\mathfrak n_a)e^{2\pi i\sum_{a=1}^3\Delta_aQ_a},\label{TTI-Omega}
\end{equation}
following (\ref{grand:ptn:fct}) of the toy model. Here $Q_a$'s are electric charges associated with the chemical potentials of flavor symmetries $\Delta_a$. Note that $\tau$ will not be fixed in terms of a string momentum, which is different from the literature \cite{Hosseini:2019lkt,Hosseini:2018tha,Zaffaroni:2019dhb,Hosseini:2020vgl}: we leave it as a free modular parameter of the torus $T^2$. 

The canonical partition function is then given by the inverse Laplace transform as
\begin{equation}
	\Omega(\tau,Q_a,\mathfrak n_a)=\int_0^1(\prod_{a=1}^3d\Re\Delta_a)\,\delta(\sum_{a=1}^3\Delta_a)\mathcal Z(\tau,\Delta_a,\mathfrak n_a)e^{-2\pi i\sum_{a=1}^3\Delta_aQ_a},\label{eq:Omega}
\end{equation}
following (\ref{micro:ptn:fct}) of the toy model. In (\ref{eq:Omega}), we put a Dirac-delta function in the integrand to impose the constraint $\sum_{a=1}^3\Delta_a\in\mathbb Z$ in (\ref{constraint:chemical}).

To compute the canonical partition function using the integral (\ref{eq:Omega}), we substitute the expression of the twisted index from the BA formula (\ref{TTI:BA:split}) into (\ref{eq:Omega}). The result can be written as
\begin{equation}
\begin{split}
	\Omega(\tau,Q_a,\mathfrak n_a)&=\Omega_\text{st}(\tau,Q_a,\mathfrak n_a)+\Omega_\text{non-st}(\tau,Q_a,\mathfrak n_a),\\
	\Omega_\text{st}(\tau,Q_a,\mathfrak n_a)&=\sum_{n|N}\sum_{r=0}^{n-1}\Omega_{\{m,n,r\}}(\tau,Q_a,\mathfrak n_a),
\end{split}\label{eq:Omega:split}
\end{equation}
where we have introduced $\Omega_{\{m,n,r\}}$ and $\Omega_\text{non-st}$ as
\begin{subequations}
\begin{align}
	\Omega_{\{m,n,r\}}(\tau,Q_a,\mathfrak n_a)&=\int_0^1(\prod_{a=1}^3d\Re\Delta_a)\,\delta(\sum_{a=1}^3\Delta_a)Z_{\{m,n,r\}}(\tau,\Delta_a,\mathfrak n_a)e^{-2\pi i\sum_{a=1}^3\Delta_aQ_a},\label{eq:Omega:mnr}\\
	\Omega_\text{non-st}(\tau,Q_a,\mathfrak n_a)&=\int_0^1(\prod_{a=1}^3d\Re\Delta_a)\,\delta(\sum_{a=1}^3\Delta_a)\mathcal Z_\text{non-st}(\tau,\Delta_a,\mathfrak n_a)e^{-2\pi i\sum_{a=1}^3\Delta_aQ_a}.\label{eq:Omega:non-st}
\end{align}
\end{subequations}
Since we do not know non-standard contributions to the twisted index $\mathcal Z_\text{non-st}$ explicitly, we leave the calculation of $\Omega_\text{non-st}$ (\ref{eq:Omega:non-st}) for future research. Instead, we focus on contributions to the canonical partition function extracted from the standard contribution to the twisted index, namely $\Omega_{\{m,n,r\}}$ (\ref{eq:Omega:mnr}). We call $\Omega_{\{m,n,r\}}$ the standard contribution to the canonical partition function.

For simplicity, from here on we turn off electric charges $Q_a$ and use the shorthand notation
\begin{equation}
	\Omega(\tau,Q_a=0,\mathfrak n_a)\quad\to\quad\Omega(\tau,\mathfrak n_a).
\end{equation}
Since we focus on purely magnetic dual AdS$_5$ black strings in the following section \ref{sec:BS}, this will be enough for our purpose. We leave a similar analysis for generic cases with non-trivial electric charges and rotations based on \cite{Hristov:2014hza,Hosseini:2019lkt,Hosseini:2020vgl} for future research.

\subsubsection{Computing the canonical partition fuction}\label{sec:TTI:micro:compute}
To evaluate the standard contribution to the canonical partition function $\Omega_{\{m,n,r\}}$ using the integral (\ref{eq:Omega:mnr}), we use the large-$N$ asymptotics of $Z_{\{m,n,r\}}$ computed in the previous subsection \ref{sec:TTI:large-N:other}. Recall that the large-$N$ asymptotics of $Z_{\{m,n,r\}}$ has been computed in two different cases separately: (\ref{TTI:mnr:large-N:m:not-finite:alter}) for an infinite $m$ and (\ref{TTI:mnr:large-N:m:finite}) for a finite $m$ with an extra assumption (\ref{m:finite}). Hence we evaluate $\Omega_{\{m,n,r\}}$ using (\ref{eq:Omega:mnr}) for these two different cases in order. The total canonical partition will then be obtained from the formula (\ref{eq:Omega:split}).

\subsubsection*{Infinite $m$}
First, we consider $\Omega_{\{m,n,r\}}$ with an infinite $m$ satisfying (\ref{m:infinite}) in the large-$N$ limit. Substituting the corresponding large-$N$ asymptotics (\ref{TTI:mnr:large-N:m:not-finite:alter}) into (\ref{eq:Omega:mnr}) gives
\begin{equation}
\begin{split}
	&\Omega_{\{m,n,r\}}(\tau,\mathfrak n_a)\\
	&=\int_0^1(\prod_{a=1}^3d\Re\Delta_a)\,\delta(\sum_{a=1}^3\Delta_a)\exp\bigg[N\pi i\sum_{a=1}^3(1-\mathfrak n_a)(m\Delta_a(1+2j_a)-j_a(1+(j_a+1)\tilde\tau))\\
	&\kern15em+o(N^2)\bigg]\qquad(m\neq\mathcal O(N^0)).
\end{split}\label{eq:Omega:mnr:m-inf}
\end{equation}
Once the $\Re\Delta_3$-integral is evaluated with the Dirac-delta function, the remaining integrals over $\Re\Delta_{1,2}$ in (\ref{eq:Omega:mnr:m-inf}) yield new Dirac-delta functions since the exponent is linear in $\Re\Delta_{1,2}$. These Dirac-delta functions from the $\Re\Delta_{1,2}$-integrals read
\begin{equation}
	\Re\Delta_a\text{-integral}\quad\to\quad\delta((1-\mathfrak n_a)(1+2j_a)-(1-\mathfrak n_3)(1+2j_3))\qquad(a=1,2).\label{Dirac-delta}
\end{equation}
For these Dirac-delta functions not to vanish, which is required for a non-zero $\Omega_{\{m,n,r\}}$, the integer $j_a$ introduced in (\ref{ja}) must be given as
\begin{equation}
	j_a=\left\lfloor\fft{\Im n\Delta_a}{\Im\tau}\right\rfloor\overset{!}{=}-\fft12\left(1+\zeta\fft{\fft{1}{1-\mathfrak n_a}}{\sum_{b=1}^3\fft{1}{1-\mathfrak n_b}}\right),\label{ja:saddle}
\end{equation}
where we have introduced $\zeta\in\{\pm1,-3\}$ as
\begin{equation}
	\sum_{a=1}^3j_a=-\fft{3+\zeta}{2}.\label{zeta}
\end{equation}
The constraint (\ref{zeta}) is from the definition (\ref{ja}) and the constraint $\sum_{a=1}^3\Delta_a\in\mathbb Z$ in (\ref{constraint:chemical}). 

The condition (\ref{ja:saddle}) cannot be satisfied, however, since the RHS of (\ref{ja:saddle}) is not an integer for generic magnetic charges $\mathfrak n_a$ under the constraint $\sum_{a=1}^3\mathfrak n_a=2$ in (\ref{constraint:chemical}). This means that at least one of the Dirac-delta functions from the $\Re\Delta_{1,2}$-integrals, (\ref{Dirac-delta}), will vanish. As a result, $\Omega_{\{m,n,r\}}$ with an infinite $m$ given in (\ref{eq:Omega:mnr:m-inf}) vanishes. To be precise, since we have been keeping track of the $N^2$-leading order only in the exponent of the integrand in (\ref{eq:Omega:mnr:m-inf}), we conclude
\begin{empheq}[box=\fbox]{equation}
	\log\Omega_{\{m,n,r\}}(\tau,\mathfrak n_a)=o(N^2)\qquad(m\neq\mathcal O(N^0)).\label{eq:Omega:mnr:m-inf:result}
\end{empheq}
%

\subsubsection*{Finite $m$}
Next, we consider $\Omega_{\{m,n,r\}}$ with a finite $m$ and a finite fraction $s$ such that a natural number $p$ defined by the relation (\ref{tau:e}) satisfies the condition (\ref{m:finite}) in the large-$N$ limit. Substituting the corresponding large-$N$ asymptotics (\ref{TTI:mnr:large-N:m:finite}) into (\ref{eq:Omega:mnr}) gives
\begin{subequations}
\begin{align}
	\Omega_{\{m,n,r\}}(\tau,\mathfrak n_a)&=\int_0^1(\prod_{a=1}^3d\Re\Delta_a)\,\delta(\sum_{a=1}^3\Delta_a)\exp[S_{\{m,n,r\}}^\text{eff}(\tau,\Delta_a,\mathfrak n_a)+o(N^2)]\nn\\
	&\quad~(m=\mathcal O(N^0),~p=o(N)),\label{eq:Omega:mnr:m-fin}\\
	S_{\{m,n,r\}}^\text{eff}(\tau,\Delta_a,\mathfrak n_a)&=-\fft{N^2\pi i}{(mp)^2(\tau+s)}\sum_{a=1}^3(1-\mathfrak n_a)\{mp\Delta_a\}_{\tau+s}(1-\{mp\Delta_a\}_{\tau+s}).\label{Seff}
\end{align}
\end{subequations}
Note that we have omitted pure imaginary terms in the effective action, since they are defined modulo $2\pi i\mathbb Z$ so that we can ignore them when we focus on the $N^2$-leading order in the large-$N$ limit. 

To evaluate the integral (\ref{eq:Omega:mnr:m-fin}), we use the saddle point approximation. The first step is to solve the following saddle point equation,
\begin{equation}
	0=\partial_{\Re\Delta_a}S_{\{m,n,r\}}^\text{eff}(\tau,\Delta_a,\mathfrak n_a)-2\pi i\Lambda,\label{eq:saddle}
\end{equation}
where $\Lambda$ is the Lagrange multiplier introduced to enforce the constraint $\sum_{a=1}^3\Delta_a\in\mathbb Z$ in (\ref{constraint:chemical}). One can solve the saddle point equation (\ref{eq:saddle}) using the relations
\begin{equation}
	\fft{\partial}{\partial\Re\Delta_b}\{mp\Delta_a\}_{\tau+s}=mp\,\delta^b{}_a,\qquad \fft{\partial k_a}{\partial\Re\Delta_b}=0,
\end{equation}
for the integer $k_a$ defined in (\ref{ka}), assuming
\begin{equation}
	\Re mp\Delta_a-\cot(\arg(\tau+s))\Im mp\Delta_a=mp(\tilde\Delta_a-\check\Delta_as)\notin\mathbb Z\label{assumption:ka}
\end{equation}
around a saddle point $\Delta_a=\Delta_a^\star$. This assumption is required for the large-$N$ asymptotics (\ref{TTI:mnr:large-N:m:finite}) to be differentiable with respect to chemical potentials $\Delta_a$ at the saddle point. The solution to the saddle point equation (\ref{eq:saddle}) is given as
\begin{subequations}
\begin{align}
	\{mp\Delta_a^\star\}_{\tau+s}&=\fft12\left(1+\fft{\fft{\eta_{mp,s}}{1-\mathfrak n_a}}{\sum_{b=1}^3\fft{1}{1-\mathfrak n_b}}\right)=\fft12\left(1+\eta_{mp,s}\left(1-\fft{\mathfrak n_a(\mathfrak n_a-1)}{1-\mathfrak n_1\mathfrak n_2-\mathfrak n_2\mathfrak n_3-\mathfrak n_3\mathfrak n_1}\right)\right),\label{Del:ext:m:finite}\\
	\Lambda&=\fft{\fft{N^2\pi i}{mp(\tau+s)}\eta_{mp,s}}{2\pi i\sum_{a=1}^3\fft{1}{1-\mathfrak n_a}},
\end{align}\label{Del:L:ext:m:finite}%
\end{subequations}
where $\eta_{mp,s}\in\{\pm1\}$ is defined as (\ref{eq:eta}). The second equation of (\ref{Del:ext:m:finite}) is from the constraint on magnetic charges $\sum_{a=1}^3\mathfrak n_a=2$ in (\ref{constraint:chemical}). 

Next, substituting the saddle point (\ref{Del:ext:m:finite}) into (\ref{Seff}) gives the effective action at the saddle point as
\begin{equation}
	S_{\{m,n,r\}}^\text{eff}(\tau,\Delta_a^\star,\mathfrak n_a)=\fft{N^2\pi i}{4(mp)^2(\tau+s)}\fft{\mathfrak n_1\mathfrak n_2\mathfrak n_3}{1-\mathfrak n_1\mathfrak n_2-\mathfrak n_2\mathfrak n_3-\mathfrak n_3\mathfrak n_1}.\label{Seff:ext}
\end{equation}
Since the effective action at the saddle point (\ref{Seff:ext}) becomes large in the large-$N$ limit due to the condition (\ref{m:finite}), one can apply the saddle point approximation to the integral (\ref{eq:Omega:mnr:m-fin}) with $\fft{N}{mp}$ as a large control parameter. The result is given as
\begin{empheq}[box=\fbox]{equation}
\begin{split}
	\log\Omega_{\{m,n,r\}}(\tau,\mathfrak n_a)&=\fft{N^2\pi i}{4(mp)^2(\tau+s)}\fft{\mathfrak n_1\mathfrak n_2\mathfrak n_3}{1-\mathfrak n_1\mathfrak n_2-\mathfrak n_2\mathfrak n_3-\mathfrak n_3\mathfrak n_1}+o(N^2)\\
	&\quad~(m=\mathcal O(N^0),~p=o(N)).\label{eq:Omega:mnr:m-fin:result}
\end{split}
\end{empheq}
In (\ref{eq:Omega:mnr:m-fin:result}), a natural number $mp$ and a finite fraction $s\in\fft{1}{mp}\mathbb Z$ determine the precise $N^2$-leading order in the logarithm of a particular $\{m,n,r\}$ standard contribution to the canonical partition function, namely $\log\Omega_{\{m,n,r\}}$. In particular, it is obvious that $mp$ has to be finite for $\log\Omega_{\{m,n,r\}}$ to have a non-vanishing $N^2$-leading order according to (\ref{eq:Omega:mnr:m-fin:result}).

\subsubsection*{The total canonical partition function}
The total canonical partition function is obtained by subsituting (\ref{eq:Omega:mnr:m-inf:result}) and (\ref{eq:Omega:mnr:m-fin:result}) into the formula (\ref{eq:Omega:split}). The result can be written as
\begin{equation}
	\Omega(\tau,\mathfrak n_a)=\sum_{n\in\mathbb N}^{N}\sum_{r=0}^{n-1}\underbrace{\exp[\fft{N^2\pi i}{4(mp)^2(\tau+s)}\fft{\mathfrak n_1\mathfrak n_2\mathfrak n_3}{1-\mathfrak n_1\mathfrak n_2-\mathfrak n_2\mathfrak n_3-\mathfrak n_3\mathfrak n_1}+o(N^2)]}_{=\Omega_{\{m,n,r\}}(\tau,\mathfrak n_a)}+\Omega_\text{non-st}(\tau,\mathfrak n_a).\label{eq:Omega:split:prep}
\end{equation}
Note that the first term in the exponent of the RHS scales like $o(N^2)$ for an infinite $m$ and therefore (\ref{eq:Omega:split:prep}) is consistent with (\ref{eq:Omega:mnr:m-inf:result}) as well as (\ref{eq:Omega:mnr:m-fin:result}). The expression (\ref{eq:Omega:split:prep}) is inconvenient to explore, however, since the large-$N$ asymptotics of $\Omega_{\{m,n,r\}}$ in (\ref{eq:Omega:split:prep}) is labelled by $(mp,s)$, which has to be determined case by case by minimizing an integer $p$ introduced together with a fraction $s$ in the relation (\ref{tau:e}) for a given configuration $\{m,n,r\}$. 

To make the expression (\ref{eq:Omega:split:prep}) more explicit, an important question is the following: what are the all possible values of finite $(mp,s)\in\mathbb N\times\fft{1}{mp}\mathbb Z$, each of which labels the large-$N$ asymptotics of $\Omega_{\{m,n,r\}}$ whose exponent has a non-vanishing $N^2$-leading order in (\ref{eq:Omega:split:prep})? At first glance, it seems like for an \underline{arbitrary} set of integers $(c,d)\in\mathbb N\times\mathbb Z$, one can always find a configuration $\{m,n,r\}$ such that the large-$N$ asymptotics of $\Omega_{\{m,n,r\}}$ is labelled by
\begin{equation}
	(mp,s)=(c,d/c).\label{claim:mpe}
\end{equation}
This expectation comes from the following configuration
\begin{equation}
	\{m,n,r\}=\{1,N,\fft{Nq+d}{c}\}\quad\to\quad\text{choose~~}s=\fft{d}{c}\text{~~then~~}mp=c.\label{mpe:cd}
\end{equation}
One subtlety of the configuration (\ref{mpe:cd}) is that, it is valid only if $r=\fft{Nq+d}{c}$ is an integer within the range $r\in\mathbb Z_n=\{0,1,\cdots,n-1\}$ as it should be. Since we are interested in the large-$N$ limit, we may circumvent this issue by setting the rank $N$ as any large natural number satisfying this condition. For example, for $(c,d)=(2,0)$, we may choose an even $N$ for the configuration (\ref{mpe:cd}) then the condition $r=\fft{Nq+d}{c}=\fft{N}{2}\in\mathbb Z_N$ is satisfied with $q=1$: for an even $N$, the large-$N$ asymptotics of $\Omega_{\{1,N,N/2\}}$ is labelled by $(mp,s)=(c,d/c)=(2,0)$. On the contrary, an odd $N$ does not provide any $\{m,n,r\}$ configuration whose corresponding contribution $\Omega_{\{m,n,r\}}$ is labelled by $(mp,s)=(c,d/c)=(2,0)$ in the large-$N$ limit. 

This logic does not make sense, however, since we are interested in all the possible values of finite labels $(mp,s)\in\mathbb N\times\fft{1}{mp}\mathbb Z$ for a `given' rank $N$ in (\ref{eq:Omega:split:prep}): $N$ cannot be chosen to yield a particular label $(mp,s)=(c,d/c)$. Then we arrive at an weird conclusion: the factorization of the rank $N$ affects the possible values of finite labels $(mp,s)\in\mathbb N\times\fft{1}{mp}\mathbb Z$ even in the large-$N$ limit. For example, just by replacing the rank originally given as an even number $N$ with an odd number $N+1$, we lose a particular standard contribution labelled by $(mp,s)=(2,0)$ in (\ref{eq:Omega:split:prep}). This is definitely unnatural in the large-$N$ limit.

We expect that this conundrum be resolved by exploring non-standard contributions. To be specific, we conjecture that \emph{for an \underline{arbitrary} set of integers $(c,d)\in\mathbb N\times\mathbb Z$, one may find a configuration $\{m,n,r\}$ whose corresonding contribution $\Omega_{\{m,n,r\}}$ is labelled by $(mp,s)=(c,d/c)$ in the large-$N$ limit as shown in (\ref{eq:Omega:split:prep}), where the label $(mp,s)$ is determined by minimizing an integer $p$ in the relation (\ref{tau:e}); otherwise, if there is no such $\{m,n,r\}$ configuration, the contribution labelled by $(mp,s)=(c,d/c)$ comes from the non-standard contribution $\Omega_\text{non-st}$ in (\ref{eq:Omega:split:prep}).} For example, according to this conjecture, the contribution labelled by $(mp,s)=(2,0)$ does exist for an odd $N$ but it comes from the non-standard contribution $\Omega_\text{non-st}$ in (\ref{eq:Omega:split:prep}). Refer to \cite{ArabiArdehali:2019orz} for both analytical and numerical evidence for this conjecture.\footnote{The authors studied the superconformal index in \cite{ArabiArdehali:2019orz}, but the BAE are the same as (\ref{BAE}). Hence we can apply the analysis in section 4.2.2 of \cite{ArabiArdehali:2019orz} to the current case; there it has been shown that a non-standard BAE solution whose contribution to the twisted index is labelled by $(mp,s)=(2,0)$ does exist for an odd $N$.}

Based on the above conjecture, we can simplify the canonical partition function given in (\ref{eq:Omega:split:prep}) further as
\begin{empheq}[box=\fbox]{equation}
	\Omega(\tau,\mathfrak n_a)=\sum_{c\in\mathbb N}\sum_{d\in\mathbb Z}\exp[\fft{N^2\pi i}{4c^2(\tau+d/c)}\fft{\mathfrak n_1\mathfrak n_2\mathfrak n_3}{1-\mathfrak n_1\mathfrak n_2-\mathfrak n_2\mathfrak n_3-\mathfrak n_3\mathfrak n_1}+o(N^2)]+\Omega_\text{non-st}'(\tau,\mathfrak n_a).\label{eq:Omega:split:result}
\end{empheq}
Note that the degeneracy for a given label $(c,d)\in\mathbb N\times\mathbb Z$ can be absorbed into $o(N^2)$. The primed non-standard contribution $\Omega_\text{non-st}'$ means that some non-standard contributions in $\Omega_\text{non-st}$ are factored out and included in the sum over $(c,d)\in\mathbb N\times\mathbb Z$ instead in (\ref{eq:Omega:split:result}). We will explore holographic duals of this canonical partition function (\ref{eq:Omega:split:result}) in the next section \ref{sec:BS} and then compare the results in section \ref{sec:compare}.

\section{Holographic dual family of 5d extremal solutions}\label{sec:BS}
In this section, we explore holographic duals of the canonical partition function derived from the twisted index in the previous section. To begin with, in subsection \ref{sec:BS:STU}, we review a supersymmetric, magnetically charged AdS$_5$ black string solution of 5d $\mathcal N=2$ gauged STU model and its near-horizon geometry. In subsection \ref{sec:BS:sl2z}, we construct a family of extremal solutions in 5d $\mathcal N=2$ gauged STU model, which includes the near-horizon limit of an AdS$_5$ black string reviewed in \ref{sec:BS:STU}, based on the black hole Farey tail \cite{Dijkgraaf:2000fq}. In subsection \ref{sec:BS:on-shell}, we compute the regularized on-shell actions of these extremal solutions and then derive a gravitational on-shell action using them. 
 
\subsection{AdS$_5$ black strings in 5d $\mathcal N=2$ gauged STU model}\label{sec:BS:STU}
We are interested in a supersymmetric magnetically charged AdS$_5$ black string solution of $\mathcal N=2$ gauged STU model reviewed in Appendix \ref{App:N=2}, which is dual to the ensemble of BPS states of $\mathcal N=4$ SU($N$) SYM theory on $T^2\times S^2$. One may find such solutions by solving the BPS equations (\ref{N=2:sugra:BPS}) with the following magnetic black string ansatz ($I\in\{1,2,3\}$)
\begin{subequations}
\begin{align}
	ds^2&=-e^{2f_1(r)}dt^2+e^{2f_2(r)}(d\phi+\Omega(r)dt)^2+e^{2f_3(r)}dr^2+e^{2f_4(r)}e^{2h_{\mathfrak g}(x,y)}(dx^2+dy^2),\label{BS:ansatz:metric}\\
	A^I&=-p^I\omega_{\mathfrak g}\quad\to\quad F^I=-p^Ie^{2h_{\mathfrak g}(x,y)}dx\wedge dy,\\
	X^I&=X^I(r),
\end{align}\label{BS:ansatz}%
\end{subequations}
since the remaining equations of motion (\ref{N=2:sugra:Einstein}), (\ref{N=2:sugra:scalar}), and (\ref{N=2:sugra:vector}) then follow automatically \cite{Gauntlett:2003fk,Gutowski:2004yv}. In this ansatz, we have generalized $S^2$ to a Riemann surface of genus $\mathfrak g$, namely $\Sigma_{\mathfrak g}$, where we set $\kappa=1,0,-1$ for $\mathfrak g=0$, $\mathfrak g=1$, and $\mathfrak g>1$ respectively for later purpose. The Riemann surface $\Sigma_{\mathfrak g}$ is supported by $(x,y)$ coordinates, and we have defined $h_{\mathfrak g}$ and $\omega_{\mathfrak g}$ on this Riemann surface as \cite{Bobev:2020pjk}
\begin{equation}
	e^{h_{\mathfrak g}(x,y)}=\begin{cases}
		\fft{2}{1+x^2+y^2} & (\mathfrak g=0)\\
		\sqrt{2\pi} & (\mathfrak g=1)\\
		\fft1y & (\mathfrak g>1)
	\end{cases},\qquad
	\omega_{\mathfrak g}=\begin{cases}
		\fft{2(-ydx+xdy)}{1+x^2+y^2} & (\mathfrak g=0)\\
		\pi(-ydx+xdy) & (\mathfrak g=1)\\
		\fft{dx}{y} & (\mathfrak g>1)
	\end{cases},
\end{equation}
which satisfy $d\omega_{\mathfrak g}=e^{2h_{\mathfrak g}(x,y)}dx\wedge dy$. The asymptotically AdS$_5$ condition of an ansatz (\ref{BS:ansatz}) can be summarized as
\begin{equation}
	f_1(r),f_2(r),-f_3(r),f_4(r)\to\log r,\qquad \Omega(r)\to0,\qquad X^I(r)\to 1\label{BS:ansatz:asymp}
\end{equation}
in the asymptotic region $r\to\infty$.

Note that the conformal boundary of an ansatz (\ref{BS:ansatz}) is given as $\mathbb R^{1,1}\times\Sigma_{\mathfrak g}$, which is replaced with $T^2\times\Sigma_{\mathfrak g}$ upon the Wick rotation $t\to-it_E$ and the periodic identification of $(t_E,\phi)$ coordinates. Hence, the Euclidean version of an ansatz (\ref{BS:ansatz}) with $\mathfrak g=0$ has the $T^2\times S^2$ conformal boundary, where we have introduced the $\mathcal N=4$ SYM theory and computed its twisted index in section \ref{sec:TTI}. This is consistent with that an AdS$_5$ black string solution of the form (\ref{BS:ansatz}) with $\mathfrak g=0$ is holographically dual to the ensemble of BPS states of $\mathcal N=4$ SYM theory on $T^2\times S^2$. 

A supersymmetric AdS$_5$ black string solution with generic magnetic charges $p^I$ has not yet been constructed analytically from the above ansatz (\ref{BS:ansatz}). In \cite{Benini:2013cda}, however, the corresponding AdS$_3\times\Sigma_{\mathfrak g}$ near-horizon solution that solves the BPS equations (\ref{N=2:sugra:BPS}) was found. Replacing the Poincar\'{e} coordinates used in \cite{Benini:2013cda} with the near-horizon extremal BTZ coordinates used in \cite{Hristov:2014eza}, the AdS$_3\times\Sigma_{\mathfrak g}$ near-horizon solution can be rewritten as
\begin{subequations}
\begin{align}
	ds^2&=\left(\fft{8p^1p^2p^3\Pi}{\Theta^3}\right)^\fft23\left(-\fft14r^2dt^2+\rho_+^2\left(d\phi+\fft{r}{2\rho_+}dt\right)^2+\fft{dr^2}{4r^2}\right)\nn\\
	&\quad+\left(\fft{(p^1p^2p^3)^2}{\Pi}\right)^\fft13e^{2h_{\mathfrak g}(x,y)}(dx^2+dy^2),\label{AdS3:NH:metric}\\
	A^I&=-p^I\omega_{\Sigma_{\mathfrak g}},\\
	X^I&=\fft{p^I(p^1+p^2+p^3-2p^I)}{(p^1p^2p^3\Pi)^\fft13},\\
	-\kappa&=p^1+p^2+p^3.
\end{align}\label{AdS3:NH}%
\end{subequations}
The constants $\Pi$ and $\Theta$ in (\ref{AdS3:NH}) are defined in terms of magnetic charges as
\begin{subequations}
\begin{align}
	\Pi&=(p^1+p^2-p^3)(p^1-p^2+p^3)(-p^1+p^2+p^3),\\
	\Theta&=-(p^1)^2-(p^2)^2-(p^3)^2+2(p^1p^2+p^2p^3+p^3p^1).
\end{align}\label{PiTheta}%
\end{subequations}
See Appendix \ref{App:N=2:BPS} to check directly that (\ref{AdS3:NH}) solves the BPS equations (\ref{N=2:sugra:BPS}). Also refer to Appendix \ref{App:AdS3} for a coordinate transformation between the Poincar\'e coordinates and the near-horizon extremal BTZ coordinates.

For some special configurations of magnetic charges, full analytic AdS$_5$ black string solutions have been constructed \cite{Klemm:2000nj,Cacciatori:2003kv,Bernamonti:2007bu,Azzola:2018sld}. For example, when all three magnetic charges are identical as $p^I=-\fft{\kappa}{3}$ and the Riemann surface is a hyperbolic plane as $\Sigma_{\mathfrak g}=H^2~(\kappa=-1)$, the corresponding magnetic black string solution is given from the ansatz (\ref{BS:ansatz}) as \cite{Bernamonti:2007bu,Azzola:2018sld}
\begin{subequations}
\begin{align}
	ds^2&=-\fft{8r}{H(r)}\left(c+\fft{4}{3r}\right)^{-\fft12}dt^2+2r\left(c+\fft{4}{3r}\right)^{-\fft12}H(r)\left(d\phi+\fft{2}{H(r)}dt\right)^2\nn\\
	&\quad+\fft{dr^2}{4r^2}+\fft{r}{4}\left(c+\fft{4}{3r}\right)\fft{dx^2+dy^2}{y^2},\label{AdS5:BS:metric}\\
	A^I&=-\fft{dx}{3y},\\
	X^I&=1.
\end{align}\label{AdS5:BS}%
\end{subequations}
Here the function $H(r)$ is given as
\begin{equation}
	H(r)=h+3q_0\left(c+\fft{4}{3r}\right)^\fft12\left(c-\fft{2}{3r}\right).
\end{equation}
See Appendix \ref{App:N=2:BPS} to check directly that (\ref{AdS5:BS}) solves the BPS equations (\ref{N=2:sugra:BPS}). For the metric (\ref{AdS5:BS:metric}) to be positive definite over $r\in(0,\infty)$, we must impose the following constraints on parameters:
\begin{equation}
	c>0,\quad q_0<0,\quad H(r\to\infty)=h+3q_0c^\fft32>0.
\end{equation}

It is straightforward to see that the full flow (\ref{AdS5:BS}) matches the near-horizon solution (\ref{AdS5:BS}) with identical magnetic charges in the near-horizon regime $r\to0$ under the identification $\rho_+^2=-9q_0$ and the following rescaling of coordinates:
\begin{equation}
	r\to r^\fft23,\qquad t\to\fft{\sqrt{-q_0}}{3\sqrt3}t.
\end{equation}
%

\subsection{Family of extremal solutions in 5d $\mathcal N=2$ gauged STU model}\label{sec:BS:sl2z}
In this subsection, we construct a family of extremal solutions to the BPS equations of 5d $\mathcal N=2$ gauged STU model (\ref{N=2:sugra:BPS}). It can be obtained by replacing the near-horizon extremal BTZ part of (\ref{AdS3:NH:metric}), namely
\begin{equation}
	ds_3^2=-\fft14r^2dt^2+\rho_+^2\left(d\phi+\fft{r}{2\rho_+}dt\right)^2+\fft{dr^2}{4r^2},\label{AdS3:local:1}
\end{equation}
with locally equivalent but globally distinguished extremal BTZ black holes. Note that the BPS equations (\ref{N=2:sugra:BPS}) are local conditions and therefore they are not sensitive to the replacement of the near-horizon extremal BTZ part of (\ref{AdS3:NH:metric}) given in (\ref{AdS3:local:1}) with any locally equivalent metric as pointed out in \cite{Hristov:2014eza}. 

In \ref{sec:BS:sl2z:farey}, we construct a family of extremal BTZ black holes that will replace (\ref{AdS3:local:1}). This family of extremal BTZ black holes is obtained by imposing the extremal limit on the SL(2,$\mathbb Z$) family of black holes, which was first constructed in \cite{Maldacena:1998bw} and then dubbed as a black hole Farey tail in \cite{Dijkgraaf:2000fq}. Then in \ref{sec:BS:sl2z:5d}, we write down the family of 5d extremal solutions obtained by a simple replacement of (\ref{AdS3:local:1}) with the family of extremal BTZ black holes.

\subsubsection{Family of extremal BTZ black holes}\label{sec:BS:sl2z:farey}
To construct the family of extremal BTZ black holes, first recall that a BTZ black hole reads \cite{Banados:1992wn} (see Appendix \ref{App:AdS3} for a coordinate transformation between (\ref{AdS3:local:1}) and (\ref{BTZ:Lorentz}))
\begin{equation}
	ds^2=-\fft{(\rho^2-\rho_+^2)(\rho^2-\rho_-^2)}{\rho^2}dt^2+\fft{\rho^2}{(\rho^2-\rho_+^2)(\rho^2-\rho_-^2)}d\rho^2+\rho^2\left(d\phi-\fft{\rho_+\rho_-}{\rho^2}dt\right)^2\label{BTZ:Lorentz}
\end{equation}
with $\rho_+>\rho_->0$ and a periodic angular coordinate \cite{Banados:1992gq}
\begin{equation}
	\phi\sim\phi+2\pi.\label{angle:period}
\end{equation}
The extremal limit corresponds to $\rho_+\to\rho_-$. The family of extremal BTZ black holes therefore consists of geometries locally equivalent as (\ref{BTZ:Lorentz}) with $\rho_+\to\rho_-$ but globally distinguished by different periods of $(t,\phi)$ coordinates. 

Since the periods of coordinates are determined in the Euclidean signature, we will construct a family of Euclidean BTZ black holes first by introducing various periods of $(t_E,\phi)$ coordinates ($t\to-it_E$) based on \cite{Maldacena:1998bw,Dijkgraaf:2000fq}. Then going back to the Lorentzian signature, we take the extremal limit to obtain the family of extremal BTZ black holes.

\subsubsection*{Periodic identifications in the Euclidean signature}
The Euclidean BTZ metric obtained from (\ref{BTZ:Lorentz}) by the Wick rotation $t\to-it_E$ reads
\begin{equation}
	ds^2=\fft{(\rho^2-\rho_+^2)(\rho^2-\rho_-^2)}{\rho^2}dt_E^2+\fft{\rho^2}{(\rho^2-\rho_+^2)(\rho^2-\rho_-^2)}d\rho^2+\rho^2\left(d\phi+\fft{i\rho_+\rho_-}{\rho^2}dt_E\right)^2.\label{BTZ:Euclid}
\end{equation}
There are two different ways to introduce periods of $(t_E,\phi)$ coordinates in (\ref{BTZ:Euclid}), and we will introduce both of them in order.

First, we use the copmlex structure of a Euclidean BTZ black hole (\ref{BTZ:Euclid}) in the asymptotic region to determine periods of $(t_E,\phi)$ coordinates. To be specific, we set the conformal boundary of a Euclidean BTZ black hole (\ref{BTZ:Euclid}) to be a torus with a complex structure specified by a modular parameter $\tau$, and then the periods of $(t_E,\phi)$ coordinates will be determined in terms of $\tau$. 

To begin with, consider the asymptotic region $(\rho\to\infty)$ of a Euclidean BTZ black hole (\ref{BTZ:Euclid}):
\begin{equation}
	ds^2\to\rho^2|d\phi+idt_E|^2+\fft{d\rho^2}{\rho^2}\qquad(\rho\to\infty).\label{BTZ:Euclid:asymp}
\end{equation}
From (\ref{BTZ:Euclid:asymp}), it is clear that the conformal boundary of a Euclidean BTZ black hole is supported by a complex coordinate
\begin{equation}
	w=\fft{\phi+it_E}{2\pi},\qquad \bar w=\fft{\phi-it_E}{2\pi}.\label{w:tphi}
\end{equation}
For the conformal boundary to be a torus, we should introduce two independent cycles along this complex coordinate $w$. The whole 3d space (\ref{BTZ:Euclid:asymp}) will then be a solid torus, and therefore only one of the two cycles can be a contractible one. Following \cite{Dijkgraaf:2000fq}, we take
\begin{subequations}
	\begin{align}
		1.~\text{unique contractible cycle: }& w\sim w+c\tau+d\quad(\bar w\sim\bar w+c\bar\tau+d),\label{period:asymp:con}\\
		2.~\text{a non-contractible cycle: }& w\sim w+a\tau+b\quad (\bar w\sim\bar w+a\bar\tau+b),\label{period:asymp:non1}
	\end{align}\label{period:asymp}%
\end{subequations}
where $\tau$ is a complex parameter and the integers $a,b,c,d$ satisfy 
\begin{equation}
	\begin{pmatrix}
		a&b\\c&d
	\end{pmatrix}\in\text{SL}(2,\mathbb Z).\label{gabcd}
\end{equation}
The conformal boundary now becomes a torus with a modular parameter $\fft{a\tau+b}{c\tau+d}$, where we choose (\ref{period:asymp:con}) and (\ref{period:asymp:non1}) as an $A$-cyle and a $B$-cycle respectively. In fact, since the complex structure is invariant under the SL(2,$\mathbb Z$) transformation of a modular parameter, namely
\begin{equation}
	\tau\quad\to\quad\fft{a\tau+b}{c\tau+d},
\end{equation}
we can say that the complex structure of a conformal boundary is specified by a modular parameter $\tau$ under the periodic identifications (\ref{period:asymp}).

One can construct various other non-contractible cycles in terms of linear combinantions of (\ref{period:asymp}). In particular, taking $a\times$(\ref{period:asymp:con})$-c\times$(\ref{period:asymp:non1}) and $-b\times$(\ref{period:asymp:con})$+d\times$(\ref{period:asymp:non1}), one can obtain the fundamental cycles
\begin{subequations}
\begin{align}
	2'.~\text{a non-contractible cycle: }& w\sim w+1\quad(\text{contractible if }(c,d)=(0,1)),\label{period:asymp:non2}\\
	2''.~\text{a non-contractible cycle: }& w\sim w+\tau\quad(\text{contractible if }(c,d)=(1,0)),\label{period:asymp:non3}
\end{align}\label{period:asymp:fund}%
\end{subequations}
respectively. For convenience, we write down the cycles (\ref{period:asymp:con}) and (\ref{period:asymp:fund}) in $(t_E,\phi)$ coordinates as
\begin{subequations}
\begin{align}
	1.&~(t_E,\phi)\sim(t_E,\phi)+2\pi(c\fft{\tau-\bar\tau}{2i},c\fft{\tau+\bar\tau}{2}+d),\label{period:asymp:tphi:con}\\
	2'.&~(t_E,\phi)\sim(t_E,\phi)+2\pi(0,1),\label{period:asymp:tphi:non2}\\
	2''.&~(t_E,\phi)\sim(t_E,\phi)+2\pi(\fft{\tau-\bar\tau}{2i},\fft{\tau+\bar\tau}{2}),\label{period:asymp:tphi:non3}
\end{align}\label{period:asymp:tphi}%
\end{subequations}
using (\ref{w:tphi}). The last two fundamental cycles (\ref{period:asymp:tphi:non2}) and (\ref{period:asymp:tphi:non3}) determine periods of $(t_E,\phi)$ coordinates. Note that (\ref{period:asymp:tphi:non2}) is consistent with the known angular period (\ref{angle:period}).

This time, we use the regularity of a Euclidean BTZ black hole (\ref{BTZ:Euclid}) in the near-horizon region to determine periods of $(t_E,\phi)$ coordinates. To begin with, consider the near-horizon region $\rho=\rho_++x^2~(x\ll\rho_+)$ upon which the Euclidean BTZ black hole (\ref{BTZ:Euclid}) reads
\begin{equation}
	ds^2\simeq\fft{2\rho_+}{\rho_+^2-\rho_-^2}\left(dx^2+\fft{(\rho_+^2-\rho_-^2)^2}{\rho_+^2}x^2dt_E^2\right)+\rho_+^2\left(d\phi+\fft{i\rho_-}{\rho_+}dt_E\right)^2.\label{BTZ:Euclid:NH}
\end{equation}
For the near-horizon metric (\ref{BTZ:Euclid:NH}) to represent a smooth $\mathbb R^2\times S^1$ horizon without a conical singularity at $x\to0$, $(t_E,\phi)$ coordinates have to be periodically identified as 
\begin{subequations}
\begin{align}
	1.\text{ temporal cycle (thermal cycle): }& (t_E,\phi)\sim(t_E,\phi)+2\pi\left(\fft{\rho_+}{\rho_+^2-\rho_-^2},-\fft{i\rho_-}{\rho_+^2-\rho_-^2}\right),\label{period:horizon:1}\\
	2'.\text{ spatial cycle: }&(t_E,\phi)\sim(t_E,\phi)+2\pi(0,1),\label{period:horizon:2}
\end{align}\label{period:horizon}%
\end{subequations}
where the spatial cycle is chosen to be consistent with the known angular period (\ref{angle:period}). These cycles (\ref{period:horizon}) determine periods of $(t_E,\phi)$ coordinates. 

Even though we have introduced periods of $(t_E,\phi)$ coordinates in two different ways, the final results (\ref{period:asymp:tphi:non2},\ref{period:asymp:tphi:non3}) and (\ref{period:horizon:1},\ref{period:horizon:2}) must be consistent with each other. One of the fundemantal cycles, (\ref{period:asymp:tphi:non2}), is already identical to the spatial cycle (\ref{period:horizon:2}). We then identify the unique contractible cycle (\ref{period:asymp:tphi:con}) to a thermal cycle (\ref{period:horizon:1}) \cite{Dijkgraaf:2000fq,Murthy:2009dq}. The modular parameter $\tau$ and the BTZ parameters $\rho_\pm$ are then related as
\begin{equation}
	\rho_\pm=\fft{i}{2}\left(\fft{1}{c\tau+d}\mp\fft{1}{c\bar\tau+d}\right).\label{rhopm:Euclid}
\end{equation}
Using (\ref{rhopm:Euclid}), one can rewrite the periods of $(t_E,\phi)$ coordinates given in (\ref{period:asymp:tphi:non2},\ref{period:asymp:tphi:non3}) as\,\footnote{For $c=0$, we have $(\rho_+,\rho_-)=(0,i/d)$ from (\ref{rhopm:Euclid}), which does not allow for the extremal limit $\rho_+=\rho_-$. Since we are interested in the family of extremal BTZ black holes after all, we exclude the case with $c=0$, which corresonds to a global AdS$_3$ \cite{Maldacena:1998bw,Dijkgraaf:2000fq,Murthy:2009dq}.\label{c=0}}
\begin{subequations}
\begin{align}
	(t_E,\phi)&\sim(t_E,\phi)+2\pi(0,1),\\
	(t_E,\phi)&\sim(t_E,\phi)+2\pi(\fft1c\fft{\rho_+}{\rho_+^2-\rho_-^2},-\fft1c\fft{i\rho_-}{\rho_+^2-\rho_-^2}-\fft{d}{c}).
\end{align}\label{period:tphi:Euclid}%
\end{subequations}
Note that not all integers $(c,d)$ yield inequivalent periodic identifications through (\ref{period:tphi:Euclid}): a set of integers $(c,d)\in\mathbb N\times\mathbb Z_c~(\mathbb Z_c=\{0,1,\cdots,c-1\})$ with $\gcd(c,d)=1$ represents all the inequivalent periodic identifications.

The family of Euclidean BTZ black holes constructed with different periodic identifications (\ref{period:tphi:Euclid}), however, is \underline{not} labelled completely by a set of integers $(c,d)\in\mathbb N\times\mathbb Z_c$ with $\gcd(c,d)=1$. This is because the metric (\ref{BTZ:Euclid}) itself depends on $\rho_\pm$, which is a function of integers $c$ and $d$ through (\ref{rhopm:Euclid}) for a given modular parameter $\tau$. To be specific, even though $(c,d)$ and $(c,d+c\mathbb Z)$ yield equivalent periodic identifications according to (\ref{period:tphi:Euclid}), they give different $\rho_\pm$ values through (\ref{rhopm:Euclid}) for a given modular parameter $\tau$ and therefore the corresponding Euclidean black holes are distinguished. Hence an element of the family of Euclidean BTZ black holes constructed with periodic identifications (\ref{period:tphi:Euclid}) should be labelled by $(c,d)\in\mathbb N\times\mathbb Z$ with $\gcd(c,d)=1$.\footnote{On the contrary, in \cite{Murthy:2009dq}, the black hole mass and the angular momentum, or equivalently $\rho_\pm$, are fixed and therefore $(c,d)$ and $(c,d+c\mathbb Z)$ label exactly the same geometry.} Note that we can still exclude the cases with a negative integer $c$ since the Euclidean BTZ metric (\ref{BTZ:Euclid}) is not sensitive to the sign flip $(c,d)\to(-c,-d)$ under (\ref{rhopm:Euclid})

Consequently, we have constructed a family of Euclidean BTZ black holes, whose elements are labelled by a set of integers $(c,d)\in\mathbb N\times\mathbb Z$ with $\gcd(c,d)=1$ and satisfy the following properties:
\begin{equation}
\begin{split}
	1)&\text{ they are locally given as (\ref{BTZ:Euclid}) where }\rho_\pm\text{ are given as (\ref{rhopm:Euclid})},\\
	2)&\text{ they have a torus conformal boundary with a modular parameter }\tau,\\
	3)&\text{ their }(t_E,\phi)\text{ coordinates are periodically identified as (\ref{period:tphi:Euclid})}.
\end{split}\label{BTZ:Euclid:family}
\end{equation}
%

\subsubsection*{Extremal limit in the Lorentzian signature}
This time we discuss the extremal limit $\rho_+\to\rho_-$. We cannot impose the extremal limit directly to the family of Euclidean BTZ black holes constructed as (\ref{BTZ:Euclid:family}), however, since $\rho_+\to\rho_-$ is not allowed for a finite modular parameter $\tau$ under the relation (\ref{rhopm:Euclid}). Here we circumvent this issue by going back to the Lorentzian signature, where one can take the extremal limit to obtain the family of extremal BTZ black holes. See the next subsection \ref{sec:BS:sl2z:5d} for why we take the extremal limit.

To begin with, one can obtain a family of Lorentzian BTZ black holes from the Euclidean family (\ref{BTZ:Euclid:family}) simply by replacing a Euclidean time coordinate $t_E$ with $t_E\to it$. An important consequence of this inverse Wick rotation is that a complex coordinate $w$ and its conjugate $\bar w$ from (\ref{w:tphi}) become real as 
\begin{equation}
	w=\fft{\phi+it_E}{2\pi}\to\fft{\phi-t}{2\pi},\qquad \bar w=\fft{\phi-it_E}{2\pi}\to\fft{\phi+t}{2\pi},
\end{equation}
and they are not complex conjugates of each other. Hence $\tau$ and $\bar\tau$ introduced in (\ref{period:asymp}) cannot also be complex conjugate of each other. In the Lorentzian signature, we therefore use $\bar w\to\tilde w$ and $\bar\tau\to\tilde\tau$ instead. Consequently, the relation between a modular parameter and $\rho_\pm$ given in (\ref{rhopm:Euclid}) becomes
\begin{equation}
	\rho_\pm=\fft{i}{2}\left(\fft{1}{c\tau+d}\mp\fft{1}{c\tilde\tau+d}\right)\label{rhopm:Lorentz}
\end{equation}
in the Lorentzian signature. The periods of $(t,\phi)$ Lorentzian coordinates are given from (\ref{period:asymp:tphi:non2},\ref{period:asymp:tphi:non3}), or equivalently from (\ref{period:tphi:Euclid}), as
\begin{subequations}
\begin{align}
	(t,\phi)&\sim(t,\phi)+2\pi(0,1),\\
	(t,\phi)&\sim(t,\phi)+2\pi(\fft{\tilde\tau-\tau}{2},\fft{\tilde\tau+\tau}{2})\sim(t,\phi)+2\pi(-\fft1c\fft{i\rho_+}{\rho_+^2-\rho_-^2},-\fft1c\fft{i\rho_-}{\rho_+^2-\rho_-^2}-\fft{d}{c}).
\end{align}\label{period:tphi:Lorentz}%
\end{subequations}

As a result, a family of Lorentzian BTZ black holes obtained from (\ref{BTZ:Euclid:family}) by the inverse Wick rotation $t_E\to it$ consists of various elements labelled by a set of integers $(c,d)\in\mathbb N\times\mathbb Z$ with $\gcd(c,d)=1$, which satisfy the following properties:
\begin{equation}
	\begin{split}
		1)&\text{ they are locally given as (\ref{BTZ:Lorentz}) where }\rho_\pm\text{ is given as (\ref{rhopm:Lorentz})},\\
		2)&\text{ their }(t,\phi)\text{ coordinates are periodically identified as (\ref{period:tphi:Lorentz})}.
	\end{split}\label{BTZ:Lorentz:family}
\end{equation}

Now one can impose the extremal limit $\rho_+\to\rho_-$, or equivalently $\tilde\tau\to\infty$ according to (\ref{rhopm:Lorentz}), to the family of Lorentzian BTZ black holes constructed as (\ref{BTZ:Lorentz:family}). Since the period of a time coordinate $t$ given in (\ref{period:tphi:Lorentz}) diverges under the extremal limit, however, we first take the following coordinate transformation from ($t,\phi,\rho$) to ($t',\phi',r$) \cite{Banerjee:2008ky,Murthy:2009dq}:
\begin{equation}
	t'=(\rho_+-\rho_-)(t+\phi),\qquad \phi'=\phi-\fft{\rho_-}{\rho_+}t,\qquad r=\fft{2\rho^2-\rho_+^2-\rho_-^2}{\rho_+^2-\rho_-^2}.\label{prime}
\end{equation}
The Lorentzian BTZ black hole (\ref{BTZ:Lorentz}) then reads
\begin{empheq}[box=\fbox]{equation}
	ds^2=-\fft14(r^2-1)dt'{}^2+\fft{dr^2}{4(r^2-1)}+\rho_+^2\left(d\phi'+\fft{r-1}{2\rho_+}dt'\right)^2.\label{BTZ:Lorentz:prime}
\end{empheq}
The periods of new $(t',\phi')$ coordinates are given from the unprimed one (\ref{period:tphi:Lorentz}) and the coordinate transformation (\ref{prime}) as
\begin{subequations}
\begin{align}
	(t',\phi')&\sim(t',\phi')+2\pi(\rho_+-\rho_-,1),\\
	(t',\phi')&\sim(t',\phi')+2\pi((\rho_+-\rho_-)\tilde\tau,\fft{(\rho_++\rho_-)\tau+(\rho_+-\rho_-)\tilde\tau}{2\rho_+}).
\end{align}\label{t'phi':period}%
\end{subequations}

In the above introduced primed coordinates, we take the extremal limit and obtain the family of extremal BTZ black holes. First, the parameter $\rho_+$ in (\ref{BTZ:Lorentz:prime}) becomes
\begin{empheq}[box=\fbox]{equation}
	\rho_+=\fft{i}{2(c\tau+d)}\label{rhop:ext}
\end{empheq}
in the extremal limit $\rho_+\to\rho_-~(\tilde\tau\to\infty)$ from (\ref{rhopm:Lorentz}). Taking the extremal limit and using the relation (\ref{rhopm:Lorentz}) also simplify the periods of $(t',\phi')$ coordinates (\ref{t'phi':period}) as
\begin{subequations}
\begin{empheq}[box=\fbox]{align}
	(t',\phi')&\sim(t',\phi')+2\pi(0,1),\\
	(t',\phi')&\sim(t',\phi')-2\pi(i/c,d/c).
\end{empheq}\label{period:t'phi':Lorentz}%
\end{subequations}
Note that a primed time coordinate $t'$ has a finite period under the extremal limit. 

Finally, the resulting family of extremal BTZ black holes consists of various elements labelled by a set of integers $(c,d)\in\mathbb N\times\mathbb Z$ with $\gcd(c,d)=1$, which satisfy the following properties:
\begin{equation}
	\begin{split}
		1)&\text{ they are locally given as (\ref{BTZ:Lorentz:prime}) where }\rho_+\text{ is given as (\ref{rhop:ext})},\\
		2)&\text{ their }(t',\phi')\text{ coordinates are periodically identified as (\ref{period:t'phi':Lorentz})}.
	\end{split}\label{AdS3:BH:farey}
\end{equation}
Note that the metric of this family of extremal BTZ black holes (\ref{BTZ:Lorentz:prime}) could be complex for a generic modular parameter $\tau$ due to the relation (\ref{rhop:ext}). We will not worry about this issue, however, since we are mainly interested in the regularized on-shell action in the Euclidean signature obtained by the Wick rotation $t'=-it_E'$\footnote{Note that this Wick rotation is distinguished from the one mentioned above (\ref{BTZ:Euclid}), namely $t=-it_E$ for unprimed coordinates.}, where complex bosonic fields are allowed in principle. See the next subsection \ref{sec:BS:on-shell} for a detailed calculation of the regularized on-shell action.

\subsubsection{Family of 5d extremal solutions}\label{sec:BS:sl2z:5d}
The family of extremal solutions to the BPS equations of 5d $\mathcal N=2$ gauged STU model (\ref{N=2:sugra:BPS}) is obtained by replacing the near-horizon extremal BTZ part of (\ref{AdS3:NH:metric}), namely (\ref{AdS3:local:1}), with the family of extremal BTZ black holes constructed as (\ref{AdS3:BH:farey}). Each element of this family is labelled by a set of integers $(c,d)\in\mathbb N\times\mathbb Z$ with $\gcd(c,d)=1$, and given explicitly as
\begin{subequations}
\begin{empheq}[box=\fbox]{align}
	ds^2&=\left(\fft{8p^1p^2p^3\Pi}{\Theta^3}\right)^\fft23\left(-\fft14(r^2-1)dt'{}^2+\rho_+^2\left(d\phi'-\fft{r-1}{2\rho_+}dt'\right)^2+\fft{dr^2}{4(r^2-1)}\right)\nn\\
	&\quad+\left(\fft{(p^1p^2p^3)^2}{\Pi}\right)^\fft13e^{2h_{\mathfrak g}(x,y)}(dx^2+dy^2),\label{family:sol:metric}\\
	A^I&=-p^I\omega_{\mathfrak g}\quad\to\quad F^I=-p^Ie^{2h_{\mathfrak g}(x,y)}dx\wedge dy,\\
	X^I&=\fft{p^I(p^1+p^2+p^3-2p^I)}{(p^1p^2p^3\Pi)^\fft13},\\
	-\kappa&=p^1+p^2+p^3.
\end{empheq}\label{family:sol}%
\end{subequations}
Here $\rho_+$ is given as (\ref{rhop:ext}) and $(t',\phi')$ coordinates are periodic as (\ref{period:t'phi':Lorentz}), and both of them depend on a label $(c,d)$. See Appendix \ref{App:N=2:BPS} to check directly that (\ref{family:sol}) solves the BPS equations (\ref{N=2:sugra:BPS}). 

For the resulting family of 5d solutions (\ref{family:sol}) to correspond to the near-horizon limit of a family of supersymmetric magnetically charged AdS$_5$ black strings, we think that the extremal limit introduced in subsection \ref{sec:BS:sl2z:farey}, which yields the parameter relation (\ref{rhop:ext}) and periodicity of coordinates (\ref{period:t'phi':Lorentz}), is crucial. The family of supersymmetric magnetically charged AdS$_5$ black strings has not yet been constructed so this expectation is not confirmed yet: however, at least in the special case where all three magnetic charges are identical, the known full black string solution (\ref{AdS5:BS}) does yield an extremal BTZ geometry in the near-horizon limit as observed in subsection \ref{sec:BS:STU} and thereby supports our expectation. Furthermore, a successful comparison in section \ref{sec:compare} relies heavily on the extremal limit that gives (\ref{rhop:ext}) and (\ref{period:t'phi':Lorentz}). This explains why we focus on the family of 5d solutions in the extremal limit, namely (\ref{family:sol}) with (\ref{rhop:ext}) and (\ref{period:t'phi':Lorentz}).

\subsection{Gravitational partition function}\label{sec:BS:on-shell}
As mentioned in subsection \ref{sec:BS:STU}, a supersymmetric magnetically charged AdS$_5$ black string solution of $\mathcal N=2$ gauged STU model is holographic dual to the ensemble of BPS states of $\mathcal N=4$ SU($N$) SYM theory on $T^2\times S^2$. The AdS/CFT correspondence then implies that the gravitational partition function $\mathcal I\simeq\exp\relax[-S^E_\text{reg,BS}]$ given in terms of the regularized on-shell action of a black string solution $S^E_\text{reg,BS}$ is supposed to match a dual canonical partition function derived from the twisted index by the inverse Laplace transform. Unfortunately, however, we have not yet figured out how to compute the regularized on-shell action of a full AdS$_5$ black string of the form (\ref{BS:ansatz}) whose near-horizon limit corresponds to the family of 5d extremal solutions (\ref{family:sol}).

Hence we propose an alternative way to reproduce the canonical partition function from the gravitational side. To begin with, note that an AdS$_3\times S^2$ near-horizon solution (\ref{AdS3:NH}) of $\mathcal N=2$ gauged STU model is holographically dual to the ensemble of BPS states of 2d SCFT on $T^2$, arising from the Kaluza-Klein (KK) compactification of $\mathcal N=4$ SU($N$) SYM theory on $S^2$ (of $T^2\times S^2$) \cite{Hosseini:2016cyf}. Based on this duality, we expect that the gravitational partition function $\mathcal I\simeq\exp\relax[-S^E_\text{reg,NH}]$ given in terms of the regularized on-shell action of a \underline{near-horizon} solution $S^E_\text{reg,NH}$ matches a dual canonical partition function derived from the elliptic genus. Since the 4d twisted index behaves like a 2d elliptic genus \cite{Hong:2018viz}, this canonical partition function is equivalent to the one derived from the twisted index by the inverse Laplace transform, which is what we want to reproduce from the gravitational side.

Since we have extended a near-horizon solution (\ref{AdS3:NH}) to the family of 5d extremal solutions (\ref{family:sol}), we will compute the regularized on-shell action of this family. The gravitational partition function derived from these on-shell actions will then be compared with the canonical partition function in section \ref{sec:compare}.

To compute the regularized on-shell action of the family of 5d extremal solutions (\ref{family:sol}), first we substitute (\ref{family:sol}) to the bulk action (\ref{N=2:sugra:action:5d}). The result is given under the Wick rotation $t'=-it'_E$ as
\begin{equation}
\begin{split}
	S^E_\text{bulk}=-iS_\text{bulk}&=-\fft{\text{vol}[\Sigma_{\mathfrak g}]}{16\pi G^{(5)}_N}\left(\fft{8p^1p^2p^3\Pi}{\Theta^3}\right)^\fft13\left(\fft{(p^1p^2p^3)^2}{\Pi}\right)^\fft13\int d^3x_E\sqrt{|g_{(3)}|}(R_{(3)}+2)\\
	&=-\fft{1}{16\pi G^{(3)}_N}\int d^3x_E\sqrt{|g_{(3)}|}(R_{(3)}+2),
\end{split}\label{SE:bulk}
\end{equation}
where $g_{(3)}$ and $R_{(3)}$ represent the metric and the Ricci scalar of an extremal BTZ part, namely (\ref{BTZ:Lorentz:prime}) in (\ref{family:sol:metric}). Note that we have introduced the 3d Newton's constant in the 2nd line of (\ref{SE:bulk}) as
\begin{equation}
	G^{(5)}_N=\text{vol}[\Sigma_{\mathfrak g}]\fft{2p^1p^2p^3}{\Theta}G^{(3)}_N.\label{G5:G3}
\end{equation}

Next we substitute (\ref{family:sol}) to the Gibbons-Hawking-York (GHY) boundary action that gives
\begin{equation}
	S^E_\text{GHY}=-iS_\text{GHY}=-\fft{1}{8\pi G^{(5)}_N}\int d^4x_E\sqrt{|\partial g|}K_{(4)}=-\fft{1}{8\pi G^{(3)}_N}\int d^2x_E\sqrt{|\partial g_{(3)}|}K_{(2)},\label{SE:GHY}
\end{equation}
where $K_{(4)}$ and $K_{(2)}$ are extrinsic curvatures on the boundaries of 5d extremal solutions (\ref{family:sol:metric}) and of an extremal BTZ metric (\ref{BTZ:Lorentz:prime}) respectively. Note that we have used
\begin{equation}
	K_{(4)}=\left(\fft{8p^1p^2p^3\Pi}{\Theta^3}\right)^{-\fft13}K_{(2)}
\end{equation}
in the last equation of (\ref{SE:GHY}), which can be derived from the definition of an extrinsic curvature
\begin{equation}
	K_{\mu\nu}=\nabla_\mu n_\nu-n_\mu n^\sigma\nabla_\sigma n_\nu\quad(K=K^\mu{}_\mu)\label{def:K}
\end{equation}
with $n^\mu$ a normal vector on the boundary (see Appendix D of \cite{Carroll:2004st} for example). In (\ref{SE:GHY}), $\partial g$ denotes a metric induced on a boundary.

Now the regularized on-shell action of a family of 5d extremal solutions (\ref{family:sol}) reads
\begin{equation}
	S^E_\text{reg}=-\fft{1}{16\pi G^{(3)}_N}\int d^3x_E\sqrt{|g_{(3)}|}(R_{(3)}+2)-\fft{1}{8\pi G^{(3)}_N}\int d^2x_E\sqrt{|\partial g_{(3)}|}K_{(2)}+S^E_\text{ct}\label{SE:reg:1}
\end{equation}
from (\ref{SE:bulk}) and (\ref{SE:GHY}), where $S^E_\text{ct}$ is designed to cancel divergent terms from the bulk action and the GHY term. Substituting the extremal BTZ part (\ref{BTZ:Lorentz:prime}) of 5d extremal solutions (\ref{family:sol:metric}) into the 3d metric $g_{(3)}$ in the regularized on-shell action (\ref{SE:reg:1}), one can compute (\ref{SE:reg:1}) explicitly as
\begin{equation}
	R_{(3)}=-6~~\&~~K_{(2)}=\fft{2r}{\sqrt{r^2-1}}\quad\to\quad S^E_\text{reg}=-\fft{\beta_{t_E'}\beta_\phi\rho_+}{16\pi G_N^{(3)}}(r_0+1)+S^E_\text{ct},\label{SE:reg:2}
\end{equation}
where we have introduced the radial cutoff at $r=r_0\,(\to\infty)$. To remove the $r_0$-divergence in (\ref{SE:reg:2}), we introduce the counter term as\,\footnote{In the 5d point of view, (\ref{SE:ct}) corresponds to the counter term in terms of a superpotential $W=g\sum_IX^I$ (refer to \cite{Liu:2007rv} for example; here we take $g=1$), which respects supersymmetry (refer to Appendix C of \cite{Bobev:2013cja} for example):
\begin{equation}
	S^E_\text{ct}=\fft{1}{16\pi G_N^{(5)}}\int d^4x_E\sqrt{|\partial g|}\,W=\fft{1}{8\pi G_N^{(3)}}\int d^2x_E\sqrt{|\partial g_{(3)}|}.\nn
\end{equation}
}
\begin{equation}
	S^E_\text{ct}=\fft{1}{8\pi G_N^{(3)}}\int d^2x_E\sqrt{|\partial g_{(3)}|}=\fft{\beta_{t_E'}\beta_\phi}{8\pi G_N^{(3)}}\fft{\rho_+\sqrt{r_0^2-1}}{2}=\fft{\beta_{t_E'}\beta_\phi\rho_+}{16\pi G_N^{(3)}}r_0+\mathcal O((r_0)^{-1}).\label{SE:ct}
\end{equation}
Substituting the counter term (\ref{SE:ct}) into the regularized on-shell action into (\ref{SE:reg:2}) and using the periods of Euclidean coordinates
\begin{equation}
	\beta_{t_E'}=\fft{2\pi}{c},\qquad\beta_\phi=2\pi,
\end{equation}
from (\ref{period:t'phi':Lorentz}) with $t_E'=it'$, we obtain
\begin{equation}
	S_E^\text{reg}=-\fft{\pi\rho_+}{4cG^{(3)}_N}=-\fft{\pi\rho_+\text{vol}[\Sigma_{\mathfrak g}]}{2cG^{(5)}_N}\fft{p^1p^2p^3}{\Theta}.\label{SE:reg:3}
\end{equation}
Here we have also used (\ref{G5:G3}) in the second equation.

Substituting $\text{vol}[\Sigma_{\mathfrak g}]=4\pi|\mathfrak g-1|$ and the expression of $\rho_+$ in terms of a modular parameter $\tau$ in the extremal limit, namely (\ref{rhop:ext}), we obtain the final expression for the regularized on-shell action of a family of 5d extremal solutions (\ref{family:sol}). For a particular extremal solution labelled by $(c,d)\in\mathbb N\times\mathbb Z$ with $\gcd(c,d)=1$, the result is given as
\begin{empheq}[box=\fbox]{equation}
	S^E_\text{reg}=-\fft{\pi^2 i|\mathfrak g-1|}{c^2(\tau+d/c)G^{(5)}_N}\fft{p^1p^2p^3}{\Theta}.\label{SE:reg:tau}
\end{empheq}
Note that this is different from the Bekenstein-Hawking entropy computed from the metric (\ref{family:sol:metric}) by a factor of $-\fft{1}{2c}$, namely $S^E_\text{reg}=-\fft{1}{2c}S_\text{BH}$.

Consequently, the gravitational partition function can be approximated from (\ref{SE:reg:tau}) as
\begin{empheq}[box=\fbox]{equation}
\begin{split}
	\mathcal I(\tau,p^a)&=\sum\exp[-S^E_\text{reg}+\mathcal O((G^{(5)}_N)^0)]\\
	&=\sum_{c\in\mathbb N}\sum_{d\in\mathbb Z}^{\gcd(c,d)=1}\exp[\fft{\pi^2 i|\mathfrak g-1|}{c^2(\tau+d/c)G^{(5)}_N}\fft{p^1p^2p^3}{\Theta}+\mathcal O((G^{(5)}_N)^0)]\label{gr:ptn:fct}
\end{split}
\end{empheq}
in the small Newton's constant limit, which corresponds to the large-$N$ limit under the AdS/CFT dictionary (\ref{dictionary}). In the first equation of (\ref{gr:ptn:fct}), the sum has to be taken over asymptotically AdS$_3\times\Sigma_{\mathfrak g}$ solutions whose conformal boundary read $T^2\times\Sigma_{\mathfrak g}$ in the Euclidean signature, where the complex structure of $T^2$ is specified by a modular parameter $\tau$.

\section{Comparison of partition functions}\label{sec:compare}
In this section, we compare the result from the field theory side in section \ref{sec:TTI} with the one from the gravity side in section \ref{sec:BS}.

In the field theory side, the canonical partition function derived from the twisted index of $\mathcal N=4$ SU($N$) SYM theory on $T^2\times S^2$ by the inverse Laplace transform is given from (\ref{eq:Omega:split:result}) as
\begin{equation}
	\Omega(\tau,\mathfrak n_a)=\sum_{c\in\mathbb N}\sum_{d\in\mathbb Z}\exp[\fft{N^2\pi i}{4c^2(\tau+d/c)}\fft{\mathfrak n_1\mathfrak n_2\mathfrak n_3}{1-\mathfrak n_1\mathfrak n_2-\mathfrak n_2\mathfrak n_3-\mathfrak n_3\mathfrak n_1}+o(N^2)]+\Omega_\text{non-st}'(\tau,\mathfrak n_a).\label{eq:Omega:split:result:repeat}
\end{equation}
In the gravity side, the gravitational partition function approximated as a sum over 5d extremal solutions (\ref{family:sol}) in $\mathcal N=2$ gauged STU model is given from (\ref{gr:ptn:fct}) as
\begin{equation}
	\mathcal I(\tau,p^a)=\sum_{c\in\mathbb N}\sum_{d\in\mathbb Z}^{\gcd(c,d)=1}\exp[-\fft{\pi^2 i}{c^2(\tau+d/c)G^{(5)}_N}\fft{p^1p^2p^3}{\Theta}+\mathcal O((G^{(5)}_N)^0)]\label{gr:ptn:fct:repeat}
\end{equation}
for $\mathfrak g=0$, where $\Theta$ is defined in (\ref{PiTheta}). Using the AdS/CFT dictionary
\begin{equation}
	\mathfrak n_a=-2p^a,\qquad N^2=\fft{\pi}{2G^{(5)}_N},\label{dictionary}
\end{equation}
and the constraint on magnetic charges $\sum_{a=1}^3\mathfrak n_a=2$ given in (\ref{constraint:chemical}), it is straightforward to check that the exponent in (\ref{eq:Omega:split:result:repeat}) matches the exponent in (\ref{gr:ptn:fct:repeat}) in the large-$N$ limit. 

However, there is a subtle issue for the precise match between a canonical partition function (\ref{eq:Omega:split:result:repeat}) and a gravitational partition function (\ref{gr:ptn:fct:repeat}): the sum of labels $(c,d)$ is taken over $(c,d)\in\mathbb N\times\mathbb Z$ without the relatively prime condition $\gcd(c,d)=1$ in (\ref{eq:Omega:split:result:repeat}), which is different from the sum in (\ref{gr:ptn:fct:repeat}). Due to this difference, some of contributions in (\ref{eq:Omega:split:result:repeat}) do not seem to have counterparts in (\ref{gr:ptn:fct:repeat}). For example, the large-$N$ asymptotics of $\Omega_{\{2,N/2,N/4+1\}}$ is given from (\ref{eq:Omega:mnr:m-fin:result}) as
\begin{equation}
	\Omega_{\{2,N/2,N/4+1\}}=\exp[\fft{N^2\pi i}{64(\tau+1/2)}\fft{\mathfrak n_1\mathfrak n_2\mathfrak n_3}{1-\mathfrak n_1\mathfrak n_2-\mathfrak n_2\mathfrak n_3-\mathfrak n_3\mathfrak n_1}+o(N^2)],
\end{equation}
where $(c,d)=(mp,mps)=(4,2)$ is determined from (\ref{tau:e}) and (\ref{claim:mpe}): 
\begin{equation}
	\{m,n,r\}=\{2,N/2,N/4+1\}~~\text{with}~~s=\fft12~~\to~~\fft{r-ms}{n}=\fft{q}{p}=\fft12.
\end{equation}
Since $\gcd(c,d)=2\neq1$ for this particular contribution, $\Omega_{\{2,N/2,N/4+1\}}$ does not have a holographic dual contribution to the gravitational partition function in (\ref{gr:ptn:fct:repeat}).

There are two different possibilities for this issue to be resolved. First one is that the contributions labelled by $(c,d)\in\mathbb N\times\mathbb Z$ with $\gcd(c,d)\neq1$ cancel each other in (\ref{eq:Omega:split:result:repeat}). Recall that we have ignored pure imaginary terms in the exponent of various contributions in (\ref{eq:Omega:split:result:repeat}) under the large-$N$ limit, which is hidden in $o(N^2)$, since they are defined modulo $2\pi i\mathbb Z$. But they can still provide non-trivial phases to various contributions labelled by $(c,d)\in\mathbb N\times\mathbb Z$. Therefore, when there is more than one contributions labelled by the same set of integers $(c,d)$ in (\ref{eq:Omega:split:result:repeat}), there is a possibility for them to cancel each other. It is highly non-trivial to prove the statement, however, that \emph{such a cancellation happens precisely for $(c,d)\in\mathbb N\times\mathbb Z$ with $\gcd(c,d)\neq1$}. This is mainly because one cannot rely on the large-$N$ asymptotics to compute the phase of a standard contribution $\Omega_{\{m,n,r\}}$ precisely and, furthermore, non-standard contributions must be taken into account for a complete proof of the statement. We leave the test of this statement for future research, together with a complete analysis of the remaining non-standard contribution $\Omega'_\text{non-st}$ in (\ref{eq:Omega:split:result:repeat}).

Another possibility is that the gravitational partition function (\ref{gr:ptn:fct:repeat}) is missing the contributions labelled by $(c,d)\in\mathbb N\times\mathbb Z$ with $\gcd(c,d)\neq1$. This possibility is reminiscent of the observation that the complete sum of contributions from the SL(2,$\mathbb Z$) family of BTZ black holes \cite{Maldacena:1998bw,Dijkgraaf:2000fq} to the path integral does not yield a physically sensible partition function \cite{Maloney:2007ud}, which has been revisited in the interesting recent work \cite{Maxfield:2020ale} that shows the existence of non-trivial contributions from a class of new topologies without classical solutions. Similarly, the family of 5d extremal solutions (\ref{family:sol}) derived from the SL(2,$\mathbb Z$) family of BTZ black holes might not be enough to yield a correct gravitational partition function that matches a dual canonical partition function (\ref{eq:Omega:split:result:repeat}) including the contributions labelled by $(c,d)\in\mathbb N\times\mathbb Z$ with $\gcd(c,d)\neq1$. This could be another interesting direction to explore in the future.

\section{Discussion}\label{sec:discussion}
Here we discuss a few remaining questions for future research.

\subsection*{Complete match of partition functions}
First of all, it is important to understand that the comparison of a canonical partition function (\ref{eq:Omega:split:result:repeat}) and a gravitational partition function (\ref{gr:ptn:fct:repeat}) is not complete yet for the following reasons.
\begin{itemize}
	\item The comparison is based on the conjecture introduced in \ref{sec:TTI:micro:compute} that, if the contribution labelled by $(c,d)$ in (\ref{eq:Omega:split:result:repeat}) does not come from a standard BAE solution, there is a non-standard BAE solution that gives the same contribution. This conjecture has not yet been proven but there is a partial evidence in the literature \cite{ArabiArdehali:2019orz,Hong:2021dja}.
	\item The sum over a set of integers $(c,d)$ are different in (\ref{eq:Omega:split:result:repeat}) and (\ref{gr:ptn:fct:repeat}). To remedy the difference, one needs to figure out if the contributions labelled by $(c,d)\in\mathbb N\times\mathbb Z$ with $\gcd(c,d)\neq1$ in (\ref{eq:Omega:split:result:repeat}) cancel each other. If there is a precise cancellation, the match between (\ref{eq:Omega:split:result:repeat}) and (\ref{gr:ptn:fct:repeat}) becomes complete upto the remaining non-standard contribution $\Omega_\text{non-st}'$ in (\ref{eq:Omega:split:result:repeat}). Otherwise holographic duals of the contributions labelled by $(c,d)\in\mathbb N\times\mathbb Z$ with $\gcd(c,d)\neq1$ have to be explained, which are currently missing in (\ref{gr:ptn:fct:repeat}).
	\item The remaining non-standard contribution $\Omega_\text{non-st}'$ in (\ref{eq:Omega:split:result:repeat}) has not yet been computed. If it turns out not to vanish, the corresponding holographic dual has to be explained, which is currently missing in (\ref{gr:ptn:fct:repeat}).
\end{itemize}
To deal with the above questions and thereby make the comparison of a canonical partition function (\ref{eq:Omega:split:result:repeat}) and a gravitational partition function (\ref{gr:ptn:fct:repeat}) more solid, systematic understanding of non-standard solutions to the BAE (\ref{BAE}) is required. We leave this analysis for future research.

\subsection*{Application to generic AdS$_5$ black strings and refined indices}
Next, one may extend what we have discussed so far to more generic AdS$_5$ black string solutions and dual topologically twisted indices refined accordingly. To be specific, one may add electric charges associated with flavor symmetries \cite{Hristov:2014hza} and rotations \cite{Hosseini:2019lkt,Hosseini:2020vgl} to a purely magnetic AdS$_5$ black string solution of $\mathcal N=2$ gauged STU model studied in this paper, where the latter was done by uplifting a 4d dyonic rotating black holes in \cite{Hristov:2018spe}. The refined topologically twisted index dual to this generalized AdS$_5$ black string has also been known \cite{Hosseini:2019lkt,Hosseini:2020vgl}. It is then natural to expect that the comparison in section \ref{sec:compare} can be generalized by matching a canonical partition function derived from the refined twisted index with a gravitational partition function derived from the near-horizon limit of a dyonic rotating AdS$_5$ black string. We leave this extension for future research. 

Similarly, one may extend our analysis to recently constructed spindle solutions where the Riemann surface $\Sigma_{\mathfrak g}$ used in this paper is replaced with a two-dimensional orbifold known as a spindle \cite{Ferrero:2020laf,Hosseini:2021fge}. In this case, however, a dual twisted index has not yet been explored enough and therefore a direct comparison between a canonical partition function and a gravitational partition function could be more demanding.

\subsection*{$\mathcal N=4$ SYM superconformal index and a black hole Farey tail}
We would also like to mention that a similar physics can be studied for the superconformal index of $\mathcal N=4$ SU($N$) SYM theory \cite{Kinney:2005ej,Romelsberger:2005eg} and its holography. Note that the $\mathcal N=4$ SYM superconformal index allows for the BA formula \cite{Benini:2018ywd} with the BAE same as the one for the topologically twisted index (\ref{BAE}), provided chemical potentials associated with angular momenta are identical (refer to \cite{Benini:2020gjh} for the BA formula with generic chemical potentials). This means that the $\mathcal N=4$ SYM superconformal index receives contributions from the same family of BAE solutions with the SL(2,$\mathbb Z$) modular structure, which we reviewed in subsection \ref{sec:TTI:BA}. Hence we expect that the $\mathcal N=4$ SYM superconformal index is closely related to the black hole Farey tail in the gravity side as is the twisted index. 

In fact, a relation between the black hole Farey tail and the superconformal index has already been anticipated in \cite{Cabo-Bizet:2019eaf}. To make such a relation more explicit, following what we have done for the twisted index, one may need to explore a locally AdS$_3$ geometry in the near-horizon limit of a rotating electrically charged AdS$_5$ black hole \cite{Kunduri:2006ek} dual to the superconformal index, which has partially been pursued in \cite{David:2020ems}. The result of this line of research is expected to improve holographic understanding of the superconformal index by relating orbifold solutions constructed in \cite{Aharony:2021zkr}, which are holographically dual to partially deconfined phases \cite{ArabiArdehali:2019orz} or equivalently to generalized Cardy-like asymptotics \cite{ArabiArdehali:2021nsx}, to the black hole Farey tail. Also refer to \cite{Jejjala:2021hlt} for a recent comment on a possible relation between the black hole Farey tail and the $\mathcal N=4$ SYM superconformal index for generic chemical potentials with SL(3,$\mathbb Z$) modularity.

\subsection*{Euclidean black saddles and the grand-canonical ensemble}
Another interesting direction is to explore Euclidean black saddles \cite{Bobev:2020pjk} that include the family of 5d extremal solutions we have constructed in subsection \ref{sec:BS:sl2z} as a special case. In \cite{Bobev:2020pjk}, the authors constructed a large class of new Euclidean solutions to the BPS equations of 4d $\mathcal N=2$ gauged STU model, dubbed as Euclidean black saddles, and matched their regularized on-shell actions with the logarithm of the topologcially twisted index of dual ABJM theory. They are matched as functions of chemical potentials, not charges: in other words, they are matched in grand-canonical ensemble. Based on this observation, we expect that similar Euclidean black saddles exist in 5d $\mathcal N=2$ gauged STU model and provide holographic duals of the $\mathcal N=4$ SYM twisted index in grand-canonical ensemble. This will extend a current holographic understanding of the $\mathcal N=4$ SYM twisted index further, which is investigated only at canonical level in this paper. For example, the properties of the twisted index as a weak Jacobi form \cite{Hong:2018viz}, including the modular property reviewed in subsection \ref{sec:TTI:large-N:other}, can be studied holographically in grand-canonical ensemble where chemical potentials $\Delta_a$ are not fixed; in this paper, we left $\tau$ as a free modular parameter but still fixed chemical potentials $\Delta_a$ for a comparison of partition functions. 

\subsection*{Generalization of a black hole Farey tail}
Lastly, it is important to reproduce the matching between a canonical partition function derived from the twisted index by the inverse Laplace transform, (\ref{eq:Omega:split:result}), and a gravitational partition function for a full AdS$_5$ black string of the form (\ref{BS:ansatz}), not for its near-horizon geometry (\ref{family:sol}). Even though we have explained in \ref{sec:BS:on-shell} why the gravitational partition function for a near-horizon solution (\ref{gr:ptn:fct}) is enough to reproduce a canonical partition function (\ref{eq:Omega:split:result}), the argument depends on the observation in the field theory side, namely the equivalence between the 2d elliptic genus and the 4d twisted index \cite{Hong:2018viz}. Hence, to reproduce the canonical partition function (\ref{eq:Omega:split:result}) purely from the gravitational side without a prior knowledge in a dual field theory, it is worthwhile to derive the final expression (\ref{gr:ptn:fct}) from the regularized on-shell action of a full black string solution.

For such a direct comparison, first we should figure out how to compute the regularized on-shell action of a full black string of the form (\ref{BS:ansatz}). Then we should construct a family of supersymmetric AdS$_5$ black strings whose conformal boundary $T^2\times S^2$ has a complex structure on $T^2$ specified by a modular parameter $\tau$, following subsection \ref{sec:BS:sl2z:farey}.\footnote{The family of 5d extremal solutions in \ref{sec:BS:sl2z:5d} is just the uplift of a set of extremal BTZ black holes and does not have AdS$_5$ asymptotics.} This step corresponds to generalizing the black hole Farey tail story to the AdS$_5$/CFT$_4$ correspondence; refer to \cite{Jejjala:2021hlt} for a relevant discussion in the context of duality between an AdS$_5$ black hole and the $\mathcal N=4$ superconformal index. Based on the above two achievements, one can compute the gravitational partition function from the regularized on-shell action of a family of supersymmetric AdS$_5$ black strings, following subsection \ref{sec:BS:on-shell}. For the reason we have explained above and in subsection \ref{sec:BS:on-shell}, we expect that the resulting gravitational partition function take the same form of (\ref{gr:ptn:fct}) and thereby match a dual canonical partition function (\ref{eq:Omega:split:result}). We leave this interesting direction for future research.

\section*{Acknowledgement}
We would like to appreciate Arash Arabi Ardehali and James T. Liu for interesting discussions in the early stage of this project. Special thanks to Arash Arabi Ardehali, Seyed Morteza Hosseini, Kiril Hristov, Sameer Murthy, Anton Nedelin, and Alberto Zaffaroni for kindly reviewing the draft and providing helpful feedback. This work is supported in part by a Grant for Doctoral Study from the Korea Foundation for Advanced Studies.

\appendix

\section{Elliptic functions}\label{App:elliptic}
The Dedekind eta function is defined as
\begin{equation}
	\eta(\tau)=e^{\fft{\pi i\tau}{12}}\prod_{k=1}^\infty(1-e^{2\pi ik\tau}).\label{def:eta}
\end{equation}
The elliptic theta function $\theta_1(\cdot;\cdot)$ has the following product form:
\begin{equation}
\begin{split}
	&\theta_1(u;\tau)\\
	&=-ie^{\fft{\pi i\tau}{4}}(e^{\pi iu}-e^{-\pi iu})\prod_{k=1}^\infty(1-e^{2\pi ik\tau})(1-e^{2\pi i(k\tau+u)})(1-e^{2\pi i(k\tau-u)})\\
	&=-i(-1)^me^{\fft{\pi i\tau}{4}}e^{\pi i[(2m+1)u+m(m+1)\tau]}\prod_{k=1}^\infty(1-e^{2\pi ik\tau})(1-e^{2\pi i((k+m)\tau+u)})(1-e^{2\pi i((k-m-1)\tau-u)})\\
	&=ie^{\fft{\pi i\tau}{4}}e^{-\pi iu}\theta_0(u;\tau)\prod_{k=1}^\infty(1-e^{2\pi ik\tau}).
\end{split}\label{def:theta1}%
\end{equation}

The elliptic theta function (\ref{def:theta1}) has a quasi-double-periodicity, namely
\begin{equation}
	\theta_1(u+m+n\tau,\tau)=(-1)^{m+n}e^{-2\pi inu}e^{-\pi in^2\tau}\theta_1(u;\tau),\label{theta1:periodic}
\end{equation}
for $m,n\in\mathbb Z$. The inversion formula can be written simply as
\begin{equation}
	\theta_1(-u;\tau)=-\theta_1(u;\tau).\label{theta1:inversion}
\end{equation}

The modular property of elliptic functions under SL(2,$\mathbb Z$) transformations are given as (see \cite{Brezhnev:2013} for example)
\begin{equation}
\begin{split}
	\eta\left(\fft{a\tau+b}{c\tau+d}\right)&=\xi\sqrt{c\tau+d}\,\eta(\tau),\\
	\theta_1\left(\fft{u}{c\tau+d};\fft{a\tau+b}{c\tau+d}\right)&=\xi^3\sqrt{c\tau+d}\,e^{\fft{\pi icu^2}{c\tau+d}}\theta_1(u;\tau),
\end{split}\label{eq:sl2z}%
\end{equation}
where $\xi$ is a 24-th root of unity and $a,b,c,d\in\mathbb Z$ with $ad-bc=1$.

Finally we investigate asymptotic behaviors of elliptic functions. To begin with, the Dedekind eta function is given in the limit where $|\tau|\to0^+$ with fixed $0<\arg\tau<\pi$ as
\begin{equation}
	\log\eta(\tau)=-\fft{\pi i}{12\tau}-\fft12\log(-i\tau)+\mathcal O(e^{-\fft{2\pi\sin(\arg\tau)}{|\tau|}}).\label{eta:asymp}
\end{equation}
To study asymptotic behaviors of an elliptic theta function, first we introduce a $\tau$-modded value of a complex number $u$, namely $\{u\}_\tau$, as
\begin{equation}
	\{u\}_\tau\equiv u-\lfloor\Re u-\cot(\arg\tau)\Im u\rfloor\quad(u\in\mathbb C).\label{tau-modded}
\end{equation}
By definition, the $\tau$-modded value satisfies 
\begin{equation}
	\{u\}_\tau=\{\tilde u\}_\tau+\check u\tau,\qquad
	\{-u\}_\tau=\begin{cases}
		1-\{u\}_\tau & (\tilde u\notin\mathbb Z)\\
		-\{u\}_\tau & (\tilde u\in\mathbb Z),
	\end{cases}
\end{equation}
where we have defined $\tilde u,\check u\in\mathbb R$ as 
\begin{equation}
	u=\tilde u+\check u\tau.\label{u:component}
\end{equation}
Note that, for a real number $x$, a $\tau$-modded value $\{x\}_\tau$ reduces to a normal modded value $\{x\}$ defined as 
\begin{equation}
	\{x\}\equiv x-\lfloor x\rfloor\quad(x\in\mathbb R).\label{modded}
\end{equation}
Now the elliptic theta function is given in the limit where $|\tau|\to0^+$ with fixed $0<\arg\tau<\pi$ as
\begin{equation}
\begin{split}
	\log\theta_1(u;\tau)&=\fft{\pi i}{\tau}\{u\}_\tau(1-\{u\}_\tau)-\fft{\pi i}{4\tau}(1-\tau)+\pi i\lfloor\Re u-\cot(\arg\tau)\Im u\rfloor-\fft12\log\tau\\
	&\quad+\log(1-e^{-\fft{2\pi i}{\tau}(1-\{u\}_\tau)})\left(1-e^{-\fft{2\pi i}{\tau}\{u\}_\tau}\right)+\mathcal O(e^{-\fft{2\pi\sin(\arg\tau)}{|\tau|}}),\label{theta1:asymp}
\end{split}
\end{equation}
based on an alternative product form ($m\in\mathbb Z$):
\begin{equation}
\begin{split}
	\theta_1(u;\tau)&=(-i\tau)^{-\fft12}e^{-\fft{\pi i}{4\tau}}e^{m\pi i}e^{\fft{\pi i}{\tau}(u-m)(1-u+m)}\\
	&\quad\times\prod_{k=1}^\infty(1-e^{-\fft{2\pi i}{\tau}k})(1-e^{-\fft{2\pi i}{\tau}(k-u+m)})(1-e^{-\fft{2\pi i}{\tau}(k-1+u-m)}).
\end{split}\label{def:theta1:alter}
\end{equation}
This product form (\ref{def:theta1:alter}) is derived by combining (\ref{def:theta1}) with the $S$-transformation, namely (\ref{eq:sl2z}) with $(a,b,c,d)=(0,-1,1,0)$.

\section{Upper bound for infinite product terms}
In this appendix, we estimate upper bounds for infinite product terms used to evaluate the large-$N$ asymptotics of $Z_{\{m,n,r\}}$ in subsection \ref{sec:TTI:large-N:other}. The base inequality reads
\begin{equation}
\begin{split}
	\left|\sum_{l=1}^\infty\log(1-e^{-lx+y})\right|&\leq-\sum_{l=1}^\infty\log(1-e^{-l\Re[x]+\Re[y]})\\
	&\leq-\fft{\Re[y]}{\Re[x]}\log(1-e^{-\Re[x]+\Re[y]})-\int_{\fft{\Re[y]}{\Re[x]}}^\infty dl\,\log(1-e^{-l\Re[x]+\Re[y]})\\
	&=-\fft{\Re[y]}{\Re[x]}\log(1-e^{-\Re[x]+\Re[y]})+\fft{\pi^2}{6\Re[x]},
\end{split}\label{base:ineq}
\end{equation}
which is valid under $\Re[x]>\Re[y]\geq0$. 

First we apply the base inquality (\ref{base:ineq}) to (\ref{m:not-finite:1}). Under the identifications
\begin{equation}
\begin{split}
	x&=-2\pi i\fft{m\tau+r}{n},\\
	y&\in\left\{0,-2\pi i\left(\fft{m\tau}{n}(1-\{\fft{n\Delta_a}{\tau}\}_{-\fft1\tau})+\fft{r}{n}(1+j_a)\right),-2\pi i\left(\fft{m\tau}{n}\{\fft{n\Delta_a}{\tau}\}_{-\fft1\tau}-\fft{r}{n}j_a\right)\right\},
\end{split}
\end{equation}
which satisfy the condition $\Re[x]>\Re[y]\geq0$ as (refer to (\ref{u:component}) for the decomposition of chemical potentials: $\Delta_a=\tilde\Delta_a+\check\Delta_a\tau$)
\begin{equation}
\begin{split}
	\Re[x]&=2\pi\fft{m}{n}|\tau|\sin(\arg\tau),\\
	\Re[y]&\in\left\{0,2\pi\fft{m}{n}|\tau|\sin(\arg\tau)(1-\{n\check\Delta_a\}),2\pi\fft{m}{n}|\tau|\sin(\arg\tau)\{n\check\Delta_a\}\right\},
\end{split}
\end{equation}
we obtain
\begin{equation}
\begin{split}
	&\left|\sum_{l=1}^\infty\log\fft{(1-e^{2\pi il\fft{m\tau+r}{n}})^2}{(1-e^{2\pi i((l-1)\fft{m\tau+r}{n}+\fft{m\tau}{n}\{\fft{n\Delta_a}{\tau}\}_{-\fft1\tau}-\fft{r}{n}j_a)})(1-e^{2\pi i(l\fft{m\tau+r}{n}-\fft{m\tau}{n}\{\fft{n\Delta_a}{\tau}\}_{-\fft1\tau}+\fft{r}{n}j_a)})}\right|\\
	&=\begin{dcases}
	\begin{rcases}
		\mathcal O(\fft{n}{m}) & (\lim_{N\to\infty}\fft{n}{m}=0)\\
		\mathcal O(1) & (\lim_{N\to\infty}\fft{n}{m}\neq0,\infty)\\
		\mathcal O(\fft{n}{m}) & (\lim_{N\to\infty}\fft{n}{m}=\infty)
	\end{rcases}
	\end{dcases}=\mathcal O(\fft{n}{m}).\label{m:not-finite:1:bound}
\end{split}
\end{equation}
Here we have assumed generic chemical potentials for $n\check\Delta_a\notin\mathbb Z$. Now it is clear that the infinite product term in (\ref{m:not-finite:1}) is bounded as in (\ref{m:not-finite:1:large-N}). 

If $\Delta_a$ is real, we have $\check\Delta_a=0$ and therefore the integer $j_a$ defined in (\ref{ja}) also vanishes. In this case, one should factor out $-\log(1-e^{2\pi im\Delta_a})$ from the LHS of (\ref{m:not-finite:1:bound}) to obtain the same upper bound in the RHS. This extra term will not affect the $N^2$-leading order in (\ref{TTI:mnr:large-N:m:not-finite:alter}) though.

Next we apply the base inequality (\ref{base:ineq}) to (\ref{m:finite:1}). Under the identifications
\begin{equation}
\begin{split}
	x&=2\pi i\left(\fft{N}{(mp)^2(\tau+s)}-\fft{a}{p}\right),\\
	y&\in\left\{0,2\pi i\left(\fft{N}{(mp)^2(\tau+s)}\{mp\Delta_a\}_{\tau+s}+\fft{a}{p}k_a\right),\right.\\
	&\qquad\left.2\pi i\left(\fft{N}{(mp)^2(\tau+s)}(1-\{mp\Delta_a\}_{\tau+s})-\fft{a}{p}(1+k_a)\right)\right\},
\end{split}
\end{equation}
which satisfy the condition $\Re[x]>\Re[y]\geq0$ as (refer to (\ref{u:component}) for the decomposition of chemical potentials: $\Delta_a=\tilde\Delta_a+\check\Delta_a\tau$)
\begin{equation}
\begin{split}
	\Re[x]&=\fft{2\pi N}{(mp)^2}\fft{\sin(\arg(\tau+s))}{|\tau+s|},\\
	\Re[y]&\in\left\{0,\fft{2\pi N}{(mp)^2}\fft{\sin(\arg(\tau+s))}{|\tau+s|}\{mp(\tilde\Delta_a-\check\Delta_as)\},\fft{2\pi N}{(mp)^2}\fft{\sin(\arg(\tau+s))}{|\tau+s|}(1-\{mp(\tilde\Delta_a-\check\Delta_as)\})\right\},
\end{split}
\end{equation}
we obtain
\begin{equation}
\begin{split}
	&\left|\sum_{l=1}^\infty\log\left[\left(1-e^{2\pi il(-\fft{N}{(mp)^2(\tau+s)}+\fft{a}{p})}\right)^2\right.\right.\\
	&\kern4em\times\left(1-e^{2\pi i(l(-\fft{N}{(mp)^2(\tau+s)}+\fft{a}{p})+\fft{N}{(mp)^2(\tau+s)}\{mp\Delta_a\}_{\tau+s}+\fft{a}{p}k_a)}\right)^{-1}\\
	&\kern4em\left.\left.\times\left(1-e^{2\pi i((l-1)(-\fft{N}{(mp)^2(\tau+s)}+\fft{a}{p})-\fft{N}{(mp)^2(\tau+s)}\{mp\Delta_a\}_{\tau+s}-\fft{a}{p}k_a)}\right)^{-1}\right]\right|\\
	&=\begin{dcases}
		\begin{rcases}
			\mathcal O(\fft{(mp)^2}{N}) & (\lim_{N\to\infty}\fft{(mp)^2}{N}=0)\\
			\mathcal O(1) & (\lim_{N\to\infty}\fft{(mp)^2}{N}\neq0,\infty)\\
			\mathcal O(\fft{(mp)^2}{N}) & (\lim_{N\to\infty}\fft{(mp)^2}{N}=\infty)
		\end{rcases}
	\end{dcases}=\mathcal O(\fft{(mp)^2}{N}).\label{m:finite:1:bound}
\end{split}
\end{equation}
Here we have assumed generic chemical potentials for $mp(\tilde\Delta_a-\check\Delta_as)\notin\mathbb Z$. Now it is clear that the infinite product term in (\ref{m:finite:1}) is bounded as in (\ref{m:finite:1:large-N}).

\section{$\mathcal N=2$ gauged STU model}\label{App:N=2}
The $\mathcal N=2$ gauged STU model refers to 5d $\mathcal N=2$ gauged supergravity coupled to two vector multiplets \cite{Gunaydin:1983bi,Gunaydin:1984ak}, which can be obtained from a consistent truncation of 10d Type \rom{2}B supergravity on AdS$_5\times S^5$ \cite{Cvetic:1999xp}. The bosonic action of this STU model is given in the convention of \cite{Cvetic:1999xp,Maldacena:2000mw} as
\begin{equation}
	\begin{split}
		S&=\fft{1}{16\pi G^{(5)}_N}\int d^5x\sqrt{|g|}\left[R+4g^2\sum_{i=1}^3\fft{1}{X^i}-\fft12\sum_{x=1}^2\partial_\mu\phi^x\partial^\mu\phi^x-\fft14\sum_{i=1}^3(X^i)^{-2}F^i_{\mu\nu}F^i{}^{\mu\nu}\right.\\
		&\kern10em\left.+\fft{1}{24}|\varepsilon_{ijk}|\varepsilon^{\mu\nu\rho\sigma\lambda}F^i_{\mu\nu}F^j_{\rho\sigma}A^k_\lambda\right],
	\end{split}\label{N=2:sugra:action:5d}
\end{equation}
where $x\in\{1,2\}$ and $i,j,k\in\{1,2,3\}$ and the Levi-Civita symbol is given as
\begin{equation}
	\varepsilon^{\mu\nu\rho\sigma\lambda}=\begin{cases}
		-|g|^{-1/2} & (\text{even permutation of }\mu\nu\rho\sigma\lambda)\\
		+|g|^{-1/2} & (\text{odd permutation of }\mu\nu\rho\sigma\lambda)
	\end{cases}.
\end{equation}
The physical scalars $\phi^x$ are parametrized by $X^i$ under the constraint $X^1X^2X^3=1$ as
\begin{equation}
	X^1=e^{-\fft{1}{\sqrt6}\phi^1-\fft{1}{\sqrt2}\phi^2},\qquad X^2=e^{-\fft{1}{\sqrt6}\phi^1+\fft{1}{\sqrt2}\phi^2},\qquad X^3=e^{\fft{2}{\sqrt6}\phi^1}.
\end{equation}
From the bosonic action (\ref{N=2:sugra:action:5d}), the Einstein equations are given as
\begin{equation}
	\begin{split}
		R_{\mu\nu}-\fft12g_{\mu\nu}(R+4g^2\sum_{i=1}^3(X^i)^{-1})&=\fft12\sum_{x=1}^2\partial_\mu\phi^x\partial_\nu\phi^x-\fft14g_{\mu\nu}\sum_{x=1}^2\partial_\rho\phi^x\partial^\rho\phi^x\\
		&\quad+\fft12\sum_{i=1}^3(X^i)^{-2}F^i_{\mu\rho}F^i{}_\nu{}^\rho-\fft18g_{\mu\nu}\sum_{i=1}^3(X^i)^{-2}F^i_{\rho\sigma}F^i{}^{\rho\sigma}.
	\end{split}\label{N=2:sugra:Einstein}
\end{equation}
The scalar equations of motion are given as
\begin{equation}
	\begin{split}
		0&=\nabla_\mu\nabla^\mu\phi^x-\fft14\sum_{i=1}^3\partial_{\phi^x}(X^i)^{-2}F^i_{\mu\nu}F^i{}^{\mu\nu}+4g^2\sum_{i=1}^3\partial_{\phi^x}(X^i)^{-1}.\label{N=2:sugra:scalar}
	\end{split}
\end{equation}
The Bianchi identity and the vector equations of motion are given as
\begin{subequations}
	\begin{align}
		0&=\partial_{[\mu} F^i_{\nu\rho]},\\
		0&=\nabla_\mu((X^i)^{-2}F^i{}^{\mu\nu})+\fft14\sqrt{|g|}|\varepsilon_{ijk}|\varepsilon^{\mu\lambda\rho\sigma\nu}F^j_{\mu\lambda}F^k_{\rho\sigma}.
	\end{align}\label{N=2:sugra:vector}%
\end{subequations}
The BPS equations are given as
\begin{subequations}
	\begin{align}
		0&=\left[\partial_\mu+\fft14\omega^{ab}_\mu\gamma_{ab}+\fft{i}{24}(\gamma_\mu{}^{\nu\rho}-4\delta_\mu^\nu\gamma^\rho)\sum_{i=1}^3(X^i)^{-1}F^i_{\nu\rho}+\fft{g}{6}\sum_{i=1}^3X^i\gamma_\mu-\fft{ig}{2}\sum_{i=1}^3A^i_\mu\right]\epsilon,\\
		0&=\left[-\fft{i}{4}\partial_\mu\phi^x\gamma^\mu+\fft18\sum_{i=1}^3(\partial_{\phi^x}(X^i)^{-1})F^i_{\mu\nu}\gamma^{\mu\nu}+\fft{ig}{2}\sum_{i=1}^3\partial_{\phi^x}X^i\right]\epsilon.
	\end{align}\label{N=2:sugra:BPS}%
\end{subequations}
%

\subsection*{BPS equations for a black string ansatz}\label{App:N=2:BPS}
For the metric in a black string ansatz (\ref{BS:ansatz:metric}), the spin connections are given as
\begin{equation}
\begin{split}
	\omega_0^{02}&=f_1'e^{-f_3},\quad\omega_0^{12}=-\omega_1^{02}=-\omega_2^{01}=\fft12\Omega'e^{-f_1+f_2-f_3},\quad\omega_1^{12}=f_2'e^{-f_3},\\
	\omega_3^{23}&=\omega_4^{24}=-f_4'e^{-f_3},\quad\omega_3^{34}=e^{-f_4-h_{\mathfrak g}}\partial_yh_{\mathfrak g}{},\quad
	\omega_4^{34}=-e^{-f_4-h_{\mathfrak g}}\partial_xh_{\mathfrak g}{}.
\end{split}
\end{equation}
Substituting the spin connections into (\ref{N=2:sugra:BPS}), assuming differential conditions $\partial_{x,y}\epsilon=0$, and imposing the projection condition
\begin{equation}
	\gamma_{34}\epsilon=i\delta\epsilon\quad(\delta\in\{\pm1\}),\label{projection:1}
\end{equation}
the BPS equations (\ref{N=2:sugra:BPS}) can be written explicitly as
\begin{subequations}
\begin{align}
	0&=\left(e^{-f_1}\partial_t-e^{-f_1}\Omega\partial_\phi+\fft12f_1'e^{-f_3}\gamma_{02}+\fft14\Omega'e^{-f_1+f_2-f_3}\gamma_{12}-\fft{i}{12}e^{-2f_4}\sum_I\fft{p^I}{X^I}\gamma_{034}+\fft{g}{6}\sum_IX^I\gamma_0\right)\epsilon,\\
	0&=\left(e^{-f_2}\partial_\phi+\fft12f_2'e^{-f_3}\gamma_{12}-\fft14\Omega'e^{-f_1+f_2-f_3}\gamma_{02}-\fft{i}{12}e^{-2f_4}\sum_I\fft{p^I}{X^I}\gamma_{134}+\fft{g}{6}\sum_IX^I\gamma_1\right)\epsilon,\\
	0&=\left(e^{-f_3}\partial_r-\fft14\Omega'e^{-f_1+f_2-f_3}\gamma_{01}-\fft{i}{12}e^{-2f_4}\sum_I\fft{p^I}{X^I}\gamma_{234}+\fft{g}{6}\sum_IX^I\gamma_2\right)\epsilon,\\
	0&=\left(-\fft12f_4'e^{-f_3}\gamma_{23}+\fft{i}{6}e^{-2f_4}\sum_I\fft{p^I}{X^I}\gamma_4+\fft{g}{6}\sum_IX^I\gamma_3\right)\epsilon,\\
	0&=\left(-\fft{i}{2}e^{-f_3}\gamma_2\partial_r\phi^x+\fft12e^{-2f_4}\gamma_{34}\sum_I\fft{p^I}{X^I}c^I{}_x+ ig\sum_Ic^I{}_xX^I\right)\epsilon,\\
	0&=\delta\kappa+g\sum_Ip^I.
\end{align}\label{BPS:BS:1}%
\end{subequations}
Using
\begin{equation}
	\gamma_4=i\beta\gamma_{0123}\quad(\beta\in\{\pm1\})\label{gamma:4}
\end{equation}
for 5d gamma matrices and an extra projection condition
\begin{equation}
	\gamma_2\epsilon=\alpha\epsilon\quad(\alpha\in\{\pm1\}),\label{projection:2}
\end{equation}
one can simplify (\ref{BPS:BS:1}) further as
\begin{subequations}
\begin{align}
	0&=\left(e^{-f_1}\partial_t-e^{-f_1}\Omega\partial_\phi+\left(\fft{\alpha}{2}f_1'e^{-f_3}+\fft{\beta\delta}{4}\Omega'e^{-f_1+f_2-f_3}+\fft{\delta}{12}e^{-2f_4}\sum_I\fft{p^I}{X^I}+\fft{g}{6}\sum_IX^I\right)\gamma_0\right)\epsilon,\label{BPS:BS:2:1}\\
	0&=\left(e^{-f_2}\partial_\phi+\left(\fft{\alpha}{2}f_2'e^{-f_3}-\fft{\beta\delta}{4}\Omega'e^{-f_1+f_2-f_3}+\fft{\delta}{12}e^{-2f_4}\sum_I\fft{p^I}{X^I}+\fft{g}{6}\sum_IX^I\right)\gamma_1\right)\epsilon,\label{BPS:BS:2:2}\\
	0&=\left(e^{-f_3}\partial_r+\alpha\left(\fft{\beta\delta}{4}\Omega'e^{-f_1+f_2-f_3}+\fft{\delta}{12}e^{-2f_4}\sum_I\fft{p^I}{X^I}+\fft{g}{6}\sum_IX^I\right)\right)\epsilon,\label{BPS:BS:2:3}\\
	0&=\alpha f_4'e^{-f_3}-\fft{\delta}{3}e^{-2f_4}\sum_I\fft{p^I}{X^I}+\fft{g}{3}\sum_IX^I,\label{BPS:BS:2:4}\\
	0&=-\fft{\alpha}{2}e^{-f_3}\partial_r\phi^x+\fft{\delta}{2}e^{-2f_4}\sum_I\fft{p^I}{X^I}c^I{}_x+g\sum_Ic^I{}_xX^I,\label{BPS:BS:2:5}\\
	0&=\delta\kappa+g\sum_Ip^I.\label{BPS:BS:2:6}
\end{align}\label{BPS:BS:2}%
\end{subequations}

Here we list several solutions of the BPS equations (\ref{BPS:BS:2}) with $g=1$ mentioned in the main text. 
\begin{itemize}
	\item A near-horizon solution, (\ref{AdS3:NH}), solves the BPS equations (\ref{BPS:BS:2}) with
	\begin{equation}
		\epsilon=e^{\fft12f_1(r)}\epsilon_0,\quad\partial_t\epsilon=\partial_\phi\epsilon=0,\quad\alpha=-1,\quad\beta=\delta=1,\label{BPS:BS:sol1}
	\end{equation}
	where $f_{1,2,3,4}(r),\Omega(r)$ can be read off by matching the solution (\ref{AdS3:NH}) with the ansatz (\ref{BS:ansatz:metric}).
	\item A full black string solution with identical magnetic charges, (\ref{AdS5:BS}), solves the BPS equations (\ref{BPS:BS:2}) with the same conditions (\ref{BPS:BS:sol1}) where $f_{1,2,3,4}(r),\Omega(r)$ can be read off by matching the solution (\ref{AdS5:BS}) with the ansatz (\ref{BS:ansatz:metric}).
	\item A near-horizon solution, (\ref{family:sol}), solves the BPS equations (\ref{BPS:BS:2}) with
	\begin{equation}
		\epsilon=e^{\fft{t'-2\rho_+\phi'}{2}\gamma_1}\epsilon_0,\quad\partial_r\epsilon=0,\quad\beta=-1,\quad\delta=1,\label{BPS:BS:sol2}
	\end{equation}
	where $f_{1,2,3,4}(r),\Omega(r)$ can be read off by matching the solution (\ref{family:sol}) with the ansatz (\ref{BS:ansatz:metric}). Note that we have used $t',\phi'$ coordinates instead of $t,\phi$ coordinates here.
\end{itemize}
%

\section{Coordinate systems of a locally AdS$_3$ geometry}\label{App:AdS3}
In this appendix, we introduce various locally AdS$_3$ coordinates used in the main text and coordinate transformations between them. The starting point is the Poincar\'{e} coordinates of an AdS$_3$, namely
\begin{equation}
	ds^2=\fft{-dw^2+dz^2+dl^2}{l^2}.\label{Poincare}
\end{equation}
%

\subsection*{BTZ metric}
Using the coordinate transformation
\begin{equation}
	\pm w+z=\sqrt{\fft{\rho^2-\rho_+^2}{\rho^2-\rho_-^2}}e^{(\rho_+\mp\rho_-)(\phi\pm t)},\qquad l=\sqrt{\fft{\rho_+^2-\rho_-^2}{\rho^2-\rho_-^2}}e^{\rho_+\phi-\rho_-t},\label{Poincare:BTZ}
\end{equation}
the Poincar\'{e} metric (\ref{Poincare}) can be rewritten in terms of a BTZ metric as
\begin{equation}
	ds^2=-\fft{(\rho^2-\rho_+^2)(\rho^2-\rho_-^2)}{\rho^2}dt^2+\rho^2\left(d\phi-\fft{\rho_+\rho_-}{\rho^2}dt\right)^2+\fft{\rho^2}{(\rho^2-\rho_+^2)(\rho^2-\rho_-^2)}d\rho^2.\label{BTZ}
\end{equation}
%

\subsection*{Near-horizon limit of an extremal BTZ metric}
Using the coordinate transformation
\begin{equation}
	-w+z=\fft12e^{2\rho_+\phi},\qquad w+z=t-\fft1r,\qquad l=r^{-\fft12}e^{\rho_+\phi},\label{Poincare:ext}
\end{equation}
the Poincar\'{e} metric (\ref{Poincare}) can be rewritten as
\begin{equation}
	ds^2=\fft14\left(-r^2dt^2+4\rho_+^2\left(d\phi+\fft{r}{2\rho_+}dt\right)^2+\fft{dr^2}{r^2}\right).\label{ext}
\end{equation}
Note that (\ref{ext}) can also be interpreted as the near-horizon limit of an extremal BTZ black metric: one can reproduce (\ref{ext}) from a BTZ metric (\ref{BTZ}) with the extremal condition $\rho_+=\rho_-$ by taking $\epsilon\to0$ after the following coordinate transformations (see \cite{Banerjee:2019vff} for example)
\begin{equation}
	t\to\fft{t}{4\epsilon},\qquad \phi\to\fft{t}{4\epsilon}+\phi,\qquad \rho=\rho_++\epsilon r.
\end{equation}
%

\subsection*{Near-horizon \& extremal limit of a BTZ metric}
Using the coordinate transformation
\begin{equation}
	w+z=\sqrt{\fft{r-1}{r+1}}e^{t},\qquad-w+z=\sqrt{\fft{r-1}{r+1}}e^{-t+2\rho_+\phi},\qquad l=\sqrt{\fft{2}{r+1}}e^{\rho_+\phi},\label{Poincare:ext-NH}
\end{equation}
the Poincar\'{e} metric (\ref{Poincare}) can be rewritten as
\begin{equation}
	ds^2=\fft14\left(-(r^2-1)dt^2+4\rho_+^2\left(d\phi+\fft{r-1}{2\rho_+}dt\right)^2+\fft{dr^2}{r^2-1}\right).\label{NH}
\end{equation}

\bibliographystyle{JHEP}
\bibliography{Large-N}

\providecommand{\href}[2]{#2}\begingroup\raggedright\begin{thebibliography}{10}

\bibitem{Dijkgraaf:2000fq}
R.~Dijkgraaf, J.~M. Maldacena, G.~W. Moore and E.~P. Verlinde, \emph{{A Black
  hole Farey tail}},  \href{https://arxiv.org/abs/hep-th/0005003}{{\ttfamily
  hep-th/0005003}}.

\bibitem{Bekenstein:1973ur}
J.~D. Bekenstein, \emph{{Black holes and entropy}},
  \href{https://doi.org/10.1103/PhysRevD.7.2333}{\emph{Phys. Rev. D} {\bfseries
  7} (1973) 2333}.

\bibitem{Hawking:1974sw}
S.~Hawking, \emph{{Particle Creation by Black Holes}},
  \href{https://doi.org/10.1007/BF02345020}{\emph{Commun. Math. Phys.}
  {\bfseries 43} (1975) 199}.

\bibitem{Strominger:1996sh}
A.~Strominger and C.~Vafa, \emph{{Microscopic origin of the Bekenstein-Hawking
  entropy}}, \href{https://doi.org/10.1016/0370-2693(96)00345-0}{\emph{Phys.
  Lett. B} {\bfseries 379} (1996) 99}
  [\href{https://arxiv.org/abs/hep-th/9601029}{{\ttfamily hep-th/9601029}}].

\bibitem{Banados:1992wn}
M.~Banados, C.~Teitelboim and J.~Zanelli, \emph{{The Black hole in
  three-dimensional space-time}},
  \href{https://doi.org/10.1103/PhysRevLett.69.1849}{\emph{Phys. Rev. Lett.}
  {\bfseries 69} (1992) 1849}
  [\href{https://arxiv.org/abs/hep-th/9204099}{{\ttfamily hep-th/9204099}}].

\bibitem{Strominger:1997eq}
A.~Strominger, \emph{{Black hole entropy from near horizon microstates}},
  \href{https://doi.org/10.1088/1126-6708/1998/02/009}{\emph{JHEP} {\bfseries
  02} (1998) 009} [\href{https://arxiv.org/abs/hep-th/9712251}{{\ttfamily
  hep-th/9712251}}].

\bibitem{Benini:2015eyy}
F.~Benini, K.~Hristov and A.~Zaffaroni, \emph{{Black hole microstates in
  AdS$_{4}$ from supersymmetric localization}},
  \href{https://doi.org/10.1007/JHEP05(2016)054}{\emph{JHEP} {\bfseries 05}
  (2016) 054} [\href{https://arxiv.org/abs/1511.04085}{{\ttfamily
  1511.04085}}].

\bibitem{Pestun:2007rz}
V.~Pestun, \emph{{Localization of gauge theory on a four-sphere and
  supersymmetric Wilson loops}},
  \href{https://doi.org/10.1007/s00220-012-1485-0}{\emph{Commun. Math. Phys.}
  {\bfseries 313} (2012) 71} [\href{https://arxiv.org/abs/0712.2824}{{\ttfamily
  0712.2824}}].

\bibitem{Benini:2015noa}
F.~Benini and A.~Zaffaroni, \emph{{A topologically twisted index for
  three-dimensional supersymmetric theories}},
  \href{https://doi.org/10.1007/JHEP07(2015)127}{\emph{JHEP} {\bfseries 07}
  (2015) 127} [\href{https://arxiv.org/abs/1504.03698}{{\ttfamily
  1504.03698}}].

\bibitem{Hosseini:2017mds}
S.~M. Hosseini, K.~Hristov and A.~Zaffaroni, \emph{{An extremization principle
  for the entropy of rotating BPS black holes in AdS$_{5}$}},
  \href{https://doi.org/10.1007/JHEP07(2017)106}{\emph{JHEP} {\bfseries 07}
  (2017) 106} [\href{https://arxiv.org/abs/1705.05383}{{\ttfamily
  1705.05383}}].

\bibitem{Benini:2016rke}
F.~Benini, K.~Hristov and A.~Zaffaroni, \emph{{Exact microstate counting for
  dyonic black holes in AdS4}},
  \href{https://doi.org/10.1016/j.physletb.2017.05.076}{\emph{Phys. Lett. B}
  {\bfseries 771} (2017) 462}
  [\href{https://arxiv.org/abs/1608.07294}{{\ttfamily 1608.07294}}].

\bibitem{Benini:2016hjo}
F.~Benini and A.~Zaffaroni, \emph{{Supersymmetric partition functions on
  Riemann surfaces}}, {\emph{Proc. Symp. Pure Math.} {\bfseries 96} (2017) 13}
  [\href{https://arxiv.org/abs/1605.06120}{{\ttfamily 1605.06120}}].

\bibitem{Cabo-Bizet:2017jsl}
A.~Cabo-Bizet, V.~I. Giraldo-Rivera and L.~A. Pando~Zayas, \emph{{Microstate
  counting of AdS$_{4}$ hyperbolic black hole entropy via the topologically
  twisted index}}, \href{https://doi.org/10.1007/JHEP08(2017)023}{\emph{JHEP}
  {\bfseries 08} (2017) 023}
  [\href{https://arxiv.org/abs/1701.07893}{{\ttfamily 1701.07893}}].

\bibitem{Hosseini:2017fjo}
S.~M. Hosseini, K.~Hristov and A.~Passias, \emph{{Holographic microstate
  counting for AdS$_{4}$ black holes in massive IIA supergravity}},
  \href{https://doi.org/10.1007/JHEP10(2017)190}{\emph{JHEP} {\bfseries 10}
  (2017) 190} [\href{https://arxiv.org/abs/1707.06884}{{\ttfamily
  1707.06884}}].

\bibitem{Benini:2017oxt}
F.~Benini, H.~Khachatryan and P.~Milan, \emph{{Black hole entropy in massive
  Type IIA}}, \href{https://doi.org/10.1088/1361-6382/aa9f5b}{\emph{Class.
  Quant. Grav.} {\bfseries 35} (2018) 035004}
  [\href{https://arxiv.org/abs/1707.06886}{{\ttfamily 1707.06886}}].

\bibitem{Hosseini:2016ume}
S.~M. Hosseini and N.~Mekareeya, \emph{{Large $N$ topologically twisted index:
  necklace quivers, dualities, and Sasaki-Einstein spaces}},
  \href{https://doi.org/10.1007/JHEP08(2016)089}{\emph{JHEP} {\bfseries 08}
  (2016) 089} [\href{https://arxiv.org/abs/1604.03397}{{\ttfamily
  1604.03397}}].

\bibitem{Hosseini:2016tor}
S.~M. Hosseini and A.~Zaffaroni, \emph{{Large $N$ matrix models for 3d ${\cal
  N}=2$ theories: twisted index, free energy and black holes}},
  \href{https://doi.org/10.1007/JHEP08(2016)064}{\emph{JHEP} {\bfseries 08}
  (2016) 064} [\href{https://arxiv.org/abs/1604.03122}{{\ttfamily
  1604.03122}}].

\bibitem{Hosseini:2018uzp}
S.~M. Hosseini, I.~Yaakov and A.~Zaffaroni, \emph{{Topologically twisted
  indices in five dimensions and holography}},
  \href{https://doi.org/10.1007/JHEP11(2018)119}{\emph{JHEP} {\bfseries 11}
  (2018) 119} [\href{https://arxiv.org/abs/1808.06626}{{\ttfamily
  1808.06626}}].

\bibitem{Fluder:2019szh}
M.~Fluder, S.~M. Hosseini and C.~F. Uhlemann, \emph{{Black hole microstate
  counting in Type IIB from 5d SCFTs}},
  \href{https://doi.org/10.1007/JHEP05(2019)134}{\emph{JHEP} {\bfseries 05}
  (2019) 134} [\href{https://arxiv.org/abs/1902.05074}{{\ttfamily
  1902.05074}}].

\bibitem{Suh:2018tul}
M.~Suh, \emph{{Supersymmetric AdS$_{6}$ black holes from F(4) gauged
  supergravity}}, \href{https://doi.org/10.1007/JHEP01(2019)035}{\emph{JHEP}
  {\bfseries 01} (2019) 035}
  [\href{https://arxiv.org/abs/1809.03517}{{\ttfamily 1809.03517}}].

\bibitem{Suh:2018szn}
M.~Suh, \emph{{Supersymmetric $AdS_6$ black holes from matter coupled $F(4)$
  gauged supergravity}},
  \href{https://doi.org/10.1007/JHEP02(2019)108}{\emph{JHEP} {\bfseries 02}
  (2019) 108} [\href{https://arxiv.org/abs/1810.00675}{{\ttfamily
  1810.00675}}].

\bibitem{Honda:2015yha}
M.~Honda and Y.~Yoshida, \emph{{Supersymmetric index on $T^2 \times S^2$ and
  elliptic genus}},  \href{https://arxiv.org/abs/1504.04355}{{\ttfamily
  1504.04355}}.

\bibitem{Hosseini:2016cyf}
S.~M. Hosseini, A.~Nedelin and A.~Zaffaroni, \emph{{The Cardy limit of the
  topologically twisted index and black strings in AdS$_{5}$}},
  \href{https://doi.org/10.1007/JHEP04(2017)014}{\emph{JHEP} {\bfseries 04}
  (2017) 014} [\href{https://arxiv.org/abs/1611.09374}{{\ttfamily
  1611.09374}}].

\bibitem{Hosseini:2019lkt}
S.~M. Hosseini, K.~Hristov and A.~Zaffaroni, \emph{{Microstates of rotating
  AdS$_{5}$ strings}},
  \href{https://doi.org/10.1007/JHEP11(2019)090}{\emph{JHEP} {\bfseries 11}
  (2019) 090} [\href{https://arxiv.org/abs/1909.08000}{{\ttfamily
  1909.08000}}].

\bibitem{Hosseini:2018tha}
S.~Hosseini, \emph{{Black hole microstates and supersymmetric localization}},
  Ph.D. thesis, Milan Bicocca U., 2, 2018.
\newblock \href{https://arxiv.org/abs/1803.01863}{{\ttfamily 1803.01863}}.

\bibitem{Zaffaroni:2019dhb}
A.~Zaffaroni, \emph{{AdS black holes, holography and localization}},
  \href{https://doi.org/10.1007/s41114-020-00027-8}{\emph{Living Rev. Rel.}
  {\bfseries 23} (2020) 2} [\href{https://arxiv.org/abs/1902.07176}{{\ttfamily
  1902.07176}}].

\bibitem{Hosseini:2020vgl}
S.~M. Hosseini, K.~Hristov, Y.~Tachikawa and A.~Zaffaroni, \emph{{Anomalies,
  Black strings and the charged Cardy formula}},
  \href{https://doi.org/10.1007/JHEP09(2020)167}{\emph{JHEP} {\bfseries 09}
  (2020) 167} [\href{https://arxiv.org/abs/2006.08629}{{\ttfamily
  2006.08629}}].

\bibitem{Hristov:2014eza}
K.~Hristov, \emph{{Dimensional reduction of BPS attractors in AdS gauged
  supergravities}}, \href{https://doi.org/10.1007/JHEP12(2014)066}{\emph{JHEP}
  {\bfseries 12} (2014) 066} [\href{https://arxiv.org/abs/1409.8504}{{\ttfamily
  1409.8504}}].

\bibitem{Hong:2021dja}
J.~Hong, \emph{{The $N$ = 4 SU($N$) Super-Yang-Mills Index and Dual AdS Black
  Holes}}, Ph.D. thesis, Michigan U., 2021.
\newblock 10.7302/1496.

\bibitem{David:2021qaa}
M.~David, A.~G. Lezcano, J.~Nian and L.~A.~P. Zayas, \emph{{Logarithmic
  Corrections to the Entropy of Rotating Black Holes and Black Strings in
  AdS$_5$}},  \href{https://arxiv.org/abs/2106.09730}{{\ttfamily 2106.09730}}.

\bibitem{Hong:2018viz}
J.~Hong and J.~T. Liu, \emph{{The topologically twisted index of $ \mathcal{N}
  $ = 4 super-Yang-Mills on T$^{2} \times S^{2}$ and the elliptic genus}},
  \href{https://doi.org/10.1007/JHEP07(2018)018}{\emph{JHEP} {\bfseries 07}
  (2018) 018} [\href{https://arxiv.org/abs/1804.04592}{{\ttfamily
  1804.04592}}].

\bibitem{Aharony:2021zkr}
O.~Aharony, F.~Benini, O.~Mamroud and P.~Milan, \emph{{A gravity interpretation
  for the Bethe Ansatz expansion of the $\mathcal{N}=4$ SYM index}},
  \href{https://arxiv.org/abs/2104.13932}{{\ttfamily 2104.13932}}.

\bibitem{ArabiArdehali:2021nsx}
A.~Arabi~Ardehali and S.~Murthy, \emph{{The 4d superconformal index near roots
  of unity and 3d Chern-Simons theory}},
  \href{https://arxiv.org/abs/2104.02051}{{\ttfamily 2104.02051}}.

\bibitem{Maldacena:1998bw}
J.~M. Maldacena and A.~Strominger, \emph{{AdS(3) black holes and a stringy
  exclusion principle}},
  \href{https://doi.org/10.1088/1126-6708/1998/12/005}{\emph{JHEP} {\bfseries
  12} (1998) 005} [\href{https://arxiv.org/abs/hep-th/9804085}{{\ttfamily
  hep-th/9804085}}].

\bibitem{Murthy:2009dq}
S.~Murthy and B.~Pioline, \emph{{A Farey tale for N=4 dyons}},
  \href{https://doi.org/10.1088/1126-6708/2009/09/022}{\emph{JHEP} {\bfseries
  09} (2009) 022} [\href{https://arxiv.org/abs/0904.4253}{{\ttfamily
  0904.4253}}].

\bibitem{Benini:2013nda}
F.~Benini, R.~Eager, K.~Hori and Y.~Tachikawa, \emph{{Elliptic genera of
  two-dimensional N=2 gauge theories with rank-one gauge groups}},
  \href{https://doi.org/10.1007/s11005-013-0673-y}{\emph{Lett. Math. Phys.}
  {\bfseries 104} (2014) 465}
  [\href{https://arxiv.org/abs/1305.0533}{{\ttfamily 1305.0533}}].

\bibitem{Benini:2013xpa}
F.~Benini, R.~Eager, K.~Hori and Y.~Tachikawa, \emph{{Elliptic Genera of 2d
  ${\mathcal{N}}$ = 2 Gauge Theories}},
  \href{https://doi.org/10.1007/s00220-014-2210-y}{\emph{Commun. Math. Phys.}
  {\bfseries 333} (2015) 1241}
  [\href{https://arxiv.org/abs/1308.4896}{{\ttfamily 1308.4896}}].

\bibitem{ArabiArdehali:2019orz}
A.~Arabi~Ardehali, J.~Hong and J.~T. Liu, \emph{{Asymptotic growth of the 4d $
  \mathcal{N} $ = 4 index and partially deconfined phases}},
  \href{https://doi.org/10.1007/JHEP07(2020)073}{\emph{JHEP} {\bfseries 07}
  (2020) 073} [\href{https://arxiv.org/abs/1912.04169}{{\ttfamily
  1912.04169}}].

\bibitem{Lezcano:2021qbj}
A.~G. Lezcano, J.~Hong, J.~T. Liu and L.~A.~P. Zayas, \emph{{The Bethe-Ansatz
  approach to the $\mathcal N=4$ superconformal index at finite rank}},
  \href{https://arxiv.org/abs/2101.12233}{{\ttfamily 2101.12233}}.

\bibitem{Benini:2021ano}
F.~Benini and G.~Rizi, \emph{{Superconformal index of low-rank gauge theories
  via the Bethe Ansatz}},
  \href{https://doi.org/10.1007/JHEP05(2021)061}{\emph{JHEP} {\bfseries 05}
  (2021) 061} [\href{https://arxiv.org/abs/2102.03638}{{\ttfamily
  2102.03638}}].

\bibitem{GonzalezLezcano:2020yeb}
A.~Gonz\'alez~Lezcano, J.~Hong, J.~T. Liu and L.~A. Pando~Zayas,
  \emph{{Sub-leading Structures in Superconformal Indices: Subdominant Saddles
  and Logarithmic Contributions}},
  \href{https://doi.org/10.1007/JHEP01(2021)001}{\emph{JHEP} {\bfseries 01}
  (2021) 001} [\href{https://arxiv.org/abs/2007.12604}{{\ttfamily
  2007.12604}}].

\bibitem{Benini:2018ywd}
F.~Benini and P.~Milan, \emph{{Black Holes in 4D $\mathcal{N}$=4
  Super-Yang-Mills Field Theory}},
  \href{https://doi.org/10.1103/PhysRevX.10.021037}{\emph{Phys. Rev. X}
  {\bfseries 10} (2020) 021037}
  [\href{https://arxiv.org/abs/1812.09613}{{\ttfamily 1812.09613}}].

\bibitem{Cabo-Bizet:2018ehj}
A.~Cabo-Bizet, D.~Cassani, D.~Martelli and S.~Murthy, \emph{{Microscopic origin
  of the Bekenstein-Hawking entropy of supersymmetric AdS$_{5}$ black holes}},
  \href{https://doi.org/10.1007/JHEP10(2019)062}{\emph{JHEP} {\bfseries 10}
  (2019) 062} [\href{https://arxiv.org/abs/1810.11442}{{\ttfamily
  1810.11442}}].

\bibitem{Bobev:2020pjk}
N.~Bobev, A.~M. Charles and V.~S. Min, \emph{{Euclidean black saddles and
  AdS$_{4}$ black holes}},
  \href{https://doi.org/10.1007/JHEP10(2020)073}{\emph{JHEP} {\bfseries 10}
  (2020) 073} [\href{https://arxiv.org/abs/2006.01148}{{\ttfamily
  2006.01148}}].

\bibitem{Hristov:2014hza}
K.~Hristov and S.~Katmadas, \emph{{Wilson lines for AdS$_{5}$ black strings}},
  \href{https://doi.org/10.1007/JHEP02(2015)009}{\emph{JHEP} {\bfseries 02}
  (2015) 009} [\href{https://arxiv.org/abs/1411.2432}{{\ttfamily 1411.2432}}].

\bibitem{Gauntlett:2003fk}
J.~P. Gauntlett and J.~B. Gutowski, \emph{{All supersymmetric solutions of
  minimal gauged supergravity in five-dimensions}},
  \href{https://doi.org/10.1103/PhysRevD.70.089901}{\emph{Phys. Rev. D}
  {\bfseries 68} (2003) 105009}
  [\href{https://arxiv.org/abs/hep-th/0304064}{{\ttfamily hep-th/0304064}}].

\bibitem{Gutowski:2004yv}
J.~B. Gutowski and H.~S. Reall, \emph{{General supersymmetric AdS(5) black
  holes}}, \href{https://doi.org/10.1088/1126-6708/2004/04/048}{\emph{JHEP}
  {\bfseries 04} (2004) 048}
  [\href{https://arxiv.org/abs/hep-th/0401129}{{\ttfamily hep-th/0401129}}].

\bibitem{Benini:2013cda}
F.~Benini and N.~Bobev, \emph{{Two-dimensional SCFTs from wrapped branes and
  c-extremization}}, \href{https://doi.org/10.1007/JHEP06(2013)005}{\emph{JHEP}
  {\bfseries 06} (2013) 005} [\href{https://arxiv.org/abs/1302.4451}{{\ttfamily
  1302.4451}}].

\bibitem{Klemm:2000nj}
D.~Klemm and W.~Sabra, \emph{{Supersymmetry of black strings in D = 5 gauged
  supergravities}},
  \href{https://doi.org/10.1103/PhysRevD.62.024003}{\emph{Phys. Rev. D}
  {\bfseries 62} (2000) 024003}
  [\href{https://arxiv.org/abs/hep-th/0001131}{{\ttfamily hep-th/0001131}}].

\bibitem{Cacciatori:2003kv}
S.~L. Cacciatori, D.~Klemm and W.~A. Sabra, \emph{{Supersymmetric domain walls
  and strings in D = 5 gauged supergravity coupled to vector multiplets}},
  \href{https://doi.org/10.1088/1126-6708/2003/03/023}{\emph{JHEP} {\bfseries
  03} (2003) 023} [\href{https://arxiv.org/abs/hep-th/0302218}{{\ttfamily
  hep-th/0302218}}].

\bibitem{Bernamonti:2007bu}
A.~Bernamonti, M.~M. Caldarelli, D.~Klemm, R.~Olea, C.~Sieg and E.~Zorzan,
  \emph{{Black strings in AdS(5)}},
  \href{https://doi.org/10.1088/1126-6708/2008/01/061}{\emph{JHEP} {\bfseries
  01} (2008) 061} [\href{https://arxiv.org/abs/0708.2402}{{\ttfamily
  0708.2402}}].

\bibitem{Azzola:2018sld}
M.~Azzola, D.~Klemm and M.~Rabbiosi, \emph{{AdS$_5$ black strings in the stu
  model of FI-gauged $N=2$ supergravity}},
  \href{https://doi.org/10.1007/JHEP10(2018)080}{\emph{JHEP} {\bfseries 10}
  (2018) 080} [\href{https://arxiv.org/abs/1803.03570}{{\ttfamily
  1803.03570}}].

\bibitem{Banados:1992gq}
M.~Banados, M.~Henneaux, C.~Teitelboim and J.~Zanelli, \emph{{Geometry of the
  (2+1) black hole}},
  \href{https://doi.org/10.1103/PhysRevD.48.1506}{\emph{Phys. Rev. D}
  {\bfseries 48} (1993) 1506}
  [\href{https://arxiv.org/abs/gr-qc/9302012}{{\ttfamily gr-qc/9302012}}].

\bibitem{Banerjee:2008ky}
N.~Banerjee, D.~P. Jatkar and A.~Sen, \emph{{Asymptotic Expansion of the N=4
  Dyon Degeneracy}},
  \href{https://doi.org/10.1088/1126-6708/2009/05/121}{\emph{JHEP} {\bfseries
  05} (2009) 121} [\href{https://arxiv.org/abs/0810.3472}{{\ttfamily
  0810.3472}}].

\bibitem{Carroll:2004st}
S.~M. Carroll, \emph{{Spacetime and Geometry}}. Cambridge University Press, 7,
  2019.

\bibitem{Liu:2007rv}
J.~T. Liu, H.~Lu, C.~N. Pope and J.~F. Vazquez-Poritz, \emph{{New
  supersymmetric solutions of N=2, D=5 gauged supergravity with hyperscalars}},
  \href{https://doi.org/10.1088/1126-6708/2007/10/093}{\emph{JHEP} {\bfseries
  10} (2007) 093} [\href{https://arxiv.org/abs/0705.2234}{{\ttfamily
  0705.2234}}].

\bibitem{Bobev:2013cja}
N.~Bobev, H.~Elvang, D.~Z. Freedman and S.~S. Pufu, \emph{{Holography for $N =
  2^*$ on $S^4$}}, \href{https://doi.org/10.1007/JHEP07(2014)001}{\emph{JHEP}
  {\bfseries 07} (2014) 001} [\href{https://arxiv.org/abs/1311.1508}{{\ttfamily
  1311.1508}}].

\bibitem{Maloney:2007ud}
A.~Maloney and E.~Witten, \emph{{Quantum Gravity Partition Functions in Three
  Dimensions}}, \href{https://doi.org/10.1007/JHEP02(2010)029}{\emph{JHEP}
  {\bfseries 02} (2010) 029} [\href{https://arxiv.org/abs/0712.0155}{{\ttfamily
  0712.0155}}].

\bibitem{Maxfield:2020ale}
H.~Maxfield and G.~J. Turiaci, \emph{{The path integral of 3D gravity near
  extremality; or, JT gravity with defects as a matrix integral}},
  \href{https://doi.org/10.1007/JHEP01(2021)118}{\emph{JHEP} {\bfseries 01}
  (2021) 118} [\href{https://arxiv.org/abs/2006.11317}{{\ttfamily
  2006.11317}}].

\bibitem{Hristov:2018spe}
K.~Hristov, S.~Katmadas and C.~Toldo, \emph{{Rotating attractors and BPS black
  holes in $AdS_4$}},
  \href{https://doi.org/10.1007/JHEP01(2019)199}{\emph{JHEP} {\bfseries 01}
  (2019) 199} [\href{https://arxiv.org/abs/1811.00292}{{\ttfamily
  1811.00292}}].

\bibitem{Ferrero:2020laf}
P.~Ferrero, J.~P. Gauntlett, J.~M. P\'erez Ipi\~na, D.~Martelli and J.~Sparks,
  \emph{{D3-Branes Wrapped on a Spindle}},
  \href{https://doi.org/10.1103/PhysRevLett.126.111601}{\emph{Phys. Rev. Lett.}
  {\bfseries 126} (2021) 111601}
  [\href{https://arxiv.org/abs/2011.10579}{{\ttfamily 2011.10579}}].

\bibitem{Hosseini:2021fge}
S.~M. Hosseini, K.~Hristov and A.~Zaffaroni, \emph{{Rotating multi-charge
  spindles and their microstates}},
  \href{https://arxiv.org/abs/2104.11249}{{\ttfamily 2104.11249}}.

\bibitem{Kinney:2005ej}
J.~Kinney, J.~M. Maldacena, S.~Minwalla and S.~Raju, \emph{{An Index for 4
  dimensional super conformal theories}},
  \href{https://doi.org/10.1007/s00220-007-0258-7}{\emph{Commun. Math. Phys.}
  {\bfseries 275} (2007) 209}
  [\href{https://arxiv.org/abs/hep-th/0510251}{{\ttfamily hep-th/0510251}}].

\bibitem{Romelsberger:2005eg}
C.~Romelsberger, \emph{{Counting chiral primaries in N = 1, d=4 superconformal
  field theories}},
  \href{https://doi.org/10.1016/j.nuclphysb.2006.03.037}{\emph{Nucl. Phys. B}
  {\bfseries 747} (2006) 329}
  [\href{https://arxiv.org/abs/hep-th/0510060}{{\ttfamily hep-th/0510060}}].

\bibitem{Benini:2020gjh}
F.~Benini, E.~Colombo, S.~Soltani, A.~Zaffaroni and Z.~Zhang,
  \emph{{Superconformal indices at large $N$ and the entropy of AdS$_5$
  $\times$ SE$_5$ black holes}},
  \href{https://doi.org/10.1088/1361-6382/abb39b}{\emph{Class. Quant. Grav.}
  {\bfseries 37} (2020) 215021}
  [\href{https://arxiv.org/abs/2005.12308}{{\ttfamily 2005.12308}}].

\bibitem{Cabo-Bizet:2019eaf}
A.~Cabo-Bizet and S.~Murthy, \emph{{Supersymmetric phases of 4d $ \mathcal{N} $
  = 4 SYM at large $N$}},
  \href{https://doi.org/10.1007/JHEP09(2020)184}{\emph{JHEP} {\bfseries 09}
  (2020) 184} [\href{https://arxiv.org/abs/1909.09597}{{\ttfamily
  1909.09597}}].

\bibitem{Kunduri:2006ek}
H.~K. Kunduri, J.~Lucietti and H.~S. Reall, \emph{{Supersymmetric multi-charge
  AdS(5) black holes}},
  \href{https://doi.org/10.1088/1126-6708/2006/04/036}{\emph{JHEP} {\bfseries
  04} (2006) 036} [\href{https://arxiv.org/abs/hep-th/0601156}{{\ttfamily
  hep-th/0601156}}].

\bibitem{David:2020ems}
M.~David, J.~Nian and L.~A. Pando~Zayas, \emph{{Gravitational Cardy Limit and
  AdS Black Hole Entropy}},
  \href{https://doi.org/10.1007/JHEP11(2020)041}{\emph{JHEP} {\bfseries 11}
  (2020) 041} [\href{https://arxiv.org/abs/2005.10251}{{\ttfamily
  2005.10251}}].

\bibitem{Jejjala:2021hlt}
V.~Jejjala, Y.~Lei, S.~van Leuven and W.~Li, \emph{{$SL(3,\mathbb{Z})$
  Modularity and New Cardy Limits of the $\mathcal{N}=4$ Superconformal
  Index}},  \href{https://arxiv.org/abs/2104.07030}{{\ttfamily 2104.07030}}.

\bibitem{Brezhnev:2013}
Y.~V. Brezhnev, \emph{{Non-canonical extension of $\theta$-functions and
  modular integrability of $\theta$-constants}},
  \href{https://doi.org/10.1017/S0308210512001023}{\emph{Proceedings of the
  Royal Society of Edinburgh Section A: Mathematics} {\bfseries 143(4)} (2013)
  689} [\href{https://arxiv.org/abs/math/1011.1643}{{\ttfamily
  math/1011.1643}}].

\bibitem{Gunaydin:1983bi}
M.~Gunaydin, G.~Sierra and P.~Townsend, \emph{{The Geometry of N=2
  Maxwell-Einstein Supergravity and Jordan Algebras}},
  \href{https://doi.org/10.1016/0550-3213(84)90142-1}{\emph{Nucl. Phys. B}
  {\bfseries 242} (1984) 244}.

\bibitem{Gunaydin:1984ak}
M.~Gunaydin, G.~Sierra and P.~Townsend, \emph{{Gauging the d = 5
  Maxwell-Einstein Supergravity Theories: More on Jordan Algebras}},
  \href{https://doi.org/10.1016/0550-3213(85)90547-4}{\emph{Nucl. Phys. B}
  {\bfseries 253} (1985) 573}.

\bibitem{Cvetic:1999xp}
M.~Cvetic, M.~Duff, P.~Hoxha, J.~T. Liu, H.~Lu, J.~Lu et~al., \emph{{Embedding
  AdS black holes in ten-dimensions and eleven-dimensions}},
  \href{https://doi.org/10.1016/S0550-3213(99)00419-8}{\emph{Nucl. Phys. B}
  {\bfseries 558} (1999) 96}
  [\href{https://arxiv.org/abs/hep-th/9903214}{{\ttfamily hep-th/9903214}}].

\bibitem{Maldacena:2000mw}
J.~M. Maldacena and C.~Nunez, \emph{{Supergravity description of field theories
  on curved manifolds and a no go theorem}},
  \href{https://doi.org/10.1142/S0217751X01003937}{\emph{Int. J. Mod. Phys. A}
  {\bfseries 16} (2001) 822}
  [\href{https://arxiv.org/abs/hep-th/0007018}{{\ttfamily hep-th/0007018}}].

\bibitem{Banerjee:2019vff}
A.~Banerjee, A.~Kundu and R.~R. Poojary, \emph{{Rotating black holes in AdS
  spacetime, extremality, and chaos}},
  \href{https://doi.org/10.1103/PhysRevD.102.106013}{\emph{Phys. Rev. D}
  {\bfseries 102} (2020) 106013}
  [\href{https://arxiv.org/abs/1912.12996}{{\ttfamily 1912.12996}}].

\end{thebibliography}\endgroup

\end{document}